\documentclass[11pt]{article}
\usepackage{fullpage}
\usepackage{times}

\newif\iflongversion
\longversionfalse
\longversiontrue

\usepackage[T1]{fontenc}
\usepackage{amsmath}
\usepackage{comment}
\usepackage{amsfonts}
\usepackage{amssymb}
\usepackage{mathtools}
\usepackage{marvosym}
\usepackage{wrapfig}
\usepackage{tikz}
\usepackage{pgfplots}
\usepackage{graphicx}
\usepackage{cite}
\usepackage{color}
\usepackage[caption=false]{subfig}
\usepackage[inline]{enumitem}

\newcommand{\FOLR}{$\text{FOL}_{\reals}$}

\usepackage{isabelle}
\usepackage{isabellesym}
\newcommand{\snip}[4]{\expandafter\newcommand\csname #1\endcsname{#4}}
\snip{freeIn}{1}{2}{%
\isacommand{fun}\isamarkupfalse%
\ variableIsRemoved\ {\isacharcolon}{\kern0pt}{\isacharcolon}{\kern0pt}\ {\isachardoublequoteopen}nat\ {\isasymRightarrow}\ atom\ fm\ {\isasymRightarrow}\ bool{\isachardoublequoteclose}\ \isakeyword{where}\isanewline
\ \ {\isachardoublequoteopen}variableIsRemoved\ var\ {\isacharparenleft}{\kern0pt}Atom{\isacharparenleft}{\kern0pt}Eq\ p{\isacharparenright}{\kern0pt}{\isacharparenright}{\kern0pt}\ {\isacharequal}{\kern0pt}\ {\isacharparenleft}{\kern0pt}var\ {\isasymnotin}\ {\isacharparenleft}{\kern0pt}vars\ p{\isacharparenright}{\kern0pt}{\isacharparenright}{\kern0pt}{\isachardoublequoteclose}{\isacharbar}{\kern0pt}\isanewline
\ \ {\isachardoublequoteopen}variableIsRemoved\ var\ {\isacharparenleft}{\kern0pt}Atom{\isacharparenleft}{\kern0pt}Less\ p{\isacharparenright}{\kern0pt}{\isacharparenright}{\kern0pt}\ {\isacharequal}{\kern0pt}\ {\isacharparenleft}{\kern0pt}var\ {\isasymnotin}\ {\isacharparenleft}{\kern0pt}vars\ p{\isacharparenright}{\kern0pt}{\isacharparenright}{\kern0pt}{\isachardoublequoteclose}{\isacharbar}{\kern0pt}\isanewline
\ \ {\isachardoublequoteopen}variableIsRemoved\ var\ {\isacharparenleft}{\kern0pt}Atom{\isacharparenleft}{\kern0pt}Leq\ p{\isacharparenright}{\kern0pt}{\isacharparenright}{\kern0pt}\ {\isacharequal}{\kern0pt}\ {\isacharparenleft}{\kern0pt}var\ {\isasymnotin}\ {\isacharparenleft}{\kern0pt}vars\ p{\isacharparenright}{\kern0pt}{\isacharparenright}{\kern0pt}{\isachardoublequoteclose}{\isacharbar}{\kern0pt}\isanewline
\ \ {\isachardoublequoteopen}variableIsRemoved\ var\ {\isacharparenleft}{\kern0pt}Atom{\isacharparenleft}{\kern0pt}Neq\ p{\isacharparenright}{\kern0pt}{\isacharparenright}{\kern0pt}\ {\isacharequal}{\kern0pt}\ {\isacharparenleft}{\kern0pt}var\ {\isasymnotin}\ {\isacharparenleft}{\kern0pt}vars\ p{\isacharparenright}{\kern0pt}{\isacharparenright}{\kern0pt}{\isachardoublequoteclose}{\isacharbar}{\kern0pt}\isanewline
\ \ {\isachardoublequoteopen}variableIsRemoved\ var\ {\isacharparenleft}{\kern0pt}TrueF{\isacharparenright}{\kern0pt}\ {\isacharequal}{\kern0pt}\ True{\isachardoublequoteclose}{\isacharbar}{\kern0pt}\isanewline
\ \ {\isachardoublequoteopen}variableIsRemoved\ var\ {\isacharparenleft}{\kern0pt}FalseF{\isacharparenright}{\kern0pt}\ {\isacharequal}{\kern0pt}\ True{\isachardoublequoteclose}{\isacharbar}{\kern0pt}\isanewline
\ \ {\isachardoublequoteopen}variableIsRemoved\ var\ {\isacharparenleft}{\kern0pt}And\ a\ b{\isacharparenright}{\kern0pt}\ {\isacharequal}{\kern0pt}\isanewline
\ \ \ \ {\isacharparenleft}{\kern0pt}{\isacharparenleft}{\kern0pt}variableIsRemoved\ var\ a{\isacharparenright}{\kern0pt}\ {\isasymand}\ {\isacharparenleft}{\kern0pt}variableIsRemoved\ var\ b{\isacharparenright}{\kern0pt}{\isacharparenright}{\kern0pt}{\isachardoublequoteclose}{\isacharbar}{\kern0pt}\isanewline
\ \ {\isachardoublequoteopen}variableIsRemoved\ var\ {\isacharparenleft}{\kern0pt}Or\ a\ b{\isacharparenright}{\kern0pt}\ {\isacharequal}{\kern0pt}\ \isanewline
\ \ \ \ {\isacharparenleft}{\kern0pt}{\isacharparenleft}{\kern0pt}variableIsRemoved\ var\ a{\isacharparenright}{\kern0pt}\ {\isasymand}\ {\isacharparenleft}{\kern0pt}variableIsRemoved\ var\ b{\isacharparenright}{\kern0pt}{\isacharparenright}{\kern0pt}{\isachardoublequoteclose}{\isacharbar}{\kern0pt}\isanewline
\ \ {\isachardoublequoteopen}variableIsRemoved\ var\ {\isacharparenleft}{\kern0pt}Neg\ a{\isacharparenright}{\kern0pt}\ {\isacharequal}{\kern0pt}\ variableIsRemoved\ var\ a{\isachardoublequoteclose}{\isacharbar}{\kern0pt}\isanewline
\ \ {\isachardoublequoteopen}variableIsRemoved\ var\ {\isacharparenleft}{\kern0pt}ExQ\ a{\isacharparenright}{\kern0pt}\ {\isacharequal}{\kern0pt}\ variableIsRemoved\ {\isacharparenleft}{\kern0pt}var{\isacharplus}{\kern0pt}{\isadigit{1}}{\isacharparenright}{\kern0pt}\ a{\isachardoublequoteclose}{\isacharbar}{\kern0pt}\isanewline
\ \ {\isachardoublequoteopen}variableIsRemoved\ var\ {\isacharparenleft}{\kern0pt}AllQ\ a{\isacharparenright}{\kern0pt}\ {\isacharequal}{\kern0pt}\ variableIsRemoved\ {\isacharparenleft}{\kern0pt}var{\isacharplus}{\kern0pt}{\isadigit{1}}{\isacharparenright}{\kern0pt}\ a{\isachardoublequoteclose}{\isacharbar}{\kern0pt}\isanewline
\ \ {\isachardoublequoteopen}variableIsRemoved\ var\ {\isacharparenleft}{\kern0pt}AllN\ i\ a{\isacharparenright}{\kern0pt}\ {\isacharequal}{\kern0pt}\ variableIsRemoved\ {\isacharparenleft}{\kern0pt}var{\isacharplus}{\kern0pt}i{\isacharparenright}{\kern0pt}\ a{\isachardoublequoteclose}{\isacharbar}{\kern0pt}\isanewline
\ \ {\isachardoublequoteopen}variableIsRemoved\ var\ {\isacharparenleft}{\kern0pt}ExN\ i\ a{\isacharparenright}{\kern0pt}\ {\isacharequal}{\kern0pt}\ variableIsRemoved\ {\isacharparenleft}{\kern0pt}var{\isacharplus}{\kern0pt}i{\isacharparenright}{\kern0pt}\ a{\isachardoublequoteclose}%
}
\snip{freeInElimVar}{1}{2}{%
\isacommand{lemma}\isamarkupfalse%
\ elimVar{\isacharunderscore}{\kern0pt}removes{\isacharunderscore}{\kern0pt}variable{\isacharcolon}{\kern0pt}\ {\isachardoublequoteopen}variableIsRemoved\ var\ {\isacharparenleft}{\kern0pt}elimVar\ var\ L\ F\ At{\isacharparenright}{\kern0pt}{\isachardoublequoteclose}%
}
\snip{genQeEval}{1}{2}{%
\isacommand{lemma}\isamarkupfalse%
\ gen{\isacharunderscore}{\kern0pt}qe{\isacharunderscore}{\kern0pt}eval{\isacharcolon}{\kern0pt}\isanewline
\ \ \isakeyword{assumes}\ {\isachardoublequoteopen}F\ {\isacharequal}{\kern0pt}\ list{\isacharunderscore}{\kern0pt}conj\ {\isacharparenleft}{\kern0pt}map\ Atom\ L{\isacharparenright}{\kern0pt}{\isachardoublequoteclose}\isanewline
\ \ \isakeyword{assumes}\ {\isachardoublequoteopen}all{\isacharunderscore}{\kern0pt}degree{\isacharunderscore}{\kern0pt}{\isadigit{2}}\ var\ L{\isachardoublequoteclose}\isanewline
\ \ \isakeyword{assumes}\ {\isachardoublequoteopen}length\ {\isasymnu}\ {\isachargreater}{\kern0pt}\ var{\isachardoublequoteclose}\isanewline
\ \ \isakeyword{shows}\ {\isachardoublequoteopen}{\isacharparenleft}{\kern0pt}{\isasymexists}x{\isachardot}{\kern0pt}\ {\isacharparenleft}{\kern0pt}eval\ F\ {\isacharparenleft}{\kern0pt}{\isasymnu}{\isacharbrackleft}{\kern0pt}var{\isacharcolon}{\kern0pt}{\isacharequal}{\kern0pt}x{\isacharbrackright}{\kern0pt}{\isacharparenright}{\kern0pt}{\isacharparenright}{\kern0pt}{\isacharparenright}{\kern0pt}\ {\isacharequal}{\kern0pt}\isanewline
\ \ \ \ \ \ \ \ {\isacharparenleft}{\kern0pt}{\isasymforall}x{\isachardot}{\kern0pt}\ {\isacharparenleft}{\kern0pt}eval\ {\isacharparenleft}{\kern0pt}list{\isacharunderscore}{\kern0pt}disj{\isacharparenleft}{\kern0pt}\isanewline
\ \ \ \ \ \ \ \ \ \ \ \ mapNegInfinity\ var\ L\ {\isacharhash}{\kern0pt}\ \isanewline
\ \ \ \ \ \ \ \ \ \ \ \ mapElimVar\ var\ L{\isacharparenright}{\kern0pt}{\isacharparenright}{\kern0pt}\isanewline
\ \ \ \ \ \ \ \ {\isacharparenleft}{\kern0pt}{\isasymnu}{\isacharbrackleft}{\kern0pt}var{\isacharcolon}{\kern0pt}{\isacharequal}{\kern0pt}x{\isacharbrackright}{\kern0pt}{\isacharparenright}{\kern0pt}{\isacharparenright}{\kern0pt}{\isacharparenright}{\kern0pt}{\isachardoublequoteclose}%
}
\snip{correctnessEquality}{1}{2}{%
\isacommand{theorem}\isamarkupfalse%
\ VSEquality{\isacharunderscore}{\kern0pt}eval{\isacharcolon}{\kern0pt}\ {\isachardoublequoteopen}{\isasymforall}{\isasymnu}{\isachardot}{\kern0pt}\ eval\ {\isacharparenleft}{\kern0pt}VSEquality\ {\isasymphi}{\isacharparenright}{\kern0pt}\ {\isasymnu}\ {\isacharequal}{\kern0pt}\ eval\ {\isasymphi}\ {\isasymnu}{\isachardoublequoteclose}%
}
\snip{correctness}{1}{2}{%
\isacommand{theorem}\isamarkupfalse%
\ TopLevelSoundessTheorems{\isacharcolon}{\kern0pt}\ \isanewline
\ \ {\isachardoublequoteopen}{\isasymforall}{\isasymnu}{\isachardot}{\kern0pt}\ {\isacharparenleft}{\kern0pt}eval\ {\isacharparenleft}{\kern0pt}VSEquality\ {\isasymphi}{\isacharparenright}{\kern0pt}\ {\isasymnu}\ {\isacharequal}{\kern0pt}\ eval\ {\isasymphi}\ {\isasymnu}{\isacharparenright}{\kern0pt}{\isachardoublequoteclose}\isanewline
\ \ {\isachardoublequoteopen}{\isasymforall}{\isasymnu}{\isachardot}{\kern0pt}\ {\isacharparenleft}{\kern0pt}eval\ {\isacharparenleft}{\kern0pt}VSGeneral\ {\isasymphi}{\isacharparenright}{\kern0pt}\ {\isasymnu}\ {\isacharequal}{\kern0pt}\ eval\ {\isasymphi}\ {\isasymnu}{\isacharparenright}{\kern0pt}{\isachardoublequoteclose}\isanewline
\ \ {\isachardoublequoteopen}{\isasymforall}{\isasymnu}{\isachardot}{\kern0pt}\ {\isacharparenleft}{\kern0pt}eval\ {\isacharparenleft}{\kern0pt}VSLucky\ {\isasymphi}{\isacharparenright}{\kern0pt}\ {\isasymnu}\ {\isacharequal}{\kern0pt}\ eval\ {\isasymphi}\ {\isasymnu}{\isacharparenright}{\kern0pt}{\isachardoublequoteclose}\isanewline
\ \ {\isachardoublequoteopen}{\isasymforall}{\isasymnu}{\isachardot}{\kern0pt}\ {\isacharparenleft}{\kern0pt}eval\ {\isacharparenleft}{\kern0pt}VSLEG\ {\isasymphi}{\isacharparenright}{\kern0pt}\ {\isasymnu}\ {\isacharequal}{\kern0pt}\ eval\ {\isasymphi}\ {\isasymnu}{\isacharparenright}{\kern0pt}{\isachardoublequoteclose}%
}
\snip{overallAlgo}{1}{2}{%
\isacommand{theorem}\isamarkupfalse%
\ QE{\isacharunderscore}{\kern0pt}dnf{\isacharunderscore}{\kern0pt}eval{\isacharcolon}{\kern0pt}\ \isanewline
\ \ \isakeyword{assumes}\ steph\ {\isacharcolon}{\kern0pt}\ {\isachardoublequoteopen}{\isasymAnd}var\ amount\ L\ F\ {\isasymnu}{\isachardot}{\kern0pt}\isanewline
\ \ \ \ amount{\isasymle}var{\isacharplus}{\kern0pt}{\isadigit{1}}\ {\isasymLongrightarrow}\isanewline
\ \ \ \ eval\ {\isacharparenleft}{\kern0pt}ExN\ {\isacharparenleft}{\kern0pt}var{\isacharplus}{\kern0pt}{\isadigit{1}}{\isacharparenright}{\kern0pt}\ {\isacharparenleft}{\kern0pt}list{\isacharunderscore}{\kern0pt}conj\ {\isacharparenleft}{\kern0pt}map\ fm{\isachardot}{\kern0pt}Atom\ L\ {\isacharat}{\kern0pt}\ F{\isacharparenright}{\kern0pt}{\isacharparenright}{\kern0pt}{\isacharparenright}{\kern0pt}\ {\isasymnu}\ \isanewline
\ \ {\isacharequal}{\kern0pt}\ eval\ {\isacharparenleft}{\kern0pt}ExN\ {\isacharparenleft}{\kern0pt}var{\isacharplus}{\kern0pt}{\isadigit{1}}{\isacharparenright}{\kern0pt}\ {\isacharparenleft}{\kern0pt}step\ amount\ var\ L\ F{\isacharparenright}{\kern0pt}{\isacharparenright}{\kern0pt}\ {\isasymnu}{\isachardoublequoteclose}\isanewline
\ \ \isakeyword{assumes}\ opth\ {\isacharcolon}{\kern0pt}\ {\isachardoublequoteopen}{\isasymAnd}{\isasymnu}\ F\ {\isachardot}{\kern0pt}\ eval\ {\isacharparenleft}{\kern0pt}opt\ F{\isacharparenright}{\kern0pt}\ {\isasymnu}\ {\isacharequal}{\kern0pt}\ eval\ F\ {\isasymnu}{\isachardoublequoteclose}\isanewline
\ \ \isakeyword{shows}\ {\isachardoublequoteopen}eval\ {\isacharparenleft}{\kern0pt}QE{\isacharunderscore}{\kern0pt}dnf\ opt\ step\ {\isasymphi}{\isacharparenright}{\kern0pt}\ {\isasymnu}\ {\isacharequal}{\kern0pt}\ eval\ {\isasymphi}\ {\isasymnu}{\isachardoublequoteclose}%
}
\snip{qednf}{1}{2}{%
\isacommand{fun}\isamarkupfalse%
\ QE{\isacharunderscore}{\kern0pt}dnf\ {\isacharcolon}{\kern0pt}{\isacharcolon}{\kern0pt}\ {\isachardoublequoteopen}{\isacharparenleft}{\kern0pt}atom\ fm\ {\isasymRightarrow}\ atom\ fm{\isacharparenright}{\kern0pt}\ {\isasymRightarrow}\ {\isacharparenleft}{\kern0pt}nat\ {\isasymRightarrow}\ nat\ {\isasymRightarrow}\ atom\ list\ {\isasymRightarrow}\ atom\ fm\ list\ {\isasymRightarrow}\ atom\ fm{\isacharparenright}{\kern0pt}\ {\isasymRightarrow}\ atom\ fm\ {\isasymRightarrow}\ atom\ fm{\isachardoublequoteclose}\ \isakeyword{where}\isanewline
\ \ {\isachardoublequoteopen}QE{\isacharunderscore}{\kern0pt}dnf\ opt\ step\ {\isacharparenleft}{\kern0pt}And\ {\isasymphi}\isactrlsub {\isadigit{1}}\ {\isasymphi}\isactrlsub {\isadigit{2}}{\isacharparenright}{\kern0pt}\ {\isacharequal}{\kern0pt}\ and\ {\isacharparenleft}{\kern0pt}QE{\isacharunderscore}{\kern0pt}dnf\ opt\ step\ {\isasymphi}\isactrlsub {\isadigit{1}}{\isacharparenright}{\kern0pt}\ {\isacharparenleft}{\kern0pt}QE{\isacharunderscore}{\kern0pt}dnf\ opt\ step\ {\isasymphi}\isactrlsub {\isadigit{2}}{\isacharparenright}{\kern0pt}{\isachardoublequoteclose}\ {\isacharbar}{\kern0pt}\isanewline
\ \ {\isachardoublequoteopen}QE{\isacharunderscore}{\kern0pt}dnf\ opt\ step\ {\isacharparenleft}{\kern0pt}Or\ {\isasymphi}\isactrlsub {\isadigit{1}}\ {\isasymphi}\isactrlsub {\isadigit{2}}{\isacharparenright}{\kern0pt}\ {\isacharequal}{\kern0pt}\ or\ {\isacharparenleft}{\kern0pt}QE{\isacharunderscore}{\kern0pt}dnf\ opt\ step\ {\isasymphi}\isactrlsub {\isadigit{1}}{\isacharparenright}{\kern0pt}\ {\isacharparenleft}{\kern0pt}QE{\isacharunderscore}{\kern0pt}dnf\ opt\ step\ {\isasymphi}\isactrlsub {\isadigit{2}}{\isacharparenright}{\kern0pt}{\isachardoublequoteclose}\ {\isacharbar}{\kern0pt}\isanewline
\ \ {\isachardoublequoteopen}QE{\isacharunderscore}{\kern0pt}dnf\ opt\ step\ {\isacharparenleft}{\kern0pt}Neg\ {\isasymphi}{\isacharparenright}{\kern0pt}\ {\isacharequal}{\kern0pt}\ neg{\isacharparenleft}{\kern0pt}QE{\isacharunderscore}{\kern0pt}dnf\ opt\ step\ {\isasymphi}{\isacharparenright}{\kern0pt}{\isachardoublequoteclose}\ {\isacharbar}{\kern0pt}\isanewline
\ \ {\isachardoublequoteopen}QE{\isacharunderscore}{\kern0pt}dnf\ opt\ step\ {\isacharparenleft}{\kern0pt}ExQ\ {\isasymphi}{\isacharparenright}{\kern0pt}\ {\isacharequal}{\kern0pt}\ list{\isacharunderscore}{\kern0pt}disj\ {\isacharbrackleft}{\kern0pt}ExN\ {\isacharparenleft}{\kern0pt}n{\isacharplus}{\kern0pt}{\isadigit{1}}{\isacharparenright}{\kern0pt}\ {\isacharparenleft}{\kern0pt}step\ {\isadigit{1}}\ n\ al\ fl{\isacharparenright}{\kern0pt}{\isachardot}{\kern0pt}\ {\isacharparenleft}{\kern0pt}al{\isacharcomma}{\kern0pt}fl{\isacharcomma}{\kern0pt}n{\isacharparenright}{\kern0pt}{\isasymleftarrow}{\isacharparenleft}{\kern0pt}dnf{\isacharunderscore}{\kern0pt}modified{\isacharparenleft}{\kern0pt}opt{\isacharparenleft}{\kern0pt}QE{\isacharunderscore}{\kern0pt}dnf\ opt\ step\ {\isasymphi}{\isacharparenright}{\kern0pt}{\isacharparenright}{\kern0pt}{\isacharparenright}{\kern0pt}{\isacharbrackright}{\kern0pt}{\isachardoublequoteclose}{\isacharbar}{\kern0pt}\isanewline
\ \ {\isachardoublequoteopen}QE{\isacharunderscore}{\kern0pt}dnf\ opt\ step\ {\isacharparenleft}{\kern0pt}TrueF{\isacharparenright}{\kern0pt}\ {\isacharequal}{\kern0pt}\ TrueF{\isachardoublequoteclose}{\isacharbar}{\kern0pt}\isanewline
\ \ {\isachardoublequoteopen}QE{\isacharunderscore}{\kern0pt}dnf\ opt\ step\ {\isacharparenleft}{\kern0pt}FalseF{\isacharparenright}{\kern0pt}\ {\isacharequal}{\kern0pt}\ FalseF{\isachardoublequoteclose}{\isacharbar}{\kern0pt}\isanewline
\ \ {\isachardoublequoteopen}QE{\isacharunderscore}{\kern0pt}dnf\ opt\ step\ {\isacharparenleft}{\kern0pt}Atom\ a{\isacharparenright}{\kern0pt}\ {\isacharequal}{\kern0pt}\ simp{\isacharunderscore}{\kern0pt}atom\ a{\isachardoublequoteclose}{\isacharbar}{\kern0pt}\isanewline
\ \ {\isachardoublequoteopen}QE{\isacharunderscore}{\kern0pt}dnf\ opt\ step\ {\isacharparenleft}{\kern0pt}AllQ\ {\isasymphi}{\isacharparenright}{\kern0pt}\ {\isacharequal}{\kern0pt}\ Neg{\isacharparenleft}{\kern0pt}list{\isacharunderscore}{\kern0pt}disj\ {\isacharbrackleft}{\kern0pt}ExN\ {\isacharparenleft}{\kern0pt}n{\isacharplus}{\kern0pt}{\isadigit{1}}{\isacharparenright}{\kern0pt}\ {\isacharparenleft}{\kern0pt}step\ {\isadigit{1}}\ n\ al\ fl{\isacharparenright}{\kern0pt}{\isachardot}{\kern0pt}\ {\isacharparenleft}{\kern0pt}al{\isacharcomma}{\kern0pt}fl{\isacharcomma}{\kern0pt}n{\isacharparenright}{\kern0pt}{\isasymleftarrow}{\isacharparenleft}{\kern0pt}dnf{\isacharunderscore}{\kern0pt}modified{\isacharparenleft}{\kern0pt}opt{\isacharparenleft}{\kern0pt}neg{\isacharparenleft}{\kern0pt}QE{\isacharunderscore}{\kern0pt}dnf\ opt\ step\ {\isasymphi}{\isacharparenright}{\kern0pt}{\isacharparenright}{\kern0pt}{\isacharparenright}{\kern0pt}{\isacharparenright}{\kern0pt}{\isacharbrackright}{\kern0pt}{\isacharparenright}{\kern0pt}{\isachardoublequoteclose}{\isacharbar}{\kern0pt}\isanewline
\ \ {\isachardoublequoteopen}QE{\isacharunderscore}{\kern0pt}dnf\ opt\ step\ {\isacharparenleft}{\kern0pt}ExN\ {\isadigit{0}}\ {\isasymphi}{\isacharparenright}{\kern0pt}\ {\isacharequal}{\kern0pt}\ QE{\isacharunderscore}{\kern0pt}dnf\ opt\ step\ {\isasymphi}{\isachardoublequoteclose}{\isacharbar}{\kern0pt}\isanewline
\ \ {\isachardoublequoteopen}QE{\isacharunderscore}{\kern0pt}dnf\ opt\ step\ {\isacharparenleft}{\kern0pt}AllN\ {\isadigit{0}}\ {\isasymphi}{\isacharparenright}{\kern0pt}\ {\isacharequal}{\kern0pt}\ QE{\isacharunderscore}{\kern0pt}dnf\ opt\ step\ {\isasymphi}{\isachardoublequoteclose}{\isacharbar}{\kern0pt}\isanewline
\ \ {\isachardoublequoteopen}QE{\isacharunderscore}{\kern0pt}dnf\ opt\ step\ {\isacharparenleft}{\kern0pt}AllN\ {\isacharparenleft}{\kern0pt}Suc\ i{\isacharparenright}{\kern0pt}\ {\isasymphi}{\isacharparenright}{\kern0pt}\ {\isacharequal}{\kern0pt}\ Neg{\isacharparenleft}{\kern0pt}list{\isacharunderscore}{\kern0pt}disj\ {\isacharbrackleft}{\kern0pt}ExN\ {\isacharparenleft}{\kern0pt}n{\isacharplus}{\kern0pt}i{\isacharplus}{\kern0pt}{\isadigit{1}}{\isacharparenright}{\kern0pt}\ {\isacharparenleft}{\kern0pt}step\ {\isacharparenleft}{\kern0pt}Suc\ i{\isacharparenright}{\kern0pt}\ {\isacharparenleft}{\kern0pt}n{\isacharplus}{\kern0pt}i{\isacharparenright}{\kern0pt}\ al\ fl{\isacharparenright}{\kern0pt}{\isachardot}{\kern0pt}\ {\isacharparenleft}{\kern0pt}al{\isacharcomma}{\kern0pt}fl{\isacharcomma}{\kern0pt}n{\isacharparenright}{\kern0pt}{\isasymleftarrow}{\isacharparenleft}{\kern0pt}dnf{\isacharunderscore}{\kern0pt}modified{\isacharparenleft}{\kern0pt}opt{\isacharparenleft}{\kern0pt}neg{\isacharparenleft}{\kern0pt}QE{\isacharunderscore}{\kern0pt}dnf\ opt\ step\ {\isasymphi}{\isacharparenright}{\kern0pt}{\isacharparenright}{\kern0pt}{\isacharparenright}{\kern0pt}{\isacharparenright}{\kern0pt}{\isacharbrackright}{\kern0pt}{\isacharparenright}{\kern0pt}{\isachardoublequoteclose}{\isacharbar}{\kern0pt}\isanewline
\ \ {\isachardoublequoteopen}QE{\isacharunderscore}{\kern0pt}dnf\ opt\ step\ {\isacharparenleft}{\kern0pt}ExN\ {\isacharparenleft}{\kern0pt}Suc\ i{\isacharparenright}{\kern0pt}\ {\isasymphi}{\isacharparenright}{\kern0pt}\ {\isacharequal}{\kern0pt}\ list{\isacharunderscore}{\kern0pt}disj\ {\isacharbrackleft}{\kern0pt}ExN\ {\isacharparenleft}{\kern0pt}n{\isacharplus}{\kern0pt}i{\isacharplus}{\kern0pt}{\isadigit{1}}{\isacharparenright}{\kern0pt}\ {\isacharparenleft}{\kern0pt}step\ {\isacharparenleft}{\kern0pt}Suc\ i{\isacharparenright}{\kern0pt}\ {\isacharparenleft}{\kern0pt}n{\isacharplus}{\kern0pt}i{\isacharparenright}{\kern0pt}\ al\ fl{\isacharparenright}{\kern0pt}{\isachardot}{\kern0pt}\ {\isacharparenleft}{\kern0pt}al{\isacharcomma}{\kern0pt}fl{\isacharcomma}{\kern0pt}n{\isacharparenright}{\kern0pt}{\isasymleftarrow}{\isacharparenleft}{\kern0pt}dnf{\isacharunderscore}{\kern0pt}modified{\isacharparenleft}{\kern0pt}opt{\isacharparenleft}{\kern0pt}QE{\isacharunderscore}{\kern0pt}dnf\ opt\ step\ {\isasymphi}{\isacharparenright}{\kern0pt}{\isacharparenright}{\kern0pt}{\isacharparenright}{\kern0pt}{\isacharbrackright}{\kern0pt}{\isachardoublequoteclose}%
}
\snip{dnf}{1}{2}{%
\isacommand{fun}\isamarkupfalse%
\ dnf\ {\isacharcolon}{\kern0pt}{\isacharcolon}{\kern0pt}\ {\isachardoublequoteopen}atom\ fm\ {\isasymRightarrow}\ {\isacharparenleft}{\kern0pt}atom\ list\ {\isacharasterisk}{\kern0pt}\ atom\ fm\ list{\isacharparenright}{\kern0pt}\ list{\isachardoublequoteclose}\ \isakeyword{where}\isanewline
\ \ {\isachardoublequoteopen}dnf\ TrueF\ {\isacharequal}{\kern0pt}\ {\isacharbrackleft}{\kern0pt}{\isacharparenleft}{\kern0pt}{\isacharbrackleft}{\kern0pt}{\isacharbrackright}{\kern0pt}{\isacharcomma}{\kern0pt}{\isacharbrackleft}{\kern0pt}{\isacharbrackright}{\kern0pt}{\isacharparenright}{\kern0pt}{\isacharbrackright}{\kern0pt}{\isachardoublequoteclose}\ {\isacharbar}{\kern0pt}\isanewline
\ \ {\isachardoublequoteopen}dnf\ FalseF\ {\isacharequal}{\kern0pt}\ {\isacharbrackleft}{\kern0pt}{\isacharbrackright}{\kern0pt}{\isachardoublequoteclose}\ {\isacharbar}{\kern0pt}\isanewline
\ \ {\isachardoublequoteopen}dnf\ {\isacharparenleft}{\kern0pt}Atom\ {\isasymphi}{\isacharparenright}{\kern0pt}\ {\isacharequal}{\kern0pt}\ {\isacharbrackleft}{\kern0pt}{\isacharparenleft}{\kern0pt}{\isacharbrackleft}{\kern0pt}{\isasymphi}{\isacharbrackright}{\kern0pt}{\isacharcomma}{\kern0pt}{\isacharbrackleft}{\kern0pt}{\isacharbrackright}{\kern0pt}{\isacharparenright}{\kern0pt}{\isacharbrackright}{\kern0pt}{\isachardoublequoteclose}\ {\isacharbar}{\kern0pt}\isanewline
\ \ {\isachardoublequoteopen}dnf\ {\isacharparenleft}{\kern0pt}And\ {\isasymphi}\isactrlsub {\isadigit{1}}\ {\isasymphi}\isactrlsub {\isadigit{2}}{\isacharparenright}{\kern0pt}\ {\isacharequal}{\kern0pt}\ {\isacharbrackleft}{\kern0pt}{\isacharparenleft}{\kern0pt}A{\isacharat}{\kern0pt}B{\isacharcomma}{\kern0pt}A{\isacharprime}{\kern0pt}{\isacharat}{\kern0pt}B{\isacharprime}{\kern0pt}{\isacharparenright}{\kern0pt}{\isachardot}{\kern0pt}{\isacharparenleft}{\kern0pt}A{\isacharcomma}{\kern0pt}A{\isacharprime}{\kern0pt}{\isacharparenright}{\kern0pt}{\isasymleftarrow}dnf\ {\isasymphi}\isactrlsub {\isadigit{1}}{\isacharcomma}{\kern0pt}{\isacharparenleft}{\kern0pt}B{\isacharcomma}{\kern0pt}B{\isacharprime}{\kern0pt}{\isacharparenright}{\kern0pt}{\isasymleftarrow}dnf\ {\isasymphi}\isactrlsub {\isadigit{2}}{\isacharbrackright}{\kern0pt}{\isachardoublequoteclose}\ {\isacharbar}{\kern0pt}\isanewline
\ \ {\isachardoublequoteopen}dnf\ {\isacharparenleft}{\kern0pt}Or\ {\isasymphi}\isactrlsub {\isadigit{1}}\ {\isasymphi}\isactrlsub {\isadigit{2}}{\isacharparenright}{\kern0pt}\ {\isacharequal}{\kern0pt}\ dnf\ {\isasymphi}\isactrlsub {\isadigit{1}}\ {\isacharat}{\kern0pt}\ dnf\ {\isasymphi}\isactrlsub {\isadigit{2}}{\isachardoublequoteclose}\ {\isacharbar}{\kern0pt}\isanewline
\ \ {\isachardoublequoteopen}dnf\ {\isacharparenleft}{\kern0pt}ExQ\ {\isasymphi}{\isacharparenright}{\kern0pt}\ {\isacharequal}{\kern0pt}\ {\isacharbrackleft}{\kern0pt}{\isacharparenleft}{\kern0pt}{\isacharbrackleft}{\kern0pt}{\isacharbrackright}{\kern0pt}{\isacharcomma}{\kern0pt}{\isacharbrackleft}{\kern0pt}ExQ\ {\isasymphi}{\isacharbrackright}{\kern0pt}{\isacharparenright}{\kern0pt}{\isacharbrackright}{\kern0pt}{\isachardoublequoteclose}\ {\isacharbar}{\kern0pt}\isanewline
\ \ {\isachardoublequoteopen}dnf\ {\isacharparenleft}{\kern0pt}Neg\ {\isasymphi}{\isacharparenright}{\kern0pt}\ {\isacharequal}{\kern0pt}\ {\isacharbrackleft}{\kern0pt}{\isacharparenleft}{\kern0pt}{\isacharbrackleft}{\kern0pt}{\isacharbrackright}{\kern0pt}{\isacharcomma}{\kern0pt}{\isacharbrackleft}{\kern0pt}Neg\ {\isasymphi}{\isacharbrackright}{\kern0pt}{\isacharparenright}{\kern0pt}{\isacharbrackright}{\kern0pt}{\isachardoublequoteclose}{\isacharbar}{\kern0pt}\isanewline
\ \ {\isachardoublequoteopen}dnf\ {\isacharparenleft}{\kern0pt}AllQ\ {\isasymphi}{\isacharparenright}{\kern0pt}\ {\isacharequal}{\kern0pt}\ {\isacharbrackleft}{\kern0pt}{\isacharparenleft}{\kern0pt}{\isacharbrackleft}{\kern0pt}{\isacharbrackright}{\kern0pt}{\isacharcomma}{\kern0pt}{\isacharbrackleft}{\kern0pt}AllQ\ {\isasymphi}{\isacharbrackright}{\kern0pt}{\isacharparenright}{\kern0pt}{\isacharbrackright}{\kern0pt}{\isachardoublequoteclose}{\isacharbar}{\kern0pt}\isanewline
\ \ {\isachardoublequoteopen}dnf\ {\isacharparenleft}{\kern0pt}AllN\ i\ {\isasymphi}{\isacharparenright}{\kern0pt}\ {\isacharequal}{\kern0pt}\ {\isacharbrackleft}{\kern0pt}{\isacharparenleft}{\kern0pt}{\isacharbrackleft}{\kern0pt}{\isacharbrackright}{\kern0pt}{\isacharcomma}{\kern0pt}{\isacharbrackleft}{\kern0pt}AllN\ i\ {\isasymphi}{\isacharbrackright}{\kern0pt}{\isacharparenright}{\kern0pt}{\isacharbrackright}{\kern0pt}{\isachardoublequoteclose}{\isacharbar}{\kern0pt}\isanewline
\ \ {\isachardoublequoteopen}dnf\ {\isacharparenleft}{\kern0pt}ExN\ i\ {\isasymphi}{\isacharparenright}{\kern0pt}\ {\isacharequal}{\kern0pt}\ {\isacharbrackleft}{\kern0pt}{\isacharparenleft}{\kern0pt}{\isacharbrackleft}{\kern0pt}{\isacharbrackright}{\kern0pt}{\isacharcomma}{\kern0pt}{\isacharbrackleft}{\kern0pt}ExN\ i\ {\isasymphi}{\isacharbrackright}{\kern0pt}{\isacharparenright}{\kern0pt}{\isacharbrackright}{\kern0pt}{\isachardoublequoteclose}%
}
\snip{elimVarEq}{1}{2}{%
\isacommand{lemma}\isamarkupfalse%
\ elimVar{\isacharunderscore}{\kern0pt}eq{\isacharcolon}{\kern0pt}\isanewline
\ \ \isakeyword{assumes}\ hlength{\isacharcolon}{\kern0pt}\ {\isachardoublequoteopen}length\ {\isacharparenleft}{\kern0pt}{\isasymnu}{\isacharcolon}{\kern0pt}{\isacharcolon}{\kern0pt}real\ list{\isacharparenright}{\kern0pt}\ {\isacharequal}{\kern0pt}\ var{\isachardoublequoteclose}\isanewline
\ \ \isakeyword{assumes}\ noVariable{\isacharcolon}{\kern0pt}\ {\isachardoublequoteopen}var\ {\isasymnotin}\ vars\ a{\isachardoublequoteclose}\ {\isachardoublequoteopen}var\ {\isasymnotin}\ vars\ b{\isachardoublequoteclose}\ {\isachardoublequoteopen}var\ {\isasymnotin}\ vars\ c{\isachardoublequoteclose}\isanewline
\ \ \isakeyword{assumes}\ inList{\isacharcolon}{\kern0pt}\ {\isachardoublequoteopen}Eq\ {\isacharparenleft}{\kern0pt}a{\isacharasterisk}{\kern0pt}{\isacharparenleft}{\kern0pt}Var\ var{\isacharparenright}{\kern0pt}{\isacharcircum}{\kern0pt}{\isadigit{2}}{\isacharplus}{\kern0pt}b{\isacharasterisk}{\kern0pt}{\isacharparenleft}{\kern0pt}Var\ var{\isacharparenright}{\kern0pt}{\isacharplus}{\kern0pt}c{\isacharparenright}{\kern0pt}\ \ {\isasymin}\ set{\isacharparenleft}{\kern0pt}L{\isacharparenright}{\kern0pt}{\isachardoublequoteclose}\isanewline
\ \ \isakeyword{assumes}\ nonzero{\isacharcolon}{\kern0pt}\ \isanewline
\ \ \ \ {\isachardoublequoteopen}insertion{\isacharprime}{\kern0pt}\ a\ {\isacharparenleft}{\kern0pt}{\isasymnu}\ {\isacharat}{\kern0pt}\ x\ {\isacharhash}{\kern0pt}\ {\isasymnu}{\isacharprime}{\kern0pt}{\isacharparenright}{\kern0pt}\ {\isasymnoteq}\ {\isadigit{0}}\ {\isasymor}\ insertion{\isacharprime}{\kern0pt}\ b\ {\isacharparenleft}{\kern0pt}{\isasymnu}\ {\isacharat}{\kern0pt}\ x\ {\isacharhash}{\kern0pt}\ {\isasymnu}{\isacharprime}{\kern0pt}{\isacharparenright}{\kern0pt}\ {\isasymnoteq}\ {\isadigit{0}}{\isachardoublequoteclose}\isanewline
\ \ \isakeyword{shows}\ {\isachardoublequoteopen}{\isacharparenleft}{\kern0pt}{\isasymexists}x{\isachardot}{\kern0pt}\ eval\ {\isacharparenleft}{\kern0pt}list{\isacharunderscore}{\kern0pt}conj\ {\isacharparenleft}{\kern0pt}map\ fm{\isachardot}{\kern0pt}Atom\ L\ {\isacharat}{\kern0pt}\ F{\isacharparenright}{\kern0pt}{\isacharparenright}{\kern0pt}\ {\isacharparenleft}{\kern0pt}{\isasymnu}\ {\isacharat}{\kern0pt}\ x\ {\isacharhash}{\kern0pt}\ {\isasymnu}{\isacharprime}{\kern0pt}{\isacharparenright}{\kern0pt}{\isacharparenright}{\kern0pt}\ {\isacharequal}{\kern0pt}\isanewline
{\isacharparenleft}{\kern0pt}{\isasymexists}x{\isachardot}{\kern0pt}\ eval{\isacharparenleft}{\kern0pt}elimVar\ var\ L\ F\ {\isacharparenleft}{\kern0pt}Eq{\isacharparenleft}{\kern0pt}a{\isacharasterisk}{\kern0pt}{\isacharparenleft}{\kern0pt}Var\ var{\isacharparenright}{\kern0pt}{\isacharcircum}{\kern0pt}{\isadigit{2}}{\isacharplus}{\kern0pt}b{\isacharasterisk}{\kern0pt}{\isacharparenleft}{\kern0pt}Var\ var{\isacharparenright}{\kern0pt}{\isacharplus}{\kern0pt}c{\isacharparenright}{\kern0pt}{\isacharparenright}{\kern0pt}{\isacharparenright}{\kern0pt}{\isacharparenleft}{\kern0pt}{\isasymnu}{\isacharat}{\kern0pt}x{\isacharhash}{\kern0pt}{\isasymnu}{\isacharprime}{\kern0pt}{\isacharparenright}{\kern0pt}{\isacharparenright}{\kern0pt}{\isachardoublequoteclose}%
}
\snip{elimVar}{1}{2}{%
\isacommand{fun}\isamarkupfalse%
\ elimVar\ {\isacharcolon}{\kern0pt}{\isacharcolon}{\kern0pt}\ {\isachardoublequoteopen}nat\ {\isasymRightarrow}\ atom\ list\ {\isasymRightarrow}\ {\isacharparenleft}{\kern0pt}atom\ fm{\isacharparenright}{\kern0pt}\ list\ {\isasymRightarrow}\ atom\ {\isasymRightarrow}\ atom\ fm{\isachardoublequoteclose}\ \isakeyword{where}\isanewline
\ \ {\isachardoublequoteopen}elimVar\ var\ L\ F\ {\isacharparenleft}{\kern0pt}Eq\ p{\isacharparenright}{\kern0pt}\ {\isacharequal}{\kern0pt}\ \ {\isacharparenleft}{\kern0pt}\isanewline
\ \ let\ {\isacharparenleft}{\kern0pt}a{\isacharcomma}{\kern0pt}b{\isacharcomma}{\kern0pt}c{\isacharparenright}{\kern0pt}\ {\isacharequal}{\kern0pt}\ get{\isacharunderscore}{\kern0pt}coeffs\ var\ p\ in\isanewline
\isanewline
\ \ \ {\isacharparenleft}{\kern0pt}Or\ \isanewline
\ \ \ \ \ \ \isanewline
\ \ \ \ \ \ {\isacharparenleft}{\kern0pt}And\ {\isacharparenleft}{\kern0pt}And\ {\isacharparenleft}{\kern0pt}Atom\ {\isacharparenleft}{\kern0pt}Eq\ a{\isacharparenright}{\kern0pt}{\isacharparenright}{\kern0pt}\ {\isacharparenleft}{\kern0pt}Atom\ {\isacharparenleft}{\kern0pt}Neq\ b{\isacharparenright}{\kern0pt}{\isacharparenright}{\kern0pt}{\isacharparenright}{\kern0pt}\isanewline
\ \ \ \ \ \ {\isacharparenleft}{\kern0pt}list{\isacharunderscore}{\kern0pt}conj\ {\isacharparenleft}{\kern0pt}\isanewline
\ \ \ \ \ \ \ \ {\isacharparenleft}{\kern0pt}map\ {\isacharparenleft}{\kern0pt}{\isasymlambda}a{\isachardot}{\kern0pt}\ Atom\ {\isacharparenleft}{\kern0pt}linear{\isacharunderscore}{\kern0pt}substitution\ var\ {\isacharparenleft}{\kern0pt}{\isacharminus}{\kern0pt}c{\isacharparenright}{\kern0pt}\ b\ a{\isacharparenright}{\kern0pt}{\isacharparenright}{\kern0pt}\ L{\isacharparenright}{\kern0pt}{\isacharat}{\kern0pt}\isanewline
\ \ \ \ \ \ \ \ {\isacharparenleft}{\kern0pt}map\ {\isacharparenleft}{\kern0pt}linear{\isacharunderscore}{\kern0pt}substitution{\isacharunderscore}{\kern0pt}fm\ var\ {\isacharparenleft}{\kern0pt}{\isacharminus}{\kern0pt}c{\isacharparenright}{\kern0pt}\ b{\isacharparenright}{\kern0pt}\ F{\isacharparenright}{\kern0pt}\isanewline
\ \ \ \ \ \ \ \ {\isacharparenright}{\kern0pt}{\isacharparenright}{\kern0pt}{\isacharparenright}{\kern0pt}\isanewline
\ \ \ \ \ \ \isanewline
\isanewline
\ \ \ \ \ \ {\isacharparenleft}{\kern0pt}And\ {\isacharparenleft}{\kern0pt}Atom\ {\isacharparenleft}{\kern0pt}Neq\ a{\isacharparenright}{\kern0pt}{\isacharparenright}{\kern0pt}\ {\isacharparenleft}{\kern0pt}And\ {\isacharparenleft}{\kern0pt}Atom{\isacharparenleft}{\kern0pt}Leq\ {\isacharparenleft}{\kern0pt}{\isacharminus}{\kern0pt}{\isacharparenleft}{\kern0pt}b{\isacharcircum}{\kern0pt}{\isadigit{2}}{\isacharparenright}{\kern0pt}{\isacharplus}{\kern0pt}{\isadigit{4}}{\isacharasterisk}{\kern0pt}a{\isacharasterisk}{\kern0pt}c{\isacharparenright}{\kern0pt}{\isacharparenright}{\kern0pt}{\isacharparenright}{\kern0pt}\isanewline
\ \ \ \ \ \ \ \ {\isacharparenleft}{\kern0pt}Or\ {\isacharparenleft}{\kern0pt}list{\isacharunderscore}{\kern0pt}conj\ {\isacharparenleft}{\kern0pt}\isanewline
\ \ \ \ \ \ \ \ {\isacharparenleft}{\kern0pt}map\ {\isacharparenleft}{\kern0pt}quadratic{\isacharunderscore}{\kern0pt}sub\ var\ {\isacharparenleft}{\kern0pt}{\isacharminus}{\kern0pt}b{\isacharparenright}{\kern0pt}\ {\isadigit{1}}\ {\isacharparenleft}{\kern0pt}b{\isacharcircum}{\kern0pt}{\isadigit{2}}{\isacharminus}{\kern0pt}{\isadigit{4}}{\isacharasterisk}{\kern0pt}a{\isacharasterisk}{\kern0pt}c{\isacharparenright}{\kern0pt}\ {\isacharparenleft}{\kern0pt}{\isadigit{2}}{\isacharasterisk}{\kern0pt}a{\isacharparenright}{\kern0pt}{\isacharparenright}{\kern0pt}\ L{\isacharparenright}{\kern0pt}{\isacharat}{\kern0pt}\isanewline
\ \ \ \ \ \ \ \ {\isacharparenleft}{\kern0pt}map\ {\isacharparenleft}{\kern0pt}quadratic{\isacharunderscore}{\kern0pt}sub{\isacharunderscore}{\kern0pt}fm\ var\ {\isacharparenleft}{\kern0pt}{\isacharminus}{\kern0pt}b{\isacharparenright}{\kern0pt}\ {\isadigit{1}}\ {\isacharparenleft}{\kern0pt}b{\isacharcircum}{\kern0pt}{\isadigit{2}}{\isacharminus}{\kern0pt}{\isadigit{4}}{\isacharasterisk}{\kern0pt}a{\isacharasterisk}{\kern0pt}c{\isacharparenright}{\kern0pt}\ {\isacharparenleft}{\kern0pt}{\isadigit{2}}{\isacharasterisk}{\kern0pt}a{\isacharparenright}{\kern0pt}{\isacharparenright}{\kern0pt}\ F{\isacharparenright}{\kern0pt}\isanewline
\ \ \ \ \ \ \ \ {\isacharparenright}{\kern0pt}{\isacharparenright}{\kern0pt}\isanewline
\ \ \ \ \ \ \ \ {\isacharparenleft}{\kern0pt}list{\isacharunderscore}{\kern0pt}conj\ {\isacharparenleft}{\kern0pt}\isanewline
\ \ \ \ \ \ \ \ {\isacharparenleft}{\kern0pt}map\ {\isacharparenleft}{\kern0pt}quadratic{\isacharunderscore}{\kern0pt}sub\ var\ {\isacharparenleft}{\kern0pt}{\isacharminus}{\kern0pt}b{\isacharparenright}{\kern0pt}\ {\isacharparenleft}{\kern0pt}{\isacharminus}{\kern0pt}{\isadigit{1}}{\isacharparenright}{\kern0pt}\ {\isacharparenleft}{\kern0pt}b{\isacharcircum}{\kern0pt}{\isadigit{2}}{\isacharminus}{\kern0pt}{\isadigit{4}}{\isacharasterisk}{\kern0pt}a{\isacharasterisk}{\kern0pt}c{\isacharparenright}{\kern0pt}\ {\isacharparenleft}{\kern0pt}{\isadigit{2}}{\isacharasterisk}{\kern0pt}a{\isacharparenright}{\kern0pt}{\isacharparenright}{\kern0pt}\ L{\isacharparenright}{\kern0pt}{\isacharat}{\kern0pt}\isanewline
\ \ \ \ \ \ \ \ {\isacharparenleft}{\kern0pt}map\ {\isacharparenleft}{\kern0pt}quadratic{\isacharunderscore}{\kern0pt}sub{\isacharunderscore}{\kern0pt}fm\ var\ {\isacharparenleft}{\kern0pt}{\isacharminus}{\kern0pt}b{\isacharparenright}{\kern0pt}\ {\isacharparenleft}{\kern0pt}{\isacharminus}{\kern0pt}{\isadigit{1}}{\isacharparenright}{\kern0pt}\ {\isacharparenleft}{\kern0pt}b{\isacharcircum}{\kern0pt}{\isadigit{2}}{\isacharminus}{\kern0pt}{\isadigit{4}}{\isacharasterisk}{\kern0pt}a{\isacharasterisk}{\kern0pt}c{\isacharparenright}{\kern0pt}\ {\isacharparenleft}{\kern0pt}{\isadigit{2}}{\isacharasterisk}{\kern0pt}a{\isacharparenright}{\kern0pt}{\isacharparenright}{\kern0pt}\ F{\isacharparenright}{\kern0pt}\isanewline
\ \ \ \ \ \ \ \ {\isacharparenright}{\kern0pt}{\isacharparenright}{\kern0pt}\isanewline
\ \ \ \ \ \ \ {\isacharparenright}{\kern0pt}{\isacharparenright}{\kern0pt}\isanewline
\ \ \ \ \ \ {\isacharparenright}{\kern0pt}{\isacharparenright}{\kern0pt}\isanewline
\isanewline
{\isacharparenright}{\kern0pt}{\isachardoublequoteclose}\ {\isacharbar}{\kern0pt}\isanewline
\ \ {\isachardoublequoteopen}elimVar\ var\ L\ F\ {\isacharparenleft}{\kern0pt}Less\ p{\isacharparenright}{\kern0pt}\ {\isacharequal}{\kern0pt}\ \ {\isacharparenleft}{\kern0pt}\isanewline
\ \ let\ {\isacharparenleft}{\kern0pt}a{\isacharcomma}{\kern0pt}b{\isacharcomma}{\kern0pt}c{\isacharparenright}{\kern0pt}\ {\isacharequal}{\kern0pt}\ get{\isacharunderscore}{\kern0pt}coeffs\ var\ p\ in\isanewline
\ \ \ \ {\isacharparenleft}{\kern0pt}Or\ \isanewline
\ \ \ \ \ \ \isanewline
\ \ \ \ \ \ {\isacharparenleft}{\kern0pt}And\ {\isacharparenleft}{\kern0pt}And\ {\isacharparenleft}{\kern0pt}Atom\ {\isacharparenleft}{\kern0pt}Eq\ a{\isacharparenright}{\kern0pt}{\isacharparenright}{\kern0pt}\ {\isacharparenleft}{\kern0pt}Atom\ {\isacharparenleft}{\kern0pt}Neq\ b{\isacharparenright}{\kern0pt}{\isacharparenright}{\kern0pt}{\isacharparenright}{\kern0pt}\isanewline
\ \ \ \ \ \ {\isacharparenleft}{\kern0pt}list{\isacharunderscore}{\kern0pt}conj\ {\isacharparenleft}{\kern0pt}\isanewline
\ \ \ \ \ \ \ \ \ \ {\isacharparenleft}{\kern0pt}map\ {\isacharparenleft}{\kern0pt}substInfinitesimalLinear\ var\ {\isacharparenleft}{\kern0pt}{\isacharminus}{\kern0pt}c{\isacharparenright}{\kern0pt}\ b{\isacharparenright}{\kern0pt}\ L{\isacharparenright}{\kern0pt}\isanewline
\ \ \ \ \ \ \ \ \ \ {\isacharat}{\kern0pt}{\isacharparenleft}{\kern0pt}map\ {\isacharparenleft}{\kern0pt}substInfinitesimalLinear{\isacharunderscore}{\kern0pt}fm\ var\ {\isacharparenleft}{\kern0pt}{\isacharminus}{\kern0pt}c{\isacharparenright}{\kern0pt}\ b{\isacharparenright}{\kern0pt}\ F{\isacharparenright}{\kern0pt}\isanewline
\ \ \ \ \ \ {\isacharparenright}{\kern0pt}{\isacharparenright}{\kern0pt}{\isacharparenright}{\kern0pt}\isanewline
\ \ \ \ \ \ \isanewline
\isanewline
\ \ \ \ \ \ {\isacharparenleft}{\kern0pt}And\ {\isacharparenleft}{\kern0pt}Atom\ {\isacharparenleft}{\kern0pt}Neq\ a{\isacharparenright}{\kern0pt}{\isacharparenright}{\kern0pt}\ {\isacharparenleft}{\kern0pt}And\ {\isacharparenleft}{\kern0pt}Atom{\isacharparenleft}{\kern0pt}Leq\ {\isacharparenleft}{\kern0pt}{\isacharminus}{\kern0pt}{\isacharparenleft}{\kern0pt}b{\isacharcircum}{\kern0pt}{\isadigit{2}}{\isacharparenright}{\kern0pt}{\isacharplus}{\kern0pt}{\isadigit{4}}{\isacharasterisk}{\kern0pt}a{\isacharasterisk}{\kern0pt}c{\isacharparenright}{\kern0pt}{\isacharparenright}{\kern0pt}{\isacharparenright}{\kern0pt}\isanewline
\ \ \ \ \ \ \ \ {\isacharparenleft}{\kern0pt}Or\ {\isacharparenleft}{\kern0pt}list{\isacharunderscore}{\kern0pt}conj\ {\isacharparenleft}{\kern0pt}\isanewline
\ \ \ \ \ \ \ \ {\isacharparenleft}{\kern0pt}map\ {\isacharparenleft}{\kern0pt}substInfinitesimalQuadratic\ var\ {\isacharparenleft}{\kern0pt}{\isacharminus}{\kern0pt}b{\isacharparenright}{\kern0pt}\ {\isadigit{1}}\ {\isacharparenleft}{\kern0pt}b{\isacharcircum}{\kern0pt}{\isadigit{2}}{\isacharminus}{\kern0pt}{\isadigit{4}}{\isacharasterisk}{\kern0pt}a{\isacharasterisk}{\kern0pt}c{\isacharparenright}{\kern0pt}\ {\isacharparenleft}{\kern0pt}{\isadigit{2}}{\isacharasterisk}{\kern0pt}a{\isacharparenright}{\kern0pt}{\isacharparenright}{\kern0pt}\ L{\isacharparenright}{\kern0pt}{\isacharat}{\kern0pt}\isanewline
\ \ \ \ \ \ \ \ {\isacharparenleft}{\kern0pt}map\ {\isacharparenleft}{\kern0pt}substInfinitesimalQuadratic{\isacharunderscore}{\kern0pt}fm\ var\ {\isacharparenleft}{\kern0pt}{\isacharminus}{\kern0pt}b{\isacharparenright}{\kern0pt}\ {\isadigit{1}}\ {\isacharparenleft}{\kern0pt}b{\isacharcircum}{\kern0pt}{\isadigit{2}}{\isacharminus}{\kern0pt}{\isadigit{4}}{\isacharasterisk}{\kern0pt}a{\isacharasterisk}{\kern0pt}c{\isacharparenright}{\kern0pt}\ {\isacharparenleft}{\kern0pt}{\isadigit{2}}{\isacharasterisk}{\kern0pt}a{\isacharparenright}{\kern0pt}{\isacharparenright}{\kern0pt}\ F{\isacharparenright}{\kern0pt}\isanewline
\ \ \ \ \ \ \ \ {\isacharparenright}{\kern0pt}{\isacharparenright}{\kern0pt}\isanewline
\ \ \ \ \ \ \ \ {\isacharparenleft}{\kern0pt}list{\isacharunderscore}{\kern0pt}conj\ {\isacharparenleft}{\kern0pt}\isanewline
\ \ \ \ \ \ \ \ {\isacharparenleft}{\kern0pt}map\ {\isacharparenleft}{\kern0pt}substInfinitesimalQuadratic\ var\ {\isacharparenleft}{\kern0pt}{\isacharminus}{\kern0pt}b{\isacharparenright}{\kern0pt}\ {\isacharparenleft}{\kern0pt}{\isacharminus}{\kern0pt}{\isadigit{1}}{\isacharparenright}{\kern0pt}\ {\isacharparenleft}{\kern0pt}b{\isacharcircum}{\kern0pt}{\isadigit{2}}{\isacharminus}{\kern0pt}{\isadigit{4}}{\isacharasterisk}{\kern0pt}a{\isacharasterisk}{\kern0pt}c{\isacharparenright}{\kern0pt}\ {\isacharparenleft}{\kern0pt}{\isadigit{2}}{\isacharasterisk}{\kern0pt}a{\isacharparenright}{\kern0pt}{\isacharparenright}{\kern0pt}\ L{\isacharparenright}{\kern0pt}{\isacharat}{\kern0pt}\isanewline
\ \ \ \ \ \ \ \ {\isacharparenleft}{\kern0pt}map\ {\isacharparenleft}{\kern0pt}substInfinitesimalQuadratic{\isacharunderscore}{\kern0pt}fm\ var\ {\isacharparenleft}{\kern0pt}{\isacharminus}{\kern0pt}b{\isacharparenright}{\kern0pt}\ {\isacharparenleft}{\kern0pt}{\isacharminus}{\kern0pt}{\isadigit{1}}{\isacharparenright}{\kern0pt}\ {\isacharparenleft}{\kern0pt}b{\isacharcircum}{\kern0pt}{\isadigit{2}}{\isacharminus}{\kern0pt}{\isadigit{4}}{\isacharasterisk}{\kern0pt}a{\isacharasterisk}{\kern0pt}c{\isacharparenright}{\kern0pt}\ {\isacharparenleft}{\kern0pt}{\isadigit{2}}{\isacharasterisk}{\kern0pt}a{\isacharparenright}{\kern0pt}{\isacharparenright}{\kern0pt}\ F{\isacharparenright}{\kern0pt}\isanewline
\ \ \ \ \ \ \ \ {\isacharparenright}{\kern0pt}{\isacharparenright}{\kern0pt}\isanewline
\ \ \ \ \ \ \ {\isacharparenright}{\kern0pt}{\isacharparenright}{\kern0pt}\isanewline
\ \ \ \ \ \ {\isacharparenright}{\kern0pt}{\isacharparenright}{\kern0pt}\isanewline
{\isacharparenright}{\kern0pt}{\isachardoublequoteclose}{\isacharbar}{\kern0pt}\isanewline
\ \ {\isachardoublequoteopen}elimVar\ var\ L\ F\ {\isacharparenleft}{\kern0pt}Neq\ p{\isacharparenright}{\kern0pt}\ {\isacharequal}{\kern0pt}\ \ {\isacharparenleft}{\kern0pt}\isanewline
\ \ let\ {\isacharparenleft}{\kern0pt}a{\isacharcomma}{\kern0pt}b{\isacharcomma}{\kern0pt}c{\isacharparenright}{\kern0pt}\ {\isacharequal}{\kern0pt}\ get{\isacharunderscore}{\kern0pt}coeffs\ var\ p\ in\isanewline
\ \ \ \ {\isacharparenleft}{\kern0pt}Or\ \isanewline
\ \ \ \ \ \ \isanewline
\ \ \ \ \ \ {\isacharparenleft}{\kern0pt}And\ {\isacharparenleft}{\kern0pt}And\ {\isacharparenleft}{\kern0pt}Atom\ {\isacharparenleft}{\kern0pt}Eq\ a{\isacharparenright}{\kern0pt}{\isacharparenright}{\kern0pt}\ {\isacharparenleft}{\kern0pt}Atom\ {\isacharparenleft}{\kern0pt}Neq\ b{\isacharparenright}{\kern0pt}{\isacharparenright}{\kern0pt}{\isacharparenright}{\kern0pt}\isanewline
\ \ \ \ \ \ {\isacharparenleft}{\kern0pt}list{\isacharunderscore}{\kern0pt}conj\ {\isacharparenleft}{\kern0pt}\isanewline
\ \ \ \ \ \ \ \ \ \ {\isacharparenleft}{\kern0pt}map\ {\isacharparenleft}{\kern0pt}substInfinitesimalLinear\ var\ {\isacharparenleft}{\kern0pt}{\isacharminus}{\kern0pt}c{\isacharparenright}{\kern0pt}\ b{\isacharparenright}{\kern0pt}\ L{\isacharparenright}{\kern0pt}\isanewline
\ \ \ \ \ \ \ \ \ \ {\isacharat}{\kern0pt}{\isacharparenleft}{\kern0pt}map\ {\isacharparenleft}{\kern0pt}substInfinitesimalLinear{\isacharunderscore}{\kern0pt}fm\ var\ {\isacharparenleft}{\kern0pt}{\isacharminus}{\kern0pt}c{\isacharparenright}{\kern0pt}\ b{\isacharparenright}{\kern0pt}\ F{\isacharparenright}{\kern0pt}\isanewline
\ \ \ \ \ \ {\isacharparenright}{\kern0pt}{\isacharparenright}{\kern0pt}{\isacharparenright}{\kern0pt}\isanewline
\ \ \ \ \ \ \isanewline
\isanewline
\ \ \ \ \ \ {\isacharparenleft}{\kern0pt}And\ {\isacharparenleft}{\kern0pt}Atom\ {\isacharparenleft}{\kern0pt}Neq\ a{\isacharparenright}{\kern0pt}{\isacharparenright}{\kern0pt}\ {\isacharparenleft}{\kern0pt}And\ {\isacharparenleft}{\kern0pt}Atom{\isacharparenleft}{\kern0pt}Leq\ {\isacharparenleft}{\kern0pt}{\isacharminus}{\kern0pt}{\isacharparenleft}{\kern0pt}b{\isacharcircum}{\kern0pt}{\isadigit{2}}{\isacharparenright}{\kern0pt}{\isacharplus}{\kern0pt}{\isadigit{4}}{\isacharasterisk}{\kern0pt}a{\isacharasterisk}{\kern0pt}c{\isacharparenright}{\kern0pt}{\isacharparenright}{\kern0pt}{\isacharparenright}{\kern0pt}\isanewline
\ \ \ \ \ \ \ \ {\isacharparenleft}{\kern0pt}Or\ {\isacharparenleft}{\kern0pt}list{\isacharunderscore}{\kern0pt}conj\ {\isacharparenleft}{\kern0pt}\isanewline
\ \ \ \ \ \ \ \ {\isacharparenleft}{\kern0pt}map\ {\isacharparenleft}{\kern0pt}substInfinitesimalQuadratic\ var\ {\isacharparenleft}{\kern0pt}{\isacharminus}{\kern0pt}b{\isacharparenright}{\kern0pt}\ {\isadigit{1}}\ {\isacharparenleft}{\kern0pt}b{\isacharcircum}{\kern0pt}{\isadigit{2}}{\isacharminus}{\kern0pt}{\isadigit{4}}{\isacharasterisk}{\kern0pt}a{\isacharasterisk}{\kern0pt}c{\isacharparenright}{\kern0pt}\ {\isacharparenleft}{\kern0pt}{\isadigit{2}}{\isacharasterisk}{\kern0pt}a{\isacharparenright}{\kern0pt}{\isacharparenright}{\kern0pt}\ L{\isacharparenright}{\kern0pt}{\isacharat}{\kern0pt}\isanewline
\ \ \ \ \ \ \ \ {\isacharparenleft}{\kern0pt}map\ {\isacharparenleft}{\kern0pt}substInfinitesimalQuadratic{\isacharunderscore}{\kern0pt}fm\ var\ {\isacharparenleft}{\kern0pt}{\isacharminus}{\kern0pt}b{\isacharparenright}{\kern0pt}\ {\isadigit{1}}\ {\isacharparenleft}{\kern0pt}b{\isacharcircum}{\kern0pt}{\isadigit{2}}{\isacharminus}{\kern0pt}{\isadigit{4}}{\isacharasterisk}{\kern0pt}a{\isacharasterisk}{\kern0pt}c{\isacharparenright}{\kern0pt}\ {\isacharparenleft}{\kern0pt}{\isadigit{2}}{\isacharasterisk}{\kern0pt}a{\isacharparenright}{\kern0pt}{\isacharparenright}{\kern0pt}\ F{\isacharparenright}{\kern0pt}\isanewline
\ \ \ \ \ \ \ \ {\isacharparenright}{\kern0pt}{\isacharparenright}{\kern0pt}\isanewline
\ \ \ \ \ \ \ \ {\isacharparenleft}{\kern0pt}list{\isacharunderscore}{\kern0pt}conj\ {\isacharparenleft}{\kern0pt}\isanewline
\ \ \ \ \ \ \ \ {\isacharparenleft}{\kern0pt}map\ {\isacharparenleft}{\kern0pt}substInfinitesimalQuadratic\ var\ {\isacharparenleft}{\kern0pt}{\isacharminus}{\kern0pt}b{\isacharparenright}{\kern0pt}\ {\isacharparenleft}{\kern0pt}{\isacharminus}{\kern0pt}{\isadigit{1}}{\isacharparenright}{\kern0pt}\ {\isacharparenleft}{\kern0pt}b{\isacharcircum}{\kern0pt}{\isadigit{2}}{\isacharminus}{\kern0pt}{\isadigit{4}}{\isacharasterisk}{\kern0pt}a{\isacharasterisk}{\kern0pt}c{\isacharparenright}{\kern0pt}\ {\isacharparenleft}{\kern0pt}{\isadigit{2}}{\isacharasterisk}{\kern0pt}a{\isacharparenright}{\kern0pt}{\isacharparenright}{\kern0pt}\ L{\isacharparenright}{\kern0pt}{\isacharat}{\kern0pt}\isanewline
\ \ \ \ \ \ \ \ {\isacharparenleft}{\kern0pt}map\ {\isacharparenleft}{\kern0pt}substInfinitesimalQuadratic{\isacharunderscore}{\kern0pt}fm\ var\ {\isacharparenleft}{\kern0pt}{\isacharminus}{\kern0pt}b{\isacharparenright}{\kern0pt}\ {\isacharparenleft}{\kern0pt}{\isacharminus}{\kern0pt}{\isadigit{1}}{\isacharparenright}{\kern0pt}\ {\isacharparenleft}{\kern0pt}b{\isacharcircum}{\kern0pt}{\isadigit{2}}{\isacharminus}{\kern0pt}{\isadigit{4}}{\isacharasterisk}{\kern0pt}a{\isacharasterisk}{\kern0pt}c{\isacharparenright}{\kern0pt}\ {\isacharparenleft}{\kern0pt}{\isadigit{2}}{\isacharasterisk}{\kern0pt}a{\isacharparenright}{\kern0pt}{\isacharparenright}{\kern0pt}\ F{\isacharparenright}{\kern0pt}\isanewline
\ \ \ \ \ \ \ \ {\isacharparenright}{\kern0pt}{\isacharparenright}{\kern0pt}\isanewline
\ \ \ \ \ \ \ {\isacharparenright}{\kern0pt}{\isacharparenright}{\kern0pt}\isanewline
\ \ \ \ \ \ {\isacharparenright}{\kern0pt}{\isacharparenright}{\kern0pt}{\isacharparenright}{\kern0pt}\isanewline
{\isachardoublequoteclose}{\isacharbar}{\kern0pt}\isanewline
\ \ {\isachardoublequoteopen}elimVar\ var\ L\ F\ {\isacharparenleft}{\kern0pt}Leq\ p{\isacharparenright}{\kern0pt}\ {\isacharequal}{\kern0pt}\ \ {\isacharparenleft}{\kern0pt}\isanewline
\ \ let\ {\isacharparenleft}{\kern0pt}a{\isacharcomma}{\kern0pt}b{\isacharcomma}{\kern0pt}c{\isacharparenright}{\kern0pt}\ {\isacharequal}{\kern0pt}\ get{\isacharunderscore}{\kern0pt}coeffs\ var\ p\ in\isanewline
\isanewline
\ \ \ {\isacharparenleft}{\kern0pt}Or\ \isanewline
\ \ \ \ \ \ \isanewline
\ \ \ \ \ \ {\isacharparenleft}{\kern0pt}And\ {\isacharparenleft}{\kern0pt}And\ {\isacharparenleft}{\kern0pt}Atom\ {\isacharparenleft}{\kern0pt}Eq\ a{\isacharparenright}{\kern0pt}{\isacharparenright}{\kern0pt}\ {\isacharparenleft}{\kern0pt}Atom\ {\isacharparenleft}{\kern0pt}Neq\ b{\isacharparenright}{\kern0pt}{\isacharparenright}{\kern0pt}{\isacharparenright}{\kern0pt}\isanewline
\ \ \ \ \ \ {\isacharparenleft}{\kern0pt}list{\isacharunderscore}{\kern0pt}conj\ {\isacharparenleft}{\kern0pt}\isanewline
\ \ \ \ \ \ \ \ {\isacharparenleft}{\kern0pt}map\ {\isacharparenleft}{\kern0pt}{\isasymlambda}a{\isachardot}{\kern0pt}\ Atom\ {\isacharparenleft}{\kern0pt}linear{\isacharunderscore}{\kern0pt}substitution\ var\ {\isacharparenleft}{\kern0pt}{\isacharminus}{\kern0pt}c{\isacharparenright}{\kern0pt}\ b\ a{\isacharparenright}{\kern0pt}{\isacharparenright}{\kern0pt}\ L{\isacharparenright}{\kern0pt}{\isacharat}{\kern0pt}\isanewline
\ \ \ \ \ \ \ \ {\isacharparenleft}{\kern0pt}map\ {\isacharparenleft}{\kern0pt}linear{\isacharunderscore}{\kern0pt}substitution{\isacharunderscore}{\kern0pt}fm\ var\ {\isacharparenleft}{\kern0pt}{\isacharminus}{\kern0pt}c{\isacharparenright}{\kern0pt}\ b{\isacharparenright}{\kern0pt}\ F{\isacharparenright}{\kern0pt}\isanewline
\ \ \ \ \ \ \ \ {\isacharparenright}{\kern0pt}{\isacharparenright}{\kern0pt}{\isacharparenright}{\kern0pt}\isanewline
\ \ \ \ \ \ \isanewline
\isanewline
\ \ \ \ \ \ {\isacharparenleft}{\kern0pt}And\ {\isacharparenleft}{\kern0pt}Atom\ {\isacharparenleft}{\kern0pt}Neq\ a{\isacharparenright}{\kern0pt}{\isacharparenright}{\kern0pt}\ {\isacharparenleft}{\kern0pt}And\ {\isacharparenleft}{\kern0pt}Atom{\isacharparenleft}{\kern0pt}Leq\ {\isacharparenleft}{\kern0pt}{\isacharminus}{\kern0pt}{\isacharparenleft}{\kern0pt}b{\isacharcircum}{\kern0pt}{\isadigit{2}}{\isacharparenright}{\kern0pt}{\isacharplus}{\kern0pt}{\isadigit{4}}{\isacharasterisk}{\kern0pt}a{\isacharasterisk}{\kern0pt}c{\isacharparenright}{\kern0pt}{\isacharparenright}{\kern0pt}{\isacharparenright}{\kern0pt}\isanewline
\ \ \ \ \ \ \ \ {\isacharparenleft}{\kern0pt}Or\ {\isacharparenleft}{\kern0pt}list{\isacharunderscore}{\kern0pt}conj\ {\isacharparenleft}{\kern0pt}\isanewline
\ \ \ \ \ \ \ \ {\isacharparenleft}{\kern0pt}map\ {\isacharparenleft}{\kern0pt}quadratic{\isacharunderscore}{\kern0pt}sub\ var\ {\isacharparenleft}{\kern0pt}{\isacharminus}{\kern0pt}b{\isacharparenright}{\kern0pt}\ {\isadigit{1}}\ {\isacharparenleft}{\kern0pt}b{\isacharcircum}{\kern0pt}{\isadigit{2}}{\isacharminus}{\kern0pt}{\isadigit{4}}{\isacharasterisk}{\kern0pt}a{\isacharasterisk}{\kern0pt}c{\isacharparenright}{\kern0pt}\ {\isacharparenleft}{\kern0pt}{\isadigit{2}}{\isacharasterisk}{\kern0pt}a{\isacharparenright}{\kern0pt}{\isacharparenright}{\kern0pt}\ L{\isacharparenright}{\kern0pt}{\isacharat}{\kern0pt}\isanewline
\ \ \ \ \ \ \ \ {\isacharparenleft}{\kern0pt}map\ {\isacharparenleft}{\kern0pt}quadratic{\isacharunderscore}{\kern0pt}sub{\isacharunderscore}{\kern0pt}fm\ var\ {\isacharparenleft}{\kern0pt}{\isacharminus}{\kern0pt}b{\isacharparenright}{\kern0pt}\ {\isadigit{1}}\ {\isacharparenleft}{\kern0pt}b{\isacharcircum}{\kern0pt}{\isadigit{2}}{\isacharminus}{\kern0pt}{\isadigit{4}}{\isacharasterisk}{\kern0pt}a{\isacharasterisk}{\kern0pt}c{\isacharparenright}{\kern0pt}\ {\isacharparenleft}{\kern0pt}{\isadigit{2}}{\isacharasterisk}{\kern0pt}a{\isacharparenright}{\kern0pt}{\isacharparenright}{\kern0pt}\ F{\isacharparenright}{\kern0pt}\isanewline
\ \ \ \ \ \ \ \ {\isacharparenright}{\kern0pt}{\isacharparenright}{\kern0pt}\isanewline
\ \ \ \ \ \ \ \ {\isacharparenleft}{\kern0pt}list{\isacharunderscore}{\kern0pt}conj\ {\isacharparenleft}{\kern0pt}\isanewline
\ \ \ \ \ \ \ \ {\isacharparenleft}{\kern0pt}map\ {\isacharparenleft}{\kern0pt}quadratic{\isacharunderscore}{\kern0pt}sub\ var\ {\isacharparenleft}{\kern0pt}{\isacharminus}{\kern0pt}b{\isacharparenright}{\kern0pt}\ {\isacharparenleft}{\kern0pt}{\isacharminus}{\kern0pt}{\isadigit{1}}{\isacharparenright}{\kern0pt}\ {\isacharparenleft}{\kern0pt}b{\isacharcircum}{\kern0pt}{\isadigit{2}}{\isacharminus}{\kern0pt}{\isadigit{4}}{\isacharasterisk}{\kern0pt}a{\isacharasterisk}{\kern0pt}c{\isacharparenright}{\kern0pt}\ {\isacharparenleft}{\kern0pt}{\isadigit{2}}{\isacharasterisk}{\kern0pt}a{\isacharparenright}{\kern0pt}{\isacharparenright}{\kern0pt}\ L{\isacharparenright}{\kern0pt}{\isacharat}{\kern0pt}\isanewline
\ \ \ \ \ \ \ \ {\isacharparenleft}{\kern0pt}map\ {\isacharparenleft}{\kern0pt}quadratic{\isacharunderscore}{\kern0pt}sub{\isacharunderscore}{\kern0pt}fm\ var\ {\isacharparenleft}{\kern0pt}{\isacharminus}{\kern0pt}b{\isacharparenright}{\kern0pt}\ {\isacharparenleft}{\kern0pt}{\isacharminus}{\kern0pt}{\isadigit{1}}{\isacharparenright}{\kern0pt}\ {\isacharparenleft}{\kern0pt}b{\isacharcircum}{\kern0pt}{\isadigit{2}}{\isacharminus}{\kern0pt}{\isadigit{4}}{\isacharasterisk}{\kern0pt}a{\isacharasterisk}{\kern0pt}c{\isacharparenright}{\kern0pt}\ {\isacharparenleft}{\kern0pt}{\isadigit{2}}{\isacharasterisk}{\kern0pt}a{\isacharparenright}{\kern0pt}{\isacharparenright}{\kern0pt}\ F{\isacharparenright}{\kern0pt}\isanewline
\ \ \ \ \ \ \ \ {\isacharparenright}{\kern0pt}{\isacharparenright}{\kern0pt}\isanewline
\ \ \ \ \ \ \ {\isacharparenright}{\kern0pt}{\isacharparenright}{\kern0pt}\isanewline
\ \ \ \ \ \ {\isacharparenright}{\kern0pt}{\isacharparenright}{\kern0pt}\isanewline
\isanewline
{\isacharparenright}{\kern0pt}{\isachardoublequoteclose}%
}
\snip{substNegInfinityFun}{1}{2}{%
\isacommand{fun}\isamarkupfalse%
\ substNegInfinity\ {\isacharcolon}{\kern0pt}{\isacharcolon}{\kern0pt}\ {\isachardoublequoteopen}nat\ {\isasymRightarrow}\ atom\ {\isasymRightarrow}\ atom\ fm{\isachardoublequoteclose}\ \isakeyword{where}\isanewline
\ \ {\isachardoublequoteopen}substNegInfinity\ var\ {\isacharparenleft}{\kern0pt}Eq\ p{\isacharparenright}{\kern0pt}\ {\isacharequal}{\kern0pt}\ allZero\ p\ var\ {\isachardoublequoteclose}\ {\isacharbar}{\kern0pt}\isanewline
\ \ {\isachardoublequoteopen}substNegInfinity\ var\ {\isacharparenleft}{\kern0pt}Less\ p{\isacharparenright}{\kern0pt}\ {\isacharequal}{\kern0pt}\ alternateNegInfinity\ p\ var{\isachardoublequoteclose}\ {\isacharbar}{\kern0pt}\isanewline
\ \ {\isachardoublequoteopen}substNegInfinity\ var\ {\isacharparenleft}{\kern0pt}Leq\ p{\isacharparenright}{\kern0pt}\ {\isacharequal}{\kern0pt}\ Or\ {\isacharparenleft}{\kern0pt}alternateNegInfinity\ p\ var{\isacharparenright}{\kern0pt}\ \isanewline
\ \ \ \ {\isacharparenleft}{\kern0pt}allZero\ p\ var{\isacharparenright}{\kern0pt}{\isachardoublequoteclose}\ {\isacharbar}{\kern0pt}\isanewline
\ \ {\isachardoublequoteopen}substNegInfinity\ var\ {\isacharparenleft}{\kern0pt}Neq\ p{\isacharparenright}{\kern0pt}\ {\isacharequal}{\kern0pt}\ Neg\ {\isacharparenleft}{\kern0pt}allZero\ p\ var{\isacharparenright}{\kern0pt}{\isachardoublequoteclose}%
}
\snip{convertsubstNegInfinityLem}{1}{2}{%
\isacommand{lemma}\isamarkupfalse%
\ convert{\isacharunderscore}{\kern0pt}substNegInfinity{\isacharcolon}{\kern0pt}\ \isanewline
\ \ \isakeyword{assumes}\ {\isachardoublequoteopen}convert{\isacharunderscore}{\kern0pt}atom\ var\ A\ {\isacharparenleft}{\kern0pt}{\isasymnu}{\isacharprime}{\kern0pt}\ {\isacharat}{\kern0pt}\ x\ {\isacharhash}{\kern0pt}\ {\isasymnu}{\isacharparenright}{\kern0pt}\ {\isacharequal}{\kern0pt}\ Some{\isacharparenleft}{\kern0pt}A{\isacharprime}{\kern0pt}{\isacharparenright}{\kern0pt}{\isachardoublequoteclose}\isanewline
\ \ \isakeyword{assumes}\ {\isachardoublequoteopen}length\ {\isasymnu}{\isacharprime}{\kern0pt}\ {\isacharequal}{\kern0pt}\ var{\isachardoublequoteclose}\isanewline
\ \ \isakeyword{shows}\ {\isachardoublequoteopen}eval\ {\isacharparenleft}{\kern0pt}substNegInfinity\ var\ A{\isacharparenright}{\kern0pt}\ {\isacharparenleft}{\kern0pt}{\isasymnu}{\isacharprime}{\kern0pt}\ {\isacharat}{\kern0pt}\ x\ {\isacharhash}{\kern0pt}\ {\isasymnu}{\isacharparenright}{\kern0pt}\ {\isacharequal}{\kern0pt}\ evalUni\ {\isacharparenleft}{\kern0pt}substNegInfinityUni\ A{\isacharprime}{\kern0pt}{\isacharparenright}{\kern0pt}\ x{\isachardoublequoteclose}%
}
\snip{infinityevalLem}{1}{2}{%
\isacommand{lemma}\isamarkupfalse%
\ infinity{\isacharunderscore}{\kern0pt}evalUni{\isacharcolon}{\kern0pt}\ \isakeyword{shows}\ {\isachardoublequoteopen}{\isacharparenleft}{\kern0pt}{\isasymexists}y{\isachardot}{\kern0pt}\ {\isasymforall}x{\isacharless}{\kern0pt}y{\isachardot}{\kern0pt}\ aEvalUni\ At\ x{\isacharparenright}{\kern0pt}\ {\isacharequal}{\kern0pt}\ \isanewline
\ \ {\isacharparenleft}{\kern0pt}evalUni\ {\isacharparenleft}{\kern0pt}substNegInfinityUni\ At{\isacharparenright}{\kern0pt}\ x{\isacharparenright}{\kern0pt}{\isachardoublequoteclose}%
}
\snip{infinitesimalquad}{1}{2}{%
\isacommand{lemma}\isamarkupfalse%
\ infinitesimal{\isacharunderscore}{\kern0pt}quad{\isacharcolon}{\kern0pt}\isanewline
\ \ \isakeyword{fixes}\ A\ B\ C\ D{\isacharcolon}{\kern0pt}{\isacharcolon}{\kern0pt}\ {\isachardoublequoteopen}real{\isachardoublequoteclose}\isanewline
\ \ \isakeyword{assumes}\ {\isachardoublequoteopen}D{\isasymnoteq}{\isadigit{0}}{\isachardoublequoteclose}\isanewline
\ \ \isakeyword{assumes}\ {\isachardoublequoteopen}C{\isasymge}{\isadigit{0}}{\isachardoublequoteclose}\isanewline
\ \ \isakeyword{shows}\ {\isachardoublequoteopen}{\isacharparenleft}{\kern0pt}{\isasymexists}y{\isacharcolon}{\kern0pt}{\isacharcolon}{\kern0pt}real{\isachargreater}{\kern0pt}{\isacharparenleft}{\kern0pt}{\isacharparenleft}{\kern0pt}A{\isacharplus}{\kern0pt}B\ {\isacharasterisk}{\kern0pt}\ sqrt{\isacharparenleft}{\kern0pt}C{\isacharparenright}{\kern0pt}{\isacharparenright}{\kern0pt}{\isacharslash}{\kern0pt}{\isacharparenleft}{\kern0pt}D{\isacharparenright}{\kern0pt}{\isacharparenright}{\kern0pt}{\isachardot}{\kern0pt}\ \isanewline
\ \ \ \ \ \ {\isasymforall}x{\isacharcolon}{\kern0pt}{\isacharcolon}{\kern0pt}real\ {\isasymin}{\isacharbraceleft}{\kern0pt}{\isacharparenleft}{\kern0pt}{\isacharparenleft}{\kern0pt}A{\isacharplus}{\kern0pt}B\ {\isacharasterisk}{\kern0pt}\ sqrt{\isacharparenleft}{\kern0pt}C{\isacharparenright}{\kern0pt}{\isacharparenright}{\kern0pt}{\isacharslash}{\kern0pt}{\isacharparenleft}{\kern0pt}D{\isacharparenright}{\kern0pt}{\isacharparenright}{\kern0pt}{\isacharless}{\kern0pt}{\isachardot}{\kern0pt}{\isachardot}{\kern0pt}y{\isacharbraceright}{\kern0pt}{\isachardot}{\kern0pt}\ aEvalUni\ At\ x{\isacharparenright}{\kern0pt}\isanewline
\ \ \ \ \ \ {\isacharequal}{\kern0pt}\ {\isacharparenleft}{\kern0pt}evalUni\ {\isacharparenleft}{\kern0pt}substInfinitesimalQuadraticUni\ A\ B\ C\ D\ At{\isacharparenright}{\kern0pt}\ x{\isacharparenright}{\kern0pt}{\isachardoublequoteclose}%
}
\snip{generalqe}{1}{2}{%
\isacommand{theorem}\isamarkupfalse%
\ general{\isacharunderscore}{\kern0pt}qe{\isacharcolon}{\kern0pt}\isanewline
\ \ \isakeyword{defines}\ {\isachardoublequoteopen}R\ {\isasymequiv}\ {\isacharbraceleft}{\kern0pt}{\isacharparenleft}{\kern0pt}{\isacharequal}{\kern0pt}{\isacharparenright}{\kern0pt}{\isacharcomma}{\kern0pt}\ {\isacharparenleft}{\kern0pt}{\isacharless}{\kern0pt}{\isacharparenright}{\kern0pt}{\isacharcomma}{\kern0pt}\ {\isacharparenleft}{\kern0pt}{\isasymle}{\isacharparenright}{\kern0pt}{\isacharcomma}{\kern0pt}\ {\isacharparenleft}{\kern0pt}{\isasymnoteq}{\isacharparenright}{\kern0pt}{\isacharbraceright}{\kern0pt}{\isachardoublequoteclose}\isanewline
\ \ \isakeyword{assumes}\ {\isachardoublequoteopen}{\isasymforall}rel{\isasymin}R{\isachardot}{\kern0pt}\ finite\ {\isacharparenleft}{\kern0pt}Atoms\ rel{\isacharparenright}{\kern0pt}{\isachardoublequoteclose}\isanewline
\ \ \isakeyword{defines}\ {\isachardoublequoteopen}F\ {\isasymequiv}\ {\isacharparenleft}{\kern0pt}{\isasymlambda}x{\isachardot}{\kern0pt}\ {\isasymforall}rel{\isasymin}R{\isachardot}{\kern0pt}\ {\isasymforall}{\isacharparenleft}{\kern0pt}a{\isacharcomma}{\kern0pt}b{\isacharcomma}{\kern0pt}c{\isacharparenright}{\kern0pt}{\isasymin}{\isacharparenleft}{\kern0pt}Atoms\ rel{\isacharparenright}{\kern0pt}{\isachardot}{\kern0pt}\ rel\ {\isacharparenleft}{\kern0pt}a{\isacharasterisk}{\kern0pt}x\isactrlsup {\isadigit{2}}{\isacharplus}{\kern0pt}b{\isacharasterisk}{\kern0pt}x{\isacharplus}{\kern0pt}c{\isacharparenright}{\kern0pt}\ {\isadigit{0}}{\isacharparenright}{\kern0pt}{\isachardoublequoteclose}\isanewline
\ \ \isakeyword{defines}\ {\isachardoublequoteopen}F{\isasymepsilon}\ {\isasymequiv}\ {\isacharparenleft}{\kern0pt}{\isasymlambda}r{\isachardot}{\kern0pt}\ {\isasymforall}rel{\isasymin}R{\isachardot}{\kern0pt}\ {\isasymforall}{\isacharparenleft}{\kern0pt}a{\isacharcomma}{\kern0pt}b{\isacharcomma}{\kern0pt}c{\isacharparenright}{\kern0pt}{\isasymin}{\isacharparenleft}{\kern0pt}Atoms\ rel{\isacharparenright}{\kern0pt}{\isachardot}{\kern0pt}\ {\isasymexists}y{\isachargreater}{\kern0pt}r{\isachardot}{\kern0pt}\ {\isasymforall}x{\isasymin}{\isacharbraceleft}{\kern0pt}r{\isacharless}{\kern0pt}{\isachardot}{\kern0pt}{\isachardot}{\kern0pt}y{\isacharbraceright}{\kern0pt}{\isachardot}{\kern0pt}\ \isanewline
\ \ \ \ rel\ {\isacharparenleft}{\kern0pt}a{\isacharasterisk}{\kern0pt}x\isactrlsup {\isadigit{2}}{\isacharplus}{\kern0pt}b{\isacharasterisk}{\kern0pt}x{\isacharplus}{\kern0pt}c{\isacharparenright}{\kern0pt}\ {\isadigit{0}}{\isacharparenright}{\kern0pt}{\isachardoublequoteclose}\isanewline
\ \ \isakeyword{defines}\ {\isachardoublequoteopen}F\isactrlsub i\isactrlsub n\isactrlsub f\ {\isasymequiv}\ {\isacharparenleft}{\kern0pt}{\isasymforall}rel{\isasymin}R{\isachardot}{\kern0pt}\ {\isasymforall}{\isacharparenleft}{\kern0pt}a{\isacharcomma}{\kern0pt}b{\isacharcomma}{\kern0pt}c{\isacharparenright}{\kern0pt}{\isasymin}{\isacharparenleft}{\kern0pt}Atoms\ rel{\isacharparenright}{\kern0pt}{\isachardot}{\kern0pt}\ {\isasymexists}x{\isachardot}{\kern0pt}\ {\isasymforall}y{\isacharless}{\kern0pt}x{\isachardot}{\kern0pt}\ \isanewline
\ \ \ \ rel\ {\isacharparenleft}{\kern0pt}a{\isacharasterisk}{\kern0pt}y\isactrlsup {\isadigit{2}}{\isacharplus}{\kern0pt}b{\isacharasterisk}{\kern0pt}y{\isacharplus}{\kern0pt}c{\isacharparenright}{\kern0pt}\ {\isadigit{0}}{\isacharparenright}{\kern0pt}{\isachardoublequoteclose}\isanewline
\ \ \isakeyword{defines}\ {\isachardoublequoteopen}roots\ {\isasymequiv}\ {\isacharparenleft}{\kern0pt}{\isasymlambda}{\isacharparenleft}{\kern0pt}a{\isacharcomma}{\kern0pt}b{\isacharcomma}{\kern0pt}c{\isacharparenright}{\kern0pt}{\isachardot}{\kern0pt}\isanewline
\ \ \ \ \ if\ a{\isacharequal}{\kern0pt}{\isadigit{0}}\ {\isasymand}\ b{\isasymnoteq}{\isadigit{0}}\ then\ {\isacharbraceleft}{\kern0pt}{\isacharminus}{\kern0pt}c{\isacharslash}{\kern0pt}b{\isacharbraceright}{\kern0pt}\ else\isanewline
\ \ \ \ \ if\ a{\isasymnoteq}{\isadigit{0}}\ {\isasymand}\ b\isactrlsup {\isadigit{2}}{\isacharminus}{\kern0pt}{\isadigit{4}}{\isacharasterisk}{\kern0pt}a{\isacharasterisk}{\kern0pt}c{\isasymge}{\isadigit{0}}\ then\ {\isacharbraceleft}{\kern0pt}{\isacharparenleft}{\kern0pt}{\isacharminus}{\kern0pt}b{\isacharplus}{\kern0pt}sqrt{\isacharparenleft}{\kern0pt}b\isactrlsup {\isadigit{2}}{\isacharminus}{\kern0pt}{\isadigit{4}}{\isacharasterisk}{\kern0pt}a{\isacharasterisk}{\kern0pt}c{\isacharparenright}{\kern0pt}{\isacharparenright}{\kern0pt}{\isacharslash}{\kern0pt}{\isacharparenleft}{\kern0pt}{\isadigit{2}}{\isacharasterisk}{\kern0pt}a{\isacharparenright}{\kern0pt}{\isacharbraceright}{\kern0pt}\isanewline
\ \ \ \ \ \ \ \ {\isasymunion}\ {\isacharbraceleft}{\kern0pt}{\isacharparenleft}{\kern0pt}{\isacharminus}{\kern0pt}b{\isacharminus}{\kern0pt}sqrt{\isacharparenleft}{\kern0pt}b\isactrlsup {\isadigit{2}}{\isacharminus}{\kern0pt}{\isadigit{4}}{\isacharasterisk}{\kern0pt}a{\isacharasterisk}{\kern0pt}c{\isacharparenright}{\kern0pt}{\isacharparenright}{\kern0pt}{\isacharslash}{\kern0pt}{\isacharparenleft}{\kern0pt}{\isadigit{2}}{\isacharasterisk}{\kern0pt}a{\isacharparenright}{\kern0pt}{\isacharbraceright}{\kern0pt}\ else\ {\isacharbraceleft}{\kern0pt}{\isacharbraceright}{\kern0pt}{\isacharparenright}{\kern0pt}{\isachardoublequoteclose}\isanewline
\ \ \isakeyword{shows}\ {\isachardoublequoteopen}{\isacharparenleft}{\kern0pt}{\isasymexists}x{\isachardot}{\kern0pt}\ F{\isacharparenleft}{\kern0pt}x{\isacharparenright}{\kern0pt}{\isacharparenright}{\kern0pt}\ {\isacharequal}{\kern0pt}\ \ {\isacharparenleft}{\kern0pt}F\isactrlsub i\isactrlsub n\isactrlsub f\ {\isasymor}\isanewline
\ \ \ \ \ \ \ \ \ \ \ \ {\isacharparenleft}{\kern0pt}{\isasymexists}r{\isasymin}{\isasymUnion}{\isacharparenleft}{\kern0pt}roots\ {\isacharbackquote}{\kern0pt}\ {\isacharparenleft}{\kern0pt}Atoms\ {\isacharparenleft}{\kern0pt}{\isacharequal}{\kern0pt}{\isacharparenright}{\kern0pt}\ {\isasymunion}\ Atoms\ {\isacharparenleft}{\kern0pt}{\isasymle}{\isacharparenright}{\kern0pt}{\isacharparenright}{\kern0pt}{\isacharparenright}{\kern0pt}{\isachardot}{\kern0pt}\ F\ r{\isacharparenright}{\kern0pt}\ {\isasymor}\isanewline
\ \ \ \ \ \ \ \ \ \ \ \ {\isacharparenleft}{\kern0pt}{\isasymexists}r{\isasymin}{\isasymUnion}{\isacharparenleft}{\kern0pt}roots\ {\isacharbackquote}{\kern0pt}\ {\isacharparenleft}{\kern0pt}Atoms\ {\isacharparenleft}{\kern0pt}{\isacharless}{\kern0pt}{\isacharparenright}{\kern0pt}\ {\isasymunion}\ Atoms\ {\isacharparenleft}{\kern0pt}{\isasymnoteq}{\isacharparenright}{\kern0pt}{\isacharparenright}{\kern0pt}{\isacharparenright}{\kern0pt}{\isachardot}{\kern0pt}\ F{\isasymepsilon}\ r{\isacharparenright}{\kern0pt}{\isacharparenright}{\kern0pt}{\isachardoublequoteclose}%
}
\snip{linearsubstitution}{1}{2}{%
\isacommand{fun}\isamarkupfalse%
\ linear{\isacharunderscore}{\kern0pt}substitution\ {\isacharcolon}{\kern0pt}{\isacharcolon}{\kern0pt}\ {\isachardoublequoteopen}nat\ {\isasymRightarrow}\ real\ mpoly\ {\isasymRightarrow}\ real\ mpoly\ {\isasymRightarrow}\ atom\ {\isasymRightarrow}\ atom{\isachardoublequoteclose}\ \isakeyword{where}\isanewline
\ \ {\isachardoublequoteopen}linear{\isacharunderscore}{\kern0pt}substitution\ var\ a\ b\ {\isacharparenleft}{\kern0pt}Eq\ p{\isacharparenright}{\kern0pt}\ {\isacharequal}{\kern0pt}\ \isanewline
\ \ {\isacharparenleft}{\kern0pt}let\ d\ {\isacharequal}{\kern0pt}\ MPoly{\isacharunderscore}{\kern0pt}Type{\isachardot}{\kern0pt}degree\ p\ var\ in\isanewline
\ \ \ \ {\isacharparenleft}{\kern0pt}Eq\ {\isacharparenleft}{\kern0pt}{\isasymSum}i{\isasymin}{\isacharbraceleft}{\kern0pt}{\isadigit{0}}{\isachardot}{\kern0pt}{\isachardot}{\kern0pt}{\isacharless}{\kern0pt}{\isacharparenleft}{\kern0pt}d{\isacharplus}{\kern0pt}{\isadigit{1}}{\isacharparenright}{\kern0pt}{\isacharbraceright}{\kern0pt}{\isachardot}{\kern0pt}\ \ isolate{\isacharunderscore}{\kern0pt}variable{\isacharunderscore}{\kern0pt}sparse\ p\ var\ i\ {\isacharasterisk}{\kern0pt}\ {\isacharparenleft}{\kern0pt}a{\isacharcircum}{\kern0pt}i{\isacharparenright}{\kern0pt}\ {\isacharasterisk}{\kern0pt}\ {\isacharparenleft}{\kern0pt}b{\isacharcircum}{\kern0pt}{\isacharparenleft}{\kern0pt}d{\isacharminus}{\kern0pt}i{\isacharparenright}{\kern0pt}{\isacharparenright}{\kern0pt}{\isacharparenright}{\kern0pt}{\isacharparenright}{\kern0pt}\isanewline
\ \ {\isacharparenright}{\kern0pt}{\isachardoublequoteclose}\ {\isacharbar}{\kern0pt}\isanewline
\ \ {\isachardoublequoteopen}linear{\isacharunderscore}{\kern0pt}substitution\ var\ a\ b\ {\isacharparenleft}{\kern0pt}Less\ p{\isacharparenright}{\kern0pt}\ {\isacharequal}{\kern0pt}\ \isanewline
\ \ {\isacharparenleft}{\kern0pt}let\ d\ {\isacharequal}{\kern0pt}\ MPoly{\isacharunderscore}{\kern0pt}Type{\isachardot}{\kern0pt}degree\ p\ var\ in\isanewline
\ \ \ \ let\ P\ {\isacharequal}{\kern0pt}\ {\isacharparenleft}{\kern0pt}{\isasymSum}i{\isasymin}{\isacharbraceleft}{\kern0pt}{\isadigit{0}}{\isachardot}{\kern0pt}{\isachardot}{\kern0pt}{\isacharless}{\kern0pt}{\isacharparenleft}{\kern0pt}d{\isacharplus}{\kern0pt}{\isadigit{1}}{\isacharparenright}{\kern0pt}{\isacharbraceright}{\kern0pt}{\isachardot}{\kern0pt}\ isolate{\isacharunderscore}{\kern0pt}variable{\isacharunderscore}{\kern0pt}sparse\ p\ var\ i\ {\isacharasterisk}{\kern0pt}\ {\isacharparenleft}{\kern0pt}a{\isacharcircum}{\kern0pt}i{\isacharparenright}{\kern0pt}\ {\isacharasterisk}{\kern0pt}\ {\isacharparenleft}{\kern0pt}b{\isacharcircum}{\kern0pt}{\isacharparenleft}{\kern0pt}d{\isacharminus}{\kern0pt}i{\isacharparenright}{\kern0pt}{\isacharparenright}{\kern0pt}{\isacharparenright}{\kern0pt}\ in\isanewline
\ \ \ \ \ \ {\isacharparenleft}{\kern0pt}Less{\isacharparenleft}{\kern0pt}P\ {\isacharasterisk}{\kern0pt}\ {\isacharparenleft}{\kern0pt}b\ {\isacharcircum}{\kern0pt}\ {\isacharparenleft}{\kern0pt}d\ mod\ {\isadigit{2}}{\isacharparenright}{\kern0pt}{\isacharparenright}{\kern0pt}{\isacharparenright}{\kern0pt}{\isacharparenright}{\kern0pt}\isanewline
\ \ \ \ {\isacharparenright}{\kern0pt}{\isachardoublequoteclose}{\isacharbar}{\kern0pt}\isanewline
\ \ {\isachardoublequoteopen}linear{\isacharunderscore}{\kern0pt}substitution\ var\ a\ b\ {\isacharparenleft}{\kern0pt}Leq\ p{\isacharparenright}{\kern0pt}\ {\isacharequal}{\kern0pt}\ \isanewline
\ \ {\isacharparenleft}{\kern0pt}let\ d\ {\isacharequal}{\kern0pt}\ MPoly{\isacharunderscore}{\kern0pt}Type{\isachardot}{\kern0pt}degree\ p\ var\ in\isanewline
\ \ \ \ let\ P\ {\isacharequal}{\kern0pt}\ {\isacharparenleft}{\kern0pt}{\isasymSum}i{\isasymin}{\isacharbraceleft}{\kern0pt}{\isadigit{0}}{\isachardot}{\kern0pt}{\isachardot}{\kern0pt}{\isacharless}{\kern0pt}{\isacharparenleft}{\kern0pt}d{\isacharplus}{\kern0pt}{\isadigit{1}}{\isacharparenright}{\kern0pt}{\isacharbraceright}{\kern0pt}{\isachardot}{\kern0pt}\ isolate{\isacharunderscore}{\kern0pt}variable{\isacharunderscore}{\kern0pt}sparse\ p\ var\ i\ {\isacharasterisk}{\kern0pt}\ {\isacharparenleft}{\kern0pt}a{\isacharcircum}{\kern0pt}i{\isacharparenright}{\kern0pt}\ {\isacharasterisk}{\kern0pt}\ {\isacharparenleft}{\kern0pt}b{\isacharcircum}{\kern0pt}{\isacharparenleft}{\kern0pt}d{\isacharminus}{\kern0pt}i{\isacharparenright}{\kern0pt}{\isacharparenright}{\kern0pt}{\isacharparenright}{\kern0pt}\ in\isanewline
\ \ \ \ \ \ {\isacharparenleft}{\kern0pt}Leq{\isacharparenleft}{\kern0pt}P\ {\isacharasterisk}{\kern0pt}\ {\isacharparenleft}{\kern0pt}b\ {\isacharcircum}{\kern0pt}\ {\isacharparenleft}{\kern0pt}d\ mod\ {\isadigit{2}}{\isacharparenright}{\kern0pt}{\isacharparenright}{\kern0pt}{\isacharparenright}{\kern0pt}{\isacharparenright}{\kern0pt}\isanewline
\ \ \ \ {\isacharparenright}{\kern0pt}{\isachardoublequoteclose}{\isacharbar}{\kern0pt}\isanewline
\ \ {\isachardoublequoteopen}linear{\isacharunderscore}{\kern0pt}substitution\ var\ a\ b\ {\isacharparenleft}{\kern0pt}Neq\ p{\isacharparenright}{\kern0pt}\ {\isacharequal}{\kern0pt}\ \isanewline
\ \ {\isacharparenleft}{\kern0pt}let\ d\ {\isacharequal}{\kern0pt}\ MPoly{\isacharunderscore}{\kern0pt}Type{\isachardot}{\kern0pt}degree\ p\ var\ in\isanewline
\ \ \ \ {\isacharparenleft}{\kern0pt}Neq\ {\isacharparenleft}{\kern0pt}{\isasymSum}i{\isasymin}{\isacharbraceleft}{\kern0pt}{\isadigit{0}}{\isachardot}{\kern0pt}{\isachardot}{\kern0pt}{\isacharless}{\kern0pt}{\isacharparenleft}{\kern0pt}d{\isacharplus}{\kern0pt}{\isadigit{1}}{\isacharparenright}{\kern0pt}{\isacharbraceright}{\kern0pt}{\isachardot}{\kern0pt}\ \ isolate{\isacharunderscore}{\kern0pt}variable{\isacharunderscore}{\kern0pt}sparse\ p\ var\ i\ {\isacharasterisk}{\kern0pt}\ {\isacharparenleft}{\kern0pt}a{\isacharcircum}{\kern0pt}i{\isacharparenright}{\kern0pt}\ {\isacharasterisk}{\kern0pt}\ {\isacharparenleft}{\kern0pt}b{\isacharcircum}{\kern0pt}{\isacharparenleft}{\kern0pt}d{\isacharminus}{\kern0pt}i{\isacharparenright}{\kern0pt}{\isacharparenright}{\kern0pt}{\isacharparenright}{\kern0pt}{\isacharparenright}{\kern0pt}\isanewline
{\isacharparenright}{\kern0pt}{\isachardoublequoteclose}%
}
\snip{quadraticpartone}{1}{2}{%
\isacommand{fun}\isamarkupfalse%
\ quadratic{\isacharunderscore}{\kern0pt}part{\isacharunderscore}{\kern0pt}{\isadigit{1}}\ {\isacharcolon}{\kern0pt}{\isacharcolon}{\kern0pt}\ {\isachardoublequoteopen}nat\ {\isasymRightarrow}\ \isanewline
\ \ \ \ real\ mpoly\ {\isasymRightarrow}\ real\ mpoly\ {\isasymRightarrow}\ real\ mpoly\ {\isasymRightarrow}\ \isanewline
\ \ \ \ atom\ {\isasymRightarrow}\ real\ mpoly{\isachardoublequoteclose}\ \isakeyword{where}\isanewline
\ \ {\isachardoublequoteopen}quadratic{\isacharunderscore}{\kern0pt}part{\isacharunderscore}{\kern0pt}{\isadigit{1}}\ var\ a\ b\ d\ {\isacharparenleft}{\kern0pt}Eq\ p{\isacharparenright}{\kern0pt}\ {\isacharequal}{\kern0pt}\ {\isacharparenleft}{\kern0pt}\isanewline
\ \ let\ deg\ {\isacharequal}{\kern0pt}\ MPoly{\isacharunderscore}{\kern0pt}Type{\isachardot}{\kern0pt}degree\ p\ var\ in\isanewline
\ \ {\isasymSum}i{\isasymin}{\isacharbraceleft}{\kern0pt}{\isadigit{0}}{\isachardot}{\kern0pt}{\isachardot}{\kern0pt}{\isacharless}{\kern0pt}{\isacharparenleft}{\kern0pt}deg{\isacharplus}{\kern0pt}{\isadigit{1}}{\isacharparenright}{\kern0pt}{\isacharbraceright}{\kern0pt}{\isachardot}{\kern0pt}\ {\isacharparenleft}{\kern0pt}isolate{\isacharunderscore}{\kern0pt}variable{\isacharunderscore}{\kern0pt}sparse\ p\ var\ i{\isacharparenright}{\kern0pt}\ {\isacharasterisk}{\kern0pt}\isanewline
\ \ \ \ \ {\isacharparenleft}{\kern0pt}{\isacharparenleft}{\kern0pt}a{\isacharplus}{\kern0pt}b{\isacharasterisk}{\kern0pt}{\isacharparenleft}{\kern0pt}Var\ var{\isacharparenright}{\kern0pt}{\isacharparenright}{\kern0pt}{\isacharcircum}{\kern0pt}i{\isacharparenright}{\kern0pt}\ {\isacharasterisk}{\kern0pt}\ {\isacharparenleft}{\kern0pt}d{\isacharcircum}{\kern0pt}{\isacharparenleft}{\kern0pt}deg\ {\isacharminus}{\kern0pt}\ i{\isacharparenright}{\kern0pt}{\isacharparenright}{\kern0pt}{\isacharparenright}{\kern0pt}{\isachardoublequoteclose}\ {\isacharbar}{\kern0pt}\isanewline
\ \ {\isachardoublequoteopen}quadratic{\isacharunderscore}{\kern0pt}part{\isacharunderscore}{\kern0pt}{\isadigit{1}}\ var\ a\ b\ d\ {\isacharparenleft}{\kern0pt}Less\ p{\isacharparenright}{\kern0pt}\ {\isacharequal}{\kern0pt}\ {\isacharparenleft}{\kern0pt}\isanewline
\ \ let\ deg\ {\isacharequal}{\kern0pt}\ MPoly{\isacharunderscore}{\kern0pt}Type{\isachardot}{\kern0pt}degree\ p\ var\ in\isanewline
\ \ let\ P\ {\isacharequal}{\kern0pt}\ {\isasymSum}i{\isasymin}{\isacharbraceleft}{\kern0pt}{\isadigit{0}}{\isachardot}{\kern0pt}{\isachardot}{\kern0pt}{\isacharless}{\kern0pt}{\isacharparenleft}{\kern0pt}deg{\isacharplus}{\kern0pt}{\isadigit{1}}{\isacharparenright}{\kern0pt}{\isacharbraceright}{\kern0pt}{\isachardot}{\kern0pt}\ {\isacharparenleft}{\kern0pt}isolate{\isacharunderscore}{\kern0pt}variable{\isacharunderscore}{\kern0pt}sparse\ p\ var\ i{\isacharparenright}{\kern0pt}\ {\isacharasterisk}{\kern0pt}\ \isanewline
\ \ \ \ {\isacharparenleft}{\kern0pt}{\isacharparenleft}{\kern0pt}a{\isacharplus}{\kern0pt}b{\isacharasterisk}{\kern0pt}{\isacharparenleft}{\kern0pt}Var\ var{\isacharparenright}{\kern0pt}{\isacharparenright}{\kern0pt}{\isacharcircum}{\kern0pt}i{\isacharparenright}{\kern0pt}\ {\isacharasterisk}{\kern0pt}\ {\isacharparenleft}{\kern0pt}d{\isacharcircum}{\kern0pt}{\isacharparenleft}{\kern0pt}deg\ {\isacharminus}{\kern0pt}\ i{\isacharparenright}{\kern0pt}{\isacharparenright}{\kern0pt}\ in\isanewline
\ \ P\ {\isacharasterisk}{\kern0pt}\ {\isacharparenleft}{\kern0pt}d\ {\isacharcircum}{\kern0pt}\ {\isacharparenleft}{\kern0pt}deg\ mod\ {\isadigit{2}}{\isacharparenright}{\kern0pt}{\isacharparenright}{\kern0pt}{\isacharparenright}{\kern0pt}{\isachardoublequoteclose}{\isacharbar}{\kern0pt}\isanewline
\ \ {\isachardoublequoteopen}quadratic{\isacharunderscore}{\kern0pt}part{\isacharunderscore}{\kern0pt}{\isadigit{1}}\ var\ a\ b\ d\ {\isacharparenleft}{\kern0pt}Leq\ p{\isacharparenright}{\kern0pt}\ {\isacharequal}{\kern0pt}\ {\isacharparenleft}{\kern0pt}\isanewline
\ \ let\ deg\ {\isacharequal}{\kern0pt}\ MPoly{\isacharunderscore}{\kern0pt}Type{\isachardot}{\kern0pt}degree\ p\ var\ in\isanewline
\ \ let\ P\ {\isacharequal}{\kern0pt}\ {\isasymSum}i{\isasymin}{\isacharbraceleft}{\kern0pt}{\isadigit{0}}{\isachardot}{\kern0pt}{\isachardot}{\kern0pt}{\isacharless}{\kern0pt}{\isacharparenleft}{\kern0pt}deg{\isacharplus}{\kern0pt}{\isadigit{1}}{\isacharparenright}{\kern0pt}{\isacharbraceright}{\kern0pt}{\isachardot}{\kern0pt}\ {\isacharparenleft}{\kern0pt}isolate{\isacharunderscore}{\kern0pt}variable{\isacharunderscore}{\kern0pt}sparse\ p\ var\ i{\isacharparenright}{\kern0pt}\ {\isacharasterisk}{\kern0pt}\ \isanewline
\ \ \ \ {\isacharparenleft}{\kern0pt}{\isacharparenleft}{\kern0pt}a{\isacharplus}{\kern0pt}b{\isacharasterisk}{\kern0pt}{\isacharparenleft}{\kern0pt}Var\ var{\isacharparenright}{\kern0pt}{\isacharparenright}{\kern0pt}{\isacharcircum}{\kern0pt}i{\isacharparenright}{\kern0pt}\ {\isacharasterisk}{\kern0pt}\ {\isacharparenleft}{\kern0pt}d{\isacharcircum}{\kern0pt}{\isacharparenleft}{\kern0pt}deg\ {\isacharminus}{\kern0pt}\ i{\isacharparenright}{\kern0pt}{\isacharparenright}{\kern0pt}\ in\isanewline
\ \ P\ {\isacharasterisk}{\kern0pt}\ {\isacharparenleft}{\kern0pt}d\ {\isacharcircum}{\kern0pt}\ {\isacharparenleft}{\kern0pt}deg\ mod\ {\isadigit{2}}{\isacharparenright}{\kern0pt}{\isacharparenright}{\kern0pt}{\isacharparenright}{\kern0pt}{\isachardoublequoteclose}{\isacharbar}{\kern0pt}\isanewline
\ \ {\isachardoublequoteopen}quadratic{\isacharunderscore}{\kern0pt}part{\isacharunderscore}{\kern0pt}{\isadigit{1}}\ var\ a\ b\ d\ {\isacharparenleft}{\kern0pt}Neq\ p{\isacharparenright}{\kern0pt}\ {\isacharequal}{\kern0pt}\ {\isacharparenleft}{\kern0pt}\isanewline
\ \ let\ deg\ {\isacharequal}{\kern0pt}\ MPoly{\isacharunderscore}{\kern0pt}Type{\isachardot}{\kern0pt}degree\ p\ var\ in\isanewline
\ \ {\isasymSum}i{\isasymin}{\isacharbraceleft}{\kern0pt}{\isadigit{0}}{\isachardot}{\kern0pt}{\isachardot}{\kern0pt}{\isacharless}{\kern0pt}{\isacharparenleft}{\kern0pt}deg{\isacharplus}{\kern0pt}{\isadigit{1}}{\isacharparenright}{\kern0pt}{\isacharbraceright}{\kern0pt}{\isachardot}{\kern0pt}\ {\isacharparenleft}{\kern0pt}isolate{\isacharunderscore}{\kern0pt}variable{\isacharunderscore}{\kern0pt}sparse\ p\ var\ i{\isacharparenright}{\kern0pt}\ {\isacharasterisk}{\kern0pt}\ \isanewline
\ \ \ \ {\isacharparenleft}{\kern0pt}{\isacharparenleft}{\kern0pt}a{\isacharplus}{\kern0pt}b{\isacharasterisk}{\kern0pt}{\isacharparenleft}{\kern0pt}Var\ var{\isacharparenright}{\kern0pt}{\isacharparenright}{\kern0pt}{\isacharcircum}{\kern0pt}i{\isacharparenright}{\kern0pt}\ {\isacharasterisk}{\kern0pt}\ {\isacharparenleft}{\kern0pt}d{\isacharcircum}{\kern0pt}{\isacharparenleft}{\kern0pt}deg\ {\isacharminus}{\kern0pt}\ i{\isacharparenright}{\kern0pt}{\isacharparenright}{\kern0pt}{\isacharparenright}{\kern0pt}{\isachardoublequoteclose}%
}
\snip{quadraticparttwo}{1}{2}{%
\isacommand{fun}\isamarkupfalse%
\ quadratic{\isacharunderscore}{\kern0pt}part{\isacharunderscore}{\kern0pt}{\isadigit{2}}\ {\isacharcolon}{\kern0pt}{\isacharcolon}{\kern0pt}\isanewline
\ \ \ \ {\isachardoublequoteopen}nat\ {\isasymRightarrow}\ real\ mpoly\ {\isasymRightarrow}\ real\ mpoly\ {\isasymRightarrow}\ real\ mpoly{\isachardoublequoteclose}\ \isakeyword{where}\isanewline
\ \ {\isachardoublequoteopen}quadratic{\isacharunderscore}{\kern0pt}part{\isacharunderscore}{\kern0pt}{\isadigit{2}}\ var\ sq\ p\ {\isacharequal}{\kern0pt}\ {\isacharparenleft}{\kern0pt}\isanewline
\ \ let\ deg\ {\isacharequal}{\kern0pt}\ MPoly{\isacharunderscore}{\kern0pt}Type{\isachardot}{\kern0pt}degree\ p\ var\ in\isanewline
\ \ {\isasymSum}i{\isasymin}{\isacharbraceleft}{\kern0pt}{\isadigit{0}}{\isachardot}{\kern0pt}{\isachardot}{\kern0pt}{\isacharless}{\kern0pt}deg{\isacharplus}{\kern0pt}{\isadigit{1}}{\isacharbraceright}{\kern0pt}{\isachardot}{\kern0pt}\ \isanewline
\ \ \ \ {\isacharparenleft}{\kern0pt}isolate{\isacharunderscore}{\kern0pt}variable{\isacharunderscore}{\kern0pt}sparse\ p\ var\ i{\isacharparenright}{\kern0pt}{\isacharasterisk}{\kern0pt}{\isacharparenleft}{\kern0pt}sq{\isacharcircum}{\kern0pt}{\isacharparenleft}{\kern0pt}i\ div\ {\isadigit{2}}{\isacharparenright}{\kern0pt}{\isacharparenright}{\kern0pt}\ {\isacharasterisk}{\kern0pt}\ \isanewline
\ \ \ \ \ \ \ \ {\isacharparenleft}{\kern0pt}Const{\isacharparenleft}{\kern0pt}of{\isacharunderscore}{\kern0pt}nat{\isacharparenleft}{\kern0pt}i\ mod\ {\isadigit{2}}{\isacharparenright}{\kern0pt}{\isacharparenright}{\kern0pt}{\isacharparenright}{\kern0pt}\ {\isacharasterisk}{\kern0pt}\ {\isacharparenleft}{\kern0pt}Var\ var{\isacharparenright}{\kern0pt}\ \isanewline
\ \ \ {\isacharplus}{\kern0pt}{\isacharparenleft}{\kern0pt}isolate{\isacharunderscore}{\kern0pt}variable{\isacharunderscore}{\kern0pt}sparse\ p\ var\ i{\isacharparenright}{\kern0pt}{\isacharasterisk}{\kern0pt}{\isacharparenleft}{\kern0pt}sq{\isacharcircum}{\kern0pt}{\isacharparenleft}{\kern0pt}i\ div\ {\isadigit{2}}{\isacharparenright}{\kern0pt}{\isacharparenright}{\kern0pt}\ {\isacharasterisk}{\kern0pt}\ \isanewline
\ \ \ \ \ \ \ \ Const{\isacharparenleft}{\kern0pt}{\isadigit{1}}\ {\isacharminus}{\kern0pt}\ of{\isacharunderscore}{\kern0pt}nat{\isacharparenleft}{\kern0pt}i\ mod\ {\isadigit{2}}{\isacharparenright}{\kern0pt}{\isacharparenright}{\kern0pt}{\isacharparenright}{\kern0pt}{\isachardoublequoteclose}%
}
\snip{quadraticsub}{1}{2}{%
\isacommand{primrec}\isamarkupfalse%
\ quadratic{\isacharunderscore}{\kern0pt}sub\ {\isacharcolon}{\kern0pt}{\isacharcolon}{\kern0pt}\ {\isachardoublequoteopen}nat\ {\isasymRightarrow}\ \isanewline
\ \ \ \ \ \ real\ mpoly\ {\isasymRightarrow}\ real\ mpoly\ {\isasymRightarrow}\ real\ mpoly\ {\isasymRightarrow}\ real\ mpoly\ {\isasymRightarrow}\isanewline
\ \ \ \ \ \ atom\ {\isasymRightarrow}\ atom\ fm{\isachardoublequoteclose}\ \isakeyword{where}\isanewline
\ \ {\isachardoublequoteopen}quadratic{\isacharunderscore}{\kern0pt}sub\ var\ a\ b\ c\ d\ {\isacharparenleft}{\kern0pt}Eq\ p{\isacharparenright}{\kern0pt}\ {\isacharequal}{\kern0pt}\ {\isacharparenleft}{\kern0pt}\isanewline
\ \ \ \ let\ {\isacharparenleft}{\kern0pt}p{\isadigit{1}}{\isacharcolon}{\kern0pt}{\isacharcolon}{\kern0pt}real\ mpoly{\isacharparenright}{\kern0pt}\ {\isacharequal}{\kern0pt}\ quadratic{\isacharunderscore}{\kern0pt}part{\isacharunderscore}{\kern0pt}{\isadigit{1}}\ var\ a\ b\ d\ {\isacharparenleft}{\kern0pt}Eq\ p{\isacharparenright}{\kern0pt}\ in\isanewline
\ \ \ \ let\ {\isacharparenleft}{\kern0pt}p{\isadigit{2}}{\isacharcolon}{\kern0pt}{\isacharcolon}{\kern0pt}real\ mpoly{\isacharparenright}{\kern0pt}\ {\isacharequal}{\kern0pt}\ quadratic{\isacharunderscore}{\kern0pt}part{\isacharunderscore}{\kern0pt}{\isadigit{2}}\ var\ c\ p{\isadigit{1}}\ in\isanewline
\ \ \ \ let\ {\isacharparenleft}{\kern0pt}A{\isacharcolon}{\kern0pt}{\isacharcolon}{\kern0pt}real\ mpoly{\isacharparenright}{\kern0pt}\ {\isacharequal}{\kern0pt}\ isolate{\isacharunderscore}{\kern0pt}variable{\isacharunderscore}{\kern0pt}sparse\ p{\isadigit{2}}\ var\ {\isadigit{0}}\ in\isanewline
\ \ \ \ let\ {\isacharparenleft}{\kern0pt}B{\isacharcolon}{\kern0pt}{\isacharcolon}{\kern0pt}real\ mpoly{\isacharparenright}{\kern0pt}\ {\isacharequal}{\kern0pt}\ isolate{\isacharunderscore}{\kern0pt}variable{\isacharunderscore}{\kern0pt}sparse\ p{\isadigit{2}}\ var\ {\isadigit{1}}\ in\isanewline
\ \ \ \ And\isanewline
\ \ \ \ \ \ {\isacharparenleft}{\kern0pt}Atom\ {\isacharparenleft}{\kern0pt}Leq\ {\isacharparenleft}{\kern0pt}A{\isacharasterisk}{\kern0pt}B{\isacharparenright}{\kern0pt}{\isacharparenright}{\kern0pt}{\isacharparenright}{\kern0pt}\isanewline
\ \ \ \ \ \ {\isacharparenleft}{\kern0pt}Atom\ {\isacharparenleft}{\kern0pt}Eq\ {\isacharparenleft}{\kern0pt}A{\isacharcircum}{\kern0pt}{\isadigit{2}}{\isacharminus}{\kern0pt}B{\isacharcircum}{\kern0pt}{\isadigit{2}}{\isacharasterisk}{\kern0pt}c{\isacharparenright}{\kern0pt}{\isacharparenright}{\kern0pt}{\isacharparenright}{\kern0pt}\isanewline
{\isacharparenright}{\kern0pt}{\isachardoublequoteclose}\ {\isacharbar}{\kern0pt}\ \isanewline
\ \ {\isachardoublequoteopen}quadratic{\isacharunderscore}{\kern0pt}sub\ var\ a\ b\ c\ d\ {\isacharparenleft}{\kern0pt}Less\ p{\isacharparenright}{\kern0pt}\ {\isacharequal}{\kern0pt}\ {\isacharparenleft}{\kern0pt}\isanewline
\ \ \ \ let\ {\isacharparenleft}{\kern0pt}p{\isadigit{1}}{\isacharcolon}{\kern0pt}{\isacharcolon}{\kern0pt}real\ mpoly{\isacharparenright}{\kern0pt}\ {\isacharequal}{\kern0pt}\ quadratic{\isacharunderscore}{\kern0pt}part{\isacharunderscore}{\kern0pt}{\isadigit{1}}\ var\ a\ b\ d\ {\isacharparenleft}{\kern0pt}Less\ p{\isacharparenright}{\kern0pt}\ in\isanewline
\ \ \ \ let\ {\isacharparenleft}{\kern0pt}p{\isadigit{2}}{\isacharcolon}{\kern0pt}{\isacharcolon}{\kern0pt}real\ mpoly{\isacharparenright}{\kern0pt}\ {\isacharequal}{\kern0pt}\ quadratic{\isacharunderscore}{\kern0pt}part{\isacharunderscore}{\kern0pt}{\isadigit{2}}\ var\ c\ p{\isadigit{1}}\ in\isanewline
\ \ \ \ let\ {\isacharparenleft}{\kern0pt}A{\isacharcolon}{\kern0pt}{\isacharcolon}{\kern0pt}real\ mpoly{\isacharparenright}{\kern0pt}\ {\isacharequal}{\kern0pt}\ isolate{\isacharunderscore}{\kern0pt}variable{\isacharunderscore}{\kern0pt}sparse\ p{\isadigit{2}}\ var\ {\isadigit{0}}\ in\isanewline
\ \ \ \ let\ {\isacharparenleft}{\kern0pt}B{\isacharcolon}{\kern0pt}{\isacharcolon}{\kern0pt}real\ mpoly{\isacharparenright}{\kern0pt}\ {\isacharequal}{\kern0pt}\ isolate{\isacharunderscore}{\kern0pt}variable{\isacharunderscore}{\kern0pt}sparse\ p{\isadigit{2}}\ var\ {\isadigit{1}}\ in\isanewline
\ \ \ \ Or\isanewline
\ \ \ \ \ \ {\isacharparenleft}{\kern0pt}And\isanewline
\ \ \ \ \ \ \ \ {\isacharparenleft}{\kern0pt}Atom\ {\isacharparenleft}{\kern0pt}Less{\isacharparenleft}{\kern0pt}A{\isacharparenright}{\kern0pt}{\isacharparenright}{\kern0pt}{\isacharparenright}{\kern0pt}\isanewline
\ \ \ \ \ \ \ \ {\isacharparenleft}{\kern0pt}Atom\ {\isacharparenleft}{\kern0pt}Less\ {\isacharparenleft}{\kern0pt}B{\isacharcircum}{\kern0pt}{\isadigit{2}}{\isacharasterisk}{\kern0pt}c{\isacharminus}{\kern0pt}A{\isacharcircum}{\kern0pt}{\isadigit{2}}{\isacharparenright}{\kern0pt}{\isacharparenright}{\kern0pt}{\isacharparenright}{\kern0pt}{\isacharparenright}{\kern0pt}\isanewline
\ \ \ \ \ \ {\isacharparenleft}{\kern0pt}And\isanewline
\ \ \ \ \ \ \ \ {\isacharparenleft}{\kern0pt}Atom\ {\isacharparenleft}{\kern0pt}Leq\ B{\isacharparenright}{\kern0pt}{\isacharparenright}{\kern0pt}\isanewline
\ \ \ \ \ \ \ \ {\isacharparenleft}{\kern0pt}Or\isanewline
\ \ \ \ \ \ \ \ \ \ {\isacharparenleft}{\kern0pt}Atom\ {\isacharparenleft}{\kern0pt}Less\ A{\isacharparenright}{\kern0pt}{\isacharparenright}{\kern0pt}\isanewline
\ \ \ \ \ \ \ \ \ \ {\isacharparenleft}{\kern0pt}Atom\ {\isacharparenleft}{\kern0pt}Less\ {\isacharparenleft}{\kern0pt}A{\isacharcircum}{\kern0pt}{\isadigit{2}}{\isacharminus}{\kern0pt}B{\isacharcircum}{\kern0pt}{\isadigit{2}}{\isacharasterisk}{\kern0pt}c{\isacharparenright}{\kern0pt}{\isacharparenright}{\kern0pt}{\isacharparenright}{\kern0pt}{\isacharparenright}{\kern0pt}{\isacharparenright}{\kern0pt}\isanewline
{\isacharparenright}{\kern0pt}{\isachardoublequoteclose}\ {\isacharbar}{\kern0pt}\isanewline
\ \ {\isachardoublequoteopen}quadratic{\isacharunderscore}{\kern0pt}sub\ var\ a\ b\ c\ d\ {\isacharparenleft}{\kern0pt}Leq\ p{\isacharparenright}{\kern0pt}\ {\isacharequal}{\kern0pt}\ {\isacharparenleft}{\kern0pt}\isanewline
\ \ \ \ let\ {\isacharparenleft}{\kern0pt}p{\isadigit{1}}{\isacharcolon}{\kern0pt}{\isacharcolon}{\kern0pt}real\ mpoly{\isacharparenright}{\kern0pt}\ {\isacharequal}{\kern0pt}\ quadratic{\isacharunderscore}{\kern0pt}part{\isacharunderscore}{\kern0pt}{\isadigit{1}}\ var\ a\ b\ d\ {\isacharparenleft}{\kern0pt}Leq\ p{\isacharparenright}{\kern0pt}\ in\isanewline
\ \ \ \ let\ {\isacharparenleft}{\kern0pt}p{\isadigit{2}}{\isacharcolon}{\kern0pt}{\isacharcolon}{\kern0pt}real\ mpoly{\isacharparenright}{\kern0pt}\ {\isacharequal}{\kern0pt}\ quadratic{\isacharunderscore}{\kern0pt}part{\isacharunderscore}{\kern0pt}{\isadigit{2}}\ var\ c\ p{\isadigit{1}}\ in\isanewline
\ \ \ \ let\ {\isacharparenleft}{\kern0pt}A{\isacharcolon}{\kern0pt}{\isacharcolon}{\kern0pt}real\ mpoly{\isacharparenright}{\kern0pt}\ {\isacharequal}{\kern0pt}\ isolate{\isacharunderscore}{\kern0pt}variable{\isacharunderscore}{\kern0pt}sparse\ p{\isadigit{2}}\ var\ {\isadigit{0}}\ in\isanewline
\ \ \ \ let\ {\isacharparenleft}{\kern0pt}B{\isacharcolon}{\kern0pt}{\isacharcolon}{\kern0pt}real\ mpoly{\isacharparenright}{\kern0pt}\ {\isacharequal}{\kern0pt}\ isolate{\isacharunderscore}{\kern0pt}variable{\isacharunderscore}{\kern0pt}sparse\ p{\isadigit{2}}\ var\ {\isadigit{1}}\ in\isanewline
\ \ \ \ Or\isanewline
\ \ \ \ \ \ {\isacharparenleft}{\kern0pt}And\isanewline
\ \ \ \ \ \ \ \ {\isacharparenleft}{\kern0pt}Atom\ {\isacharparenleft}{\kern0pt}Leq{\isacharparenleft}{\kern0pt}A{\isacharparenright}{\kern0pt}{\isacharparenright}{\kern0pt}{\isacharparenright}{\kern0pt}\isanewline
\ \ \ \ \ \ \ \ {\isacharparenleft}{\kern0pt}Atom\ {\isacharparenleft}{\kern0pt}Leq{\isacharparenleft}{\kern0pt}B{\isacharcircum}{\kern0pt}{\isadigit{2}}{\isacharasterisk}{\kern0pt}c{\isacharminus}{\kern0pt}A{\isacharcircum}{\kern0pt}{\isadigit{2}}{\isacharparenright}{\kern0pt}{\isacharparenright}{\kern0pt}{\isacharparenright}{\kern0pt}{\isacharparenright}{\kern0pt}\isanewline
\ \ \ \ \ \ {\isacharparenleft}{\kern0pt}And\isanewline
\ \ \ \ \ \ \ \ {\isacharparenleft}{\kern0pt}Atom\ {\isacharparenleft}{\kern0pt}Leq\ B{\isacharparenright}{\kern0pt}{\isacharparenright}{\kern0pt}\isanewline
\ \ \ \ \ \ \ \ {\isacharparenleft}{\kern0pt}Atom\ {\isacharparenleft}{\kern0pt}Leq\ {\isacharparenleft}{\kern0pt}A{\isacharcircum}{\kern0pt}{\isadigit{2}}{\isacharminus}{\kern0pt}B{\isacharcircum}{\kern0pt}{\isadigit{2}}{\isacharasterisk}{\kern0pt}c{\isacharparenright}{\kern0pt}{\isacharparenright}{\kern0pt}{\isacharparenright}{\kern0pt}{\isacharparenright}{\kern0pt}\isanewline
{\isacharparenright}{\kern0pt}{\isachardoublequoteclose}\ {\isacharbar}{\kern0pt}\isanewline
\ \ {\isachardoublequoteopen}quadratic{\isacharunderscore}{\kern0pt}sub\ var\ a\ b\ c\ d\ {\isacharparenleft}{\kern0pt}Neq\ p{\isacharparenright}{\kern0pt}\ {\isacharequal}{\kern0pt}\ {\isacharparenleft}{\kern0pt}\isanewline
\ \ \ \ let\ {\isacharparenleft}{\kern0pt}p{\isadigit{1}}{\isacharcolon}{\kern0pt}{\isacharcolon}{\kern0pt}real\ mpoly{\isacharparenright}{\kern0pt}\ {\isacharequal}{\kern0pt}\ quadratic{\isacharunderscore}{\kern0pt}part{\isacharunderscore}{\kern0pt}{\isadigit{1}}\ var\ a\ b\ d\ {\isacharparenleft}{\kern0pt}Neq\ p{\isacharparenright}{\kern0pt}\ in\isanewline
\ \ \ \ let\ {\isacharparenleft}{\kern0pt}p{\isadigit{2}}{\isacharcolon}{\kern0pt}{\isacharcolon}{\kern0pt}real\ mpoly{\isacharparenright}{\kern0pt}\ {\isacharequal}{\kern0pt}\ quadratic{\isacharunderscore}{\kern0pt}part{\isacharunderscore}{\kern0pt}{\isadigit{2}}\ var\ c\ p{\isadigit{1}}\ in\isanewline
\ \ \ \ let\ {\isacharparenleft}{\kern0pt}A{\isacharcolon}{\kern0pt}{\isacharcolon}{\kern0pt}real\ mpoly{\isacharparenright}{\kern0pt}\ {\isacharequal}{\kern0pt}\ isolate{\isacharunderscore}{\kern0pt}variable{\isacharunderscore}{\kern0pt}sparse\ p{\isadigit{2}}\ var\ {\isadigit{0}}\ in\isanewline
\ \ \ \ let\ {\isacharparenleft}{\kern0pt}B{\isacharcolon}{\kern0pt}{\isacharcolon}{\kern0pt}real\ mpoly{\isacharparenright}{\kern0pt}\ {\isacharequal}{\kern0pt}\ isolate{\isacharunderscore}{\kern0pt}variable{\isacharunderscore}{\kern0pt}sparse\ p{\isadigit{2}}\ var\ {\isadigit{1}}\ in\isanewline
\ \ \ \ Or\isanewline
\ \ \ \ \ \ {\isacharparenleft}{\kern0pt}Atom\ {\isacharparenleft}{\kern0pt}Less{\isacharparenleft}{\kern0pt}{\isacharminus}{\kern0pt}A{\isacharasterisk}{\kern0pt}B{\isacharparenright}{\kern0pt}{\isacharparenright}{\kern0pt}{\isacharparenright}{\kern0pt}\isanewline
\ \ \ \ \ \ {\isacharparenleft}{\kern0pt}Atom\ {\isacharparenleft}{\kern0pt}Neq{\isacharparenleft}{\kern0pt}A{\isacharcircum}{\kern0pt}{\isadigit{2}}{\isacharminus}{\kern0pt}B{\isacharcircum}{\kern0pt}{\isadigit{2}}{\isacharasterisk}{\kern0pt}c{\isacharparenright}{\kern0pt}{\isacharparenright}{\kern0pt}{\isacharparenright}{\kern0pt}\isanewline
{\isacharparenright}{\kern0pt}{\isachardoublequoteclose}%
}
\snip{eval}{1}{2}{%
\isacommand{fun}\isamarkupfalse%
\ eval\ {\isacharcolon}{\kern0pt}{\isacharcolon}{\kern0pt}\ {\isachardoublequoteopen}atom\ fm\ {\isasymRightarrow}\ real\ list\ {\isasymRightarrow}\ bool{\isachardoublequoteclose}\ \isakeyword{where}\isanewline
\ \ {\isachardoublequoteopen}eval\ {\isacharparenleft}{\kern0pt}Atom\ a{\isacharparenright}{\kern0pt}\ {\isasymnu}\ {\isacharequal}{\kern0pt}\ aEval\ a\ {\isasymnu}{\isachardoublequoteclose}\ {\isacharbar}{\kern0pt}\isanewline
\ \ {\isachardoublequoteopen}eval\ {\isacharparenleft}{\kern0pt}TrueF{\isacharparenright}{\kern0pt}\ {\isacharunderscore}{\kern0pt}\ {\isacharequal}{\kern0pt}\ True{\isachardoublequoteclose}\ {\isacharbar}{\kern0pt}\isanewline
\ \ {\isachardoublequoteopen}eval\ {\isacharparenleft}{\kern0pt}FalseF{\isacharparenright}{\kern0pt}\ {\isacharunderscore}{\kern0pt}\ {\isacharequal}{\kern0pt}\ False{\isachardoublequoteclose}\ {\isacharbar}{\kern0pt}\isanewline
\ \ {\isachardoublequoteopen}eval\ {\isacharparenleft}{\kern0pt}And\ {\isasymphi}\ {\isasympsi}{\isacharparenright}{\kern0pt}\ {\isasymnu}\ {\isacharequal}{\kern0pt}\ {\isacharparenleft}{\kern0pt}{\isacharparenleft}{\kern0pt}eval\ {\isasymphi}\ {\isasymnu}{\isacharparenright}{\kern0pt}\ {\isasymand}\ {\isacharparenleft}{\kern0pt}eval\ {\isasympsi}\ {\isasymnu}{\isacharparenright}{\kern0pt}{\isacharparenright}{\kern0pt}{\isachardoublequoteclose}\ {\isacharbar}{\kern0pt}\isanewline
\ \ {\isachardoublequoteopen}eval\ {\isacharparenleft}{\kern0pt}Or\ {\isasymphi}\ {\isasympsi}{\isacharparenright}{\kern0pt}\ {\isasymnu}\ {\isacharequal}{\kern0pt}\ {\isacharparenleft}{\kern0pt}{\isacharparenleft}{\kern0pt}eval\ {\isasymphi}\ {\isasymnu}{\isacharparenright}{\kern0pt}\ {\isasymor}\ {\isacharparenleft}{\kern0pt}eval\ {\isasympsi}\ {\isasymnu}{\isacharparenright}{\kern0pt}{\isacharparenright}{\kern0pt}{\isachardoublequoteclose}\ {\isacharbar}{\kern0pt}\isanewline
\ \ {\isachardoublequoteopen}eval\ {\isacharparenleft}{\kern0pt}Neg\ {\isasymphi}{\isacharparenright}{\kern0pt}\ {\isasymnu}\ {\isacharequal}{\kern0pt}\ {\isacharparenleft}{\kern0pt}{\isasymnot}\ {\isacharparenleft}{\kern0pt}eval\ {\isasymphi}\ {\isasymnu}{\isacharparenright}{\kern0pt}{\isacharparenright}{\kern0pt}{\isachardoublequoteclose}\ {\isacharbar}{\kern0pt}\isanewline
\ \ {\isachardoublequoteopen}eval\ {\isacharparenleft}{\kern0pt}ExQ\ {\isasymphi}{\isacharparenright}{\kern0pt}\ {\isasymnu}\ {\isacharequal}{\kern0pt}\ {\isacharparenleft}{\kern0pt}{\isasymexists}x{\isachardot}{\kern0pt}\ {\isacharparenleft}{\kern0pt}eval\ {\isasymphi}\ {\isacharparenleft}{\kern0pt}x{\isacharhash}{\kern0pt}{\isasymnu}{\isacharparenright}{\kern0pt}{\isacharparenright}{\kern0pt}{\isacharparenright}{\kern0pt}{\isachardoublequoteclose}\ {\isacharbar}{\kern0pt}\isanewline
\ \ {\isachardoublequoteopen}eval\ {\isacharparenleft}{\kern0pt}AllQ\ {\isasymphi}{\isacharparenright}{\kern0pt}\ {\isasymnu}\ {\isacharequal}{\kern0pt}\ {\isacharparenleft}{\kern0pt}{\isasymforall}x{\isachardot}{\kern0pt}\ {\isacharparenleft}{\kern0pt}eval\ {\isasymphi}\ {\isacharparenleft}{\kern0pt}x{\isacharhash}{\kern0pt}{\isasymnu}{\isacharparenright}{\kern0pt}{\isacharparenright}{\kern0pt}{\isacharparenright}{\kern0pt}{\isachardoublequoteclose}{\isacharbar}{\kern0pt}\isanewline
\ \ {\isachardoublequoteopen}eval\ {\isacharparenleft}{\kern0pt}AllN\ i\ {\isasymphi}{\isacharparenright}{\kern0pt}\ {\isasymnu}\ {\isacharequal}{\kern0pt}\ {\isacharparenleft}{\kern0pt}{\isasymforall}l{\isachardot}{\kern0pt}\ length\ l\ {\isacharequal}{\kern0pt}\ i\ {\isasymlongrightarrow}\ {\isacharparenleft}{\kern0pt}eval\ {\isasymphi}\ {\isacharparenleft}{\kern0pt}l\ {\isacharat}{\kern0pt}\ {\isasymnu}{\isacharparenright}{\kern0pt}{\isacharparenright}{\kern0pt}{\isacharparenright}{\kern0pt}{\isachardoublequoteclose}{\isacharbar}{\kern0pt}\isanewline
\ \ {\isachardoublequoteopen}eval\ {\isacharparenleft}{\kern0pt}ExN\ i\ {\isasymphi}{\isacharparenright}{\kern0pt}\ {\isasymnu}\ {\isacharequal}{\kern0pt}\ {\isacharparenleft}{\kern0pt}{\isasymexists}l{\isachardot}{\kern0pt}\ length\ l\ {\isacharequal}{\kern0pt}\ i\ {\isasymand}\ {\isacharparenleft}{\kern0pt}eval\ {\isasymphi}\ {\isacharparenleft}{\kern0pt}l\ {\isacharat}{\kern0pt}\ {\isasymnu}{\isacharparenright}{\kern0pt}{\isacharparenright}{\kern0pt}{\isacharparenright}{\kern0pt}{\isachardoublequoteclose}%
}
\snip{aEval}{1}{2}{%
\isacommand{fun}\isamarkupfalse%
\ aEval\ {\isacharcolon}{\kern0pt}{\isacharcolon}{\kern0pt}\ {\isachardoublequoteopen}atom\ {\isasymRightarrow}\ real\ list\ {\isasymRightarrow}\ bool{\isachardoublequoteclose}\ \isakeyword{where}\isanewline
\ \ {\isachardoublequoteopen}aEval\ {\isacharparenleft}{\kern0pt}Eq\ p{\isacharparenright}{\kern0pt}\ {\isasymnu}\ {\isacharequal}{\kern0pt}\ {\isacharparenleft}{\kern0pt}insertion\ {\isacharparenleft}{\kern0pt}nth{\isacharunderscore}{\kern0pt}default\ {\isadigit{0}}\ {\isasymnu}{\isacharparenright}{\kern0pt}\ p\ {\isacharequal}{\kern0pt}\ {\isadigit{0}}{\isacharparenright}{\kern0pt}{\isachardoublequoteclose}\ {\isacharbar}{\kern0pt}\isanewline
\ \ {\isachardoublequoteopen}aEval\ {\isacharparenleft}{\kern0pt}Less\ p{\isacharparenright}{\kern0pt}\ {\isasymnu}\ {\isacharequal}{\kern0pt}\ {\isacharparenleft}{\kern0pt}insertion\ {\isacharparenleft}{\kern0pt}nth{\isacharunderscore}{\kern0pt}default\ {\isadigit{0}}\ {\isasymnu}{\isacharparenright}{\kern0pt}\ p\ {\isacharless}{\kern0pt}\ {\isadigit{0}}{\isacharparenright}{\kern0pt}{\isachardoublequoteclose}\ {\isacharbar}{\kern0pt}\isanewline
\ \ {\isachardoublequoteopen}aEval\ {\isacharparenleft}{\kern0pt}Leq\ p{\isacharparenright}{\kern0pt}\ {\isasymnu}\ {\isacharequal}{\kern0pt}\ {\isacharparenleft}{\kern0pt}insertion\ {\isacharparenleft}{\kern0pt}nth{\isacharunderscore}{\kern0pt}default\ {\isadigit{0}}\ {\isasymnu}{\isacharparenright}{\kern0pt}\ p\ {\isasymle}\ {\isadigit{0}}{\isacharparenright}{\kern0pt}{\isachardoublequoteclose}\ {\isacharbar}{\kern0pt}\isanewline
\ \ {\isachardoublequoteopen}aEval\ {\isacharparenleft}{\kern0pt}Neq\ p{\isacharparenright}{\kern0pt}\ {\isasymnu}\ {\isacharequal}{\kern0pt}\ {\isacharparenleft}{\kern0pt}insertion\ {\isacharparenleft}{\kern0pt}nth{\isacharunderscore}{\kern0pt}default\ {\isadigit{0}}\ {\isasymnu}{\isacharparenright}{\kern0pt}\ p\ {\isasymnoteq}\ {\isadigit{0}}{\isacharparenright}{\kern0pt}{\isachardoublequoteclose}%
}
\snip{datatypefm}{1}{2}{%
\isacommand{datatype}\isamarkupfalse%
\ {\isacharparenleft}{\kern0pt}atoms{\isacharcolon}{\kern0pt}\ {\isacharprime}{\kern0pt}a{\isacharparenright}{\kern0pt}\ fm\ {\isacharequal}{\kern0pt}\ \ TrueF\ {\isacharbar}{\kern0pt}\ FalseF\ {\isacharbar}{\kern0pt}\ Atom\ {\isacharprime}{\kern0pt}a\ \isanewline
\ \ {\isacharbar}{\kern0pt}\ And\ {\isachardoublequoteopen}{\isacharprime}{\kern0pt}a\ fm{\isachardoublequoteclose}\ {\isachardoublequoteopen}{\isacharprime}{\kern0pt}a\ fm{\isachardoublequoteclose}\ {\isacharbar}{\kern0pt}\ Or\ {\isachardoublequoteopen}{\isacharprime}{\kern0pt}a\ fm{\isachardoublequoteclose}\ {\isachardoublequoteopen}{\isacharprime}{\kern0pt}a\ fm{\isachardoublequoteclose}\ {\isacharbar}{\kern0pt}\ Neg\ {\isachardoublequoteopen}{\isacharprime}{\kern0pt}a\ fm{\isachardoublequoteclose}\ \isanewline
\ \ {\isacharbar}{\kern0pt}\ ExQ\ {\isachardoublequoteopen}{\isacharprime}{\kern0pt}a\ fm{\isachardoublequoteclose}\ {\isacharbar}{\kern0pt}\ AllQ\ {\isachardoublequoteopen}{\isacharprime}{\kern0pt}a\ fm{\isachardoublequoteclose}\ {\isacharbar}{\kern0pt}\ ExN\ {\isachardoublequoteopen}nat{\isachardoublequoteclose}\ {\isachardoublequoteopen}{\isacharprime}{\kern0pt}a\ fm{\isachardoublequoteclose}\ {\isacharbar}{\kern0pt}\ AllN\ {\isachardoublequoteopen}nat{\isachardoublequoteclose}\ {\isachardoublequoteopen}{\isacharprime}{\kern0pt}a\ fm{\isachardoublequoteclose}%
}
\snip{datatypeatom}{1}{2}{%
\isacommand{datatype}\isamarkupfalse%
\ atom\ {\isacharequal}{\kern0pt}\ Less\ {\isachardoublequoteopen}real\ mpoly{\isachardoublequoteclose}\ {\isacharbar}{\kern0pt}\ Eq\ {\isachardoublequoteopen}real\ mpoly{\isachardoublequoteclose}\ {\isacharbar}{\kern0pt}\ Leq\ {\isachardoublequoteopen}real\ mpoly{\isachardoublequoteclose}\ \isanewline
\ \ {\isacharbar}{\kern0pt}\ Neq\ {\isachardoublequoteopen}real\ mpoly{\isachardoublequoteclose}%
}
\snip{sumOverDegree}{1}{2}{%
\isacommand{lemma}\isamarkupfalse%
\ sum{\isacharunderscore}{\kern0pt}over{\isacharunderscore}{\kern0pt}degree{\isacharcolon}{\kern0pt}\ {\isachardoublequoteopen}{\isacharparenleft}{\kern0pt}p\ {\isacharcolon}{\kern0pt}{\isacharcolon}{\kern0pt}\ real\ mpoly{\isacharparenright}{\kern0pt}\ \isanewline
\ \ {\isacharequal}{\kern0pt}\ {\isacharparenleft}{\kern0pt}{\isasymSum}i{\isasymle}degree\ p\ x{\isachardot}{\kern0pt}\ isolate{\isacharunderscore}{\kern0pt}variable{\isacharunderscore}{\kern0pt}sparse\ p\ x\ i\ {\isacharasterisk}{\kern0pt}\ Var\ x{\isacharcircum}{\kern0pt}i{\isacharparenright}{\kern0pt}{\isachardoublequoteclose}%
}
\snip{datatypeatomprime}{1}{2}{%
\isacommand{datatype}\isamarkupfalse%
\ atomUni\ {\isacharequal}{\kern0pt}\ LessUni\ {\isachardoublequoteopen}real{\isacharasterisk}{\kern0pt}real{\isacharasterisk}{\kern0pt}real{\isachardoublequoteclose}\ {\isacharbar}{\kern0pt}\ EqUni\ {\isachardoublequoteopen}real{\isacharasterisk}{\kern0pt}real{\isacharasterisk}{\kern0pt}real{\isachardoublequoteclose}\ \isanewline
\ \ {\isacharbar}{\kern0pt}\ LeqUni\ {\isachardoublequoteopen}real{\isacharasterisk}{\kern0pt}real{\isacharasterisk}{\kern0pt}real{\isachardoublequoteclose}\ {\isacharbar}{\kern0pt}\ NeqUni\ {\isachardoublequoteopen}real{\isacharasterisk}{\kern0pt}real{\isacharasterisk}{\kern0pt}real{\isachardoublequoteclose}%
}

\usepackage{hyperref}

\usepackage{blkarray}
\usepackage{xspace}
\usepackage{graphicx}
\newcommand{\reals}{\mathbb{R}}

\usepackage{prettyref}

\newcommand{\rref}[2][]{\prettyref{#2}}
\newrefformat{sec}{Sect.\,\ref{#1}}
\newrefformat{def}{Def.\,\ref{#1}}
\newrefformat{thm}{Theorem\,\ref{#1}}
\newrefformat{prop}{Proposition\,\ref{#1}}
\newrefformat{rem}{Remark\,\ref{#1}}
\newrefformat{lem}{Lemma\,\ref{#1}}
\newrefformat{cor}{Cor.\,\ref{#1}}
\newrefformat{ex}{Example\,\ref{#1}}
\newrefformat{tab}{Table\,\ref{#1}}
\newrefformat{fig}{Fig.\,\ref{#1}}
\newrefformat{case}{case\,\ref{#1}}
\newrefformat{foot}{Footnote\,\ref{#1}}
\newrefformat{app}{Appendix\,\ref{#1}}

\newtheorem{remark}{Remark}
\newtheorem{example}{Example}

\usepgfplotslibrary{colormaps}
\usepgfplotslibrary{colorbrewer}
\pgfplotsset{compat=1.14}
\pgfplotsset{select coords between index/.style 2 args={
    x filter/.code={
        \ifnum\coordindex<#1\fi
        \ifnum\coordindex>#2\fi
    }
}}

\newcommand{\caderwv}{CADE09\xspace}

\newcommand{\economics}{Economics\xspace}
\newcommand{\eqvs}{E\xspace}
\newcommand{\genvs}{G\xspace}

\newcommand{\luckyvs}{L\xspace}
\newcommand{\legvs}{LEG\xspace}

\newcommand{\smtratcad}{S-QE$_\checkmark$\xspace}
\newcommand{\smtratsat}{S-SAT$_\text{\Lightning}$\xspace}
\newcommand{\zthree}{Z3\xspace}

\newcommand{\redlogold}{R$_\text{\Lightning}$\xspace}
\newcommand{\redlognew}{R$_\checkmark$\xspace}
\newcommand{\mathematica}{W\xspace}
\newcommand{\mathematicacad}{\mathematica-QE\xspace}
\newcommand{\mathematicavs}{\mathematica-VS\xspace}

\newcommand{\equalityvs}{VSEquality\xspace}
\newcommand{\generalvs}{VSGeneral\xspace}

\newcommand{\luckyvslong}{VSLucky\xspace}
\newcommand{\legvslong}{VSLEG\xspace}

\newcommand{\captioncolorsquare}[2]{{\scriptsize(}\protect\tikz{\protect\draw[index of colormap={#1 of colormap/#2},fill=.] rectangle++(1ex,1ex);}{\scriptsize)}}

\title{Verified Quadratic Virtual Substitution for \\ Real Arithmetic}
\author{Matias Scharager \and Katherine Cordwell \and
Stefan Mitsch \and Andr\'{e} Platzer\thanks{Computer Science Department, Carnegie Mellon University, Pittsburgh, USA\newline $\{$mscharag,kcordwel,smitsch,aplatzer$\}$@cs.cmu.edu}}
\date{}

\usepackage{ifpdf}
\ifpdf
\fi

\begin{document}
\maketitle
\allowdisplaybreaks
\thispagestyle{empty}
\begin{abstract}
This paper presents a formally verified quantifier elimination (QE) algorithm for first-order real arithmetic by linear and quadratic virtual substitution (VS) in Isabelle/HOL.
The Tarski-Seidenberg theorem established that the first-order logic of real arithmetic is decidable by QE.
However, in practice, QE algorithms are highly complicated and often combine multiple methods for performance.
VS is a practically successful method for QE that targets formulas with low-degree polynomials.
To our knowledge, this is the first work to formalize VS for quadratic real arithmetic including inequalities.
The proofs necessitate various contributions to the existing multivariate polynomial libraries in Isabelle/HOL.
Our framework is modularized and easily expandable (to facilitate integrating future optimizations), and could serve as a basis for developing practical general-purpose QE algorithms.
Further, as our formalization is designed with practicality in mind, we export our development to SML and test the resulting code on 378 benchmarks from the literature, comparing to Redlog, Z3, Wolfram Engine, and SMT-RAT.
This identified inconsistencies in some tools, underscoring the significance of a verified approach for the intricacies of real arithmetic.
\end{abstract}

\section{Introduction}
\textit{Quantifier elimination (QE)} is the process of transforming quantified formulas into logically equivalent quantifier-free formulas.
In this paper, we consider QE for the first-order logic of real arithmetic (\FOLR), so quantifiers range over the real numbers.
The Tarski-Seidenberg theorem proves that QE is admissible for the theory of real-closed fields \cite{Tarski, Seidenberg}.
Real quantified statements arise in a number of application domains, including geometry, chemistry, life sciences, and the verification of cyber-physical systems (CPS) \cite{DBLP:journals/mics/Sturm17}.
Many of the applications which require QE are safety-critical \cite{Platzer10,Platzer18}; accordingly, it is crucial to have both efficient and formally verified support for QE to trust the resulting decisions.

Unfortunately, QE algorithms are quite intricate, which makes it difficult to formally verify their correctness.
In practice, this necessitates the use of unverified tools.
For example, the theorem prover KeYmaera~X \cite{fulton2015keymaera}, which is designed to formally verify models of CPS (such as planes and surgical robots) uses Mathematica/Wolfram Engine and/or Z3 as blackbox solvers for QE.
While these are admirable tools, they are unverified, and their use introduces a weak link \cite{Misfortunes} into what would otherwise be a (fully verified \cite{DBLP:conf/cpp/BohrerRVVP17}) trustworthy proof.

To help fill this gap, we formally verify linear and quadratic \textit{virtual substitution (VS)} due to Weispfenning \cite{weispfenning1988complexity, weispfenning1997quantifier}, which focuses on QE for a quantified variable $x$ occurring in polynomials $f(x)$ of at most degree 2 in $x$, although variations \cite{kosta2016new,DBLP:conf/issac/Weispfenning94} handle higher degree polynomials.
Linear and quadratic VS are of practical significance. They serve to improve QE \cite{Passmore} and SMT tools and are the basis of the experimentally successful \cite{sturm2018thirty} Redlog solver \cite{DBLP:journals/cca/Dolzmann097}.
To our knowledge, ours is the first formally verified algorithm for VS with quadratic inequalities.

As we focus on correct and practical VS, we export our verified Isabelle/HOL code to SML for experimentation.
We test our exported formalization of the equality VS algorithm (\rref{sec:eqVsubst}) and of the general VS algorithm (\rref{sec:genVsubst}).
We compare to four tools that implement real QE: Redlog, SMT-RAT \cite{DBLP:conf/sat/CorziliusKJSA15}, Z3 \cite{DBLP:conf/tacas/MouraB08}, and Wolfram Engine.
With 304 examples, we solve more examples than SMT-RAT in quantifier elimination mode (solves 191) and come close to virtual substitution in Wolfram Engine (solves 322).
The remaining tools solve almost all examples; this is to be expected given that those tools have been optimized and fine-tuned (some for decades) and use efficient general-purpose fallback QE algorithms when VS does not succeed.
However, as we found 137 inconsistencies in other solvers, it is significant that ours is the only VS implementation with associated correctness proofs (assuming the orthogonal challenge of correct code generation from Isabelle \cite{DBLP:conf/esop/HupelN18}).

Our formalization is approximately 23,000 lines in Isabelle/HOL and is available on the Archive of Formal Proofs (AFP) \cite{VS_AFP}.

\section{Related Work}
The fastest known QE algorithm is Cylindrical Algebraic Decomposition (CAD)\cite{collins1975quantifier}, which has not yet been fully formally verified.
There are few general-purpose formally verified QE algorithms, and there appears to exist a tradeoff between the practicality of an algorithm and the ease of formalization.
Mahboubi and Cohen verified Tarski's original QE algorithm \cite{AssiaQE} and McLaughlin and Harrison have a proof-producing QE procedure based on Cohen-H\"ormander \cite{mclaughlin2005proof}; unfortunately, Tarski's algorithm and Cohen-H\"ormander both have non-elementary complexity, which limits the computational feasibility of these formalizations.

There has already been some work on formally verified VS: Nipkow \cite{nipkow2010linear} formally verified a VS procedure for \emph{linear} equations and inequalities.
The building blocks of \FOLR~formulas, or ``atoms," in Nipkow's work only allow for linear polynomials $\sum_i a_i x_i\sim c$, where $\sim\ \in \{=,<\}$, the $x_i$'s are quantified variables and $c$ and the $a_i$'s are real numbers.
These restrictions ensure that linear QE can always be performed, and they also simplify the substitution procedure and associated proofs.
Nipkow additionally provides a generic framework that can be applied to several different kinds of atoms (each new atom requires implementing several new code theorems in order to create an exportable algorithm).
While this is an excellent theoretical framework---we utilize several similar constructs in our formulation---we create an independent formalization that is specific to general \FOLR~formulas, as our main focus is to provide an efficient algorithm in this domain.
Specializing to one type of atom allows us to implement several optimizations, such as our modified DNF algorithm, which would be unwieldy to develop in a generic setting.

Chaieb \cite{chaieb2008automated} extends Nipkow's work to quadratic equalities.
His formalizations are not publicly available, but he generously provided us with the code.
While this was helpful for reference, we chose to build on a newer Isabelle/HOL polynomial library, and we focus on VS as an exportable standalone procedure, whereas Chaieb intrinsically links VS with an auxiliary QE procedure.

Other related work includes some unverified solvers.
For example, some work has been done in constraint solving with falsification: RSolver \cite{DBLP:conf/adhs/RatschanS06} was designed for hybrid systems verification and can find concrete counterexamples for fully quantified existential QE problems on \emph{compact} domains.
dReal \cite{DBLP:conf/cade/GaoKC13} is based on similar ideas and slightly relaxes the notion of satisfiability to $\delta$-satisfiability.
Constraint solving has also been considered in SMT-solving with Z3's nlsat \cite{DBLP:conf/cade/JovanovicM12}, which uses CDCL to decide systems of nonlinear inequalities and equations.

\section{The Virtual Substitution Algorithm}
Informally (and broadly) speaking, VS discretizes the QE problem by solving for the roots of one or more low-degree polynomials $f_1(x), \dots, f_n(x)$.
VS focuses on these roots and the intervals around them to identify and substitute appropriate representative ``sample points'' for $x$ into the rest of the formula.
However, these sample points may contain fractions, square roots, and/or other extensions of the logical language, and so they must be substituted ``virtually'': That is, VS creates a formula \textit{in \FOLR\ proper} that models the behavior of the direct substitution, which would be outside of \FOLR.
VS applies in two cases: an equality case and a general case.
We formalize both, and discuss each in turn.

\begin{remark}\label{rem:trick}
The VS algorithms need to work for \emph{multivariate} polynomials.
But as the VS correctness proofs show the equivalence is true for every real value of the free variables, they often implicitly treat all but one variable as having fixed (but arbitrary) real values.
That is why most correctness lemmas (but not the top-level algorithmic constructions) suffice for \emph{univariate} polynomials with \emph{real coefficients}.
We utilize this trick to simplify difficult proofs for general VS. 
\end{remark}

\subsection{Example}
\begin{example}\label{ex:VS}
Say that we want to perform QE on the formula $\exists x. (x^2 = 2\land xy^2 + 2y + 1  = 0)$.
One might notice that $x^2 = 2$ forces $x =\pm \sqrt{2}$ and accordingly wish to substitute.
Direct substitution yields the following expression: $(\sqrt{2}y^2 + 2y + 1 = 0\allowbreak\lor-\sqrt{2}y^2 + 2y + 1 = 0)$.
However, as its mention of the $\sqrt{\cdot}$ operator makes it an illegal \FOLR~formula, we will need some further tricks.

Cleverly, VS finds that $\sqrt{2}y^2 + 2y + 1 = 0$ is logically equivalent to $y^2\cdot(2y + 1) \leq 0\allowbreak \land 2y^4 - (2y + 1)^2= 0$, which is a \FOLR~formula\footnote{Notice that if $y = 0$, then both $\sqrt{2}y^2 + 2y + 1 = 0$ and $y^2\cdot(2y + 1) \leq 0 \land 2y^4 - (2y + 1)^2= 0$ are false.
If instead $y \neq 0$, then $\sqrt{2}y^2 + 2y + 1 = 0$ is true exactly when $\sqrt{2} = -(2y + 1)/y^2$, or exactly when $-(2y + 1)/y^2 \geq 0 \land 2y^4 - (2y + 1)^2= 0$, which is logically equivalent to $y^2\cdot(2y + 1) \leq 0 \allowbreak\land 2y^4 - (2y + 1)^2= 0$, as desired.}.
Similarly, VS identifies a \FOLR~formula that is logically equivalent to $-\sqrt{2}y^2 + 2y + 1 = 0$.
Then, VS returns the following quantifier-free \FOLR~formula which is logically equivalent to $\exists x. (x^2 = 2\land xy^2 + 2y + 1  = 0)$:
\begin{equation*}
\begin{aligned}
&\big( (y^2\cdot(2y + 1) \leq 0 \land 2y^4 - (2y + 1)^2= 0) \\
&\lor (-y^2\cdot(2y + 1) \leq 0 \land 2y^4 - (2y + 1)^2= 0)  \big).
\end{aligned}
\end{equation*}
\end{example}

\begin{remark}
If instead our starting formula were $\exists x. \exists y. (x^2 = 2\land xy^2 + 2y + 1  = 0)$, where now $y$ is quantified, then (following the same method as above) VS would identify the following logically equivalent \FOLR~formula with fewer variables:
\begin{equation}\label{eq:result}
\begin{aligned}
\exists y. &\big( (y^2\cdot(2y + 1) \leq 0 \land 2y^4 - (2y + 1)^2= 0) \\
&\phantom{\big(} \lor (-y^2\cdot(2y + 1) \leq 0 \land 2y^4 - (2y + 1)^2= 0)  \big).
\end{aligned}
\end{equation}
Unfortunately, here we are left with a quantified formula with no linear or quadratic equations or inequalities.
As we are thus outside of the fragment of \FOLR~that standard VS applies to, at this point we would want to outsource \rref{eq:result} to a general-purpose QE algorithm (like CAD) to eliminate the quantifier on $y$.
\end{remark}

\rref{ex:VS} was relatively simple, because it involved a quadratic equation with constant coefficients for $x$.
However, nothing in our reasoning was limited to constant coefficients: To perform QE on $\exists x. (x^2 = c\land xy^2 + 2y + 1 = 0)$, where $c$ is a polynomial in the variable $z$, we could handle substituting $x = \pm \sqrt{c}$ in the exact same way as for $x = \pm \sqrt{2}$, but the answer must distinguish the case of $c\geq 0$ symbolically.
More difficult is the generalization to inequalities, which seemingly require uncountably infinitely many values to be virtually substituted.
We first turn to the general equality case, and then discuss inequalities.

\subsection{Equality Virtual Substitution Algorithm}\label{sec:eqVsubst}
Let $a, b$ and $c$ be arbitrary polynomials with real coefficients that do not mention the variable $x$.
Consider the formula $\exists x. (ax^2+bx+c=0 \land F).$
There are three possible cases: Either $a \neq 0$, or $a = 0$ and $b$ is nonzero, or all of $a, b, c$ are zero (so $ax^2+bx+c=0$ is uninformative).
Letting $F_x^r$ denote the substitution of $x=r$ for $x$ in $F$, and solving for the roots of $ax^2 + bx + c$, we have the following:
\begin{align*}
&\exists x. (ax^2+bx+c=0 \land F) \longleftrightarrow \\
&\Big( (a=0\land b=0 \land c=0 \land \exists x. F)~\lor \\
&\phantom{\Big(}
(a=0 \land b\neq0 \land F_x^{-c/b})~\lor \\
&\phantom{\Big(}
(a\neq 0\land b^2-4ac\geq 0\land (F_x^{(-b+\sqrt{b^2-4ac})/(2a)}\lor F_x^{(-b-\sqrt{b^2-4ac})/(2a)}))  \Big).
\end{align*}
Conditions such as $b^2-4ac\geq 0$ are needed to ensure $(-b \pm \sqrt{b^2-4ac})/(2a)$ are well-defined; these are symbolic formulas unless $a,b,c$ are concrete numbers.

Similarly as in \rref{ex:VS}, if we were to substitute $F_x^{-c/b}$, $F_x^{(-b+\sqrt{b^2-4ac})/(2a)}$, and $F_x^{(-b-\sqrt{b^2-4ac})/(2a)}$ directly (for polynomials $a$, $b$, and $c$ that do not involve $x$), the resulting formula would no longer be in \FOLR.
Instead, VS avoids directly dividing polynomials or taking square roots with equivalent rewritings in \FOLR.
This involves two procedures: one for fractions, and one for square roots.

To virtually substitute a fraction $p/q$ of polynomials where $q \neq 0$ into the atom $\sum_{i=0}^n a_ix^i\sim0$, where $\sim\ \in \{=, <, \leq, \neq\}$ and each $a_i$ is an arbitrary polynomial expression not involving $x$, it suffices to normalize the denominator of the LHS, with the caveat that we must not flip the direction of the inequality for $<$ and $\leq$ atoms by normalizing by a value that might be negative.
When $n$ is even, $q^n\geq 0$ under any possible valuation, so normalizing by $q^n$ does not flip the inequality.
Alternatively, if $n$ is odd, $q^{n+1}\geq 0$.
We formalize this in our \isa{linear\isacharunderscore substitution} function (see \rref{app:linsubst}).

Next, we consider substituting $x = \sqrt{c}$ into an atom $\sum_{i=0}^n a_ix^i \sim 0$, where $c$ is an arbitrary polynomial expression not involving $x$ that satisfies $c\geq 0$, each $a_i$ is an arbitrary polynomial expression not involving $x$, and ${\sim} \in \{=, <, \leq, \neq\}$.
Its direct substitution can be separated out into even and odd exponents:
\begin{equation*}
\sum_{i=0}^n a_i\cdot (\sqrt{c})^i = \sum_{i=0}^{n/2}a_{2i}c^{i}+\sum_{i=0}^{n/2}a_{2i+1}c^{i}\sqrt{c}
\end{equation*}
Now our polynomial has the form $A+B\sqrt{c}$, where $A$ and $B$ and $c$ are symbolic polynomial expressions not involving $x$.
Then, we have the following cases:
\begin{align*}
A+B\sqrt{c}=0&\longleftrightarrow AB\leq 0 \land A^2-B^2c=0\\
A+B\sqrt{c}<0&\longleftrightarrow (A<0\land B^2c-A^2<0) \lor (B\leq 0 \land (A<0\lor A^2-B^2c<0))\\
A+B\sqrt{c}\leq 0&\longleftrightarrow (A\leq 0\land B^2c-A^2\leq 0) \lor (B\leq 0 \land A^2-B^2c\leq 0)\\
A+B\sqrt{c}\neq 0&\longleftrightarrow -AB<0 \lor A^2-B^2c\neq 0
\end{align*}
The equivalences for $=$ and $\neq$ atoms are derived from the observation that if $B \neq 0$, $A+B\sqrt{c}=0$ can be solved to find $\sqrt{c} = -A/B$, which holds iff $A^2 = B^2c$ and $-A/B \geq 0$.
The inequality cases involve casework to determine when polynomial $A$ is negative and dominates $B\sqrt{c}$ as $A^2>B^2c$, and when $B$ is negative and $B\sqrt{c}$ dominates $A$ as $B^2c>A$.
We formalize the VS procedure for quadratic roots in \isa{quadratic\isacharunderscore sub} (see \rref{app:quadsubst}).

\subsection{General Virtual Substitution Algorithm}\label{sec:genVsubst}
As we have seen, QE very naturally leads to finitely many cases (discretizes) for formulas that involve quadratic equality atoms (we call this the \textit{equality case}).
The VS algorithm for the \textit{general case}, which also handles inequality atoms, is more involved, because, unlike equalities, inequalities may have uncountably many solutions.
General VS only directly applies to a very specific fragment of \FOLR~formulas: conjunctions of polynomials that are at most quadratic in the variable of interest.
However, we can extend general VS to apply to more formulas with the help of a disjunctive normal form (DNF) transformation.

\begin{wrapfigure}[9]{R}{0.44\textwidth}
\vspace{-2\baselineskip}
\includegraphics[width=0.4\textwidth]{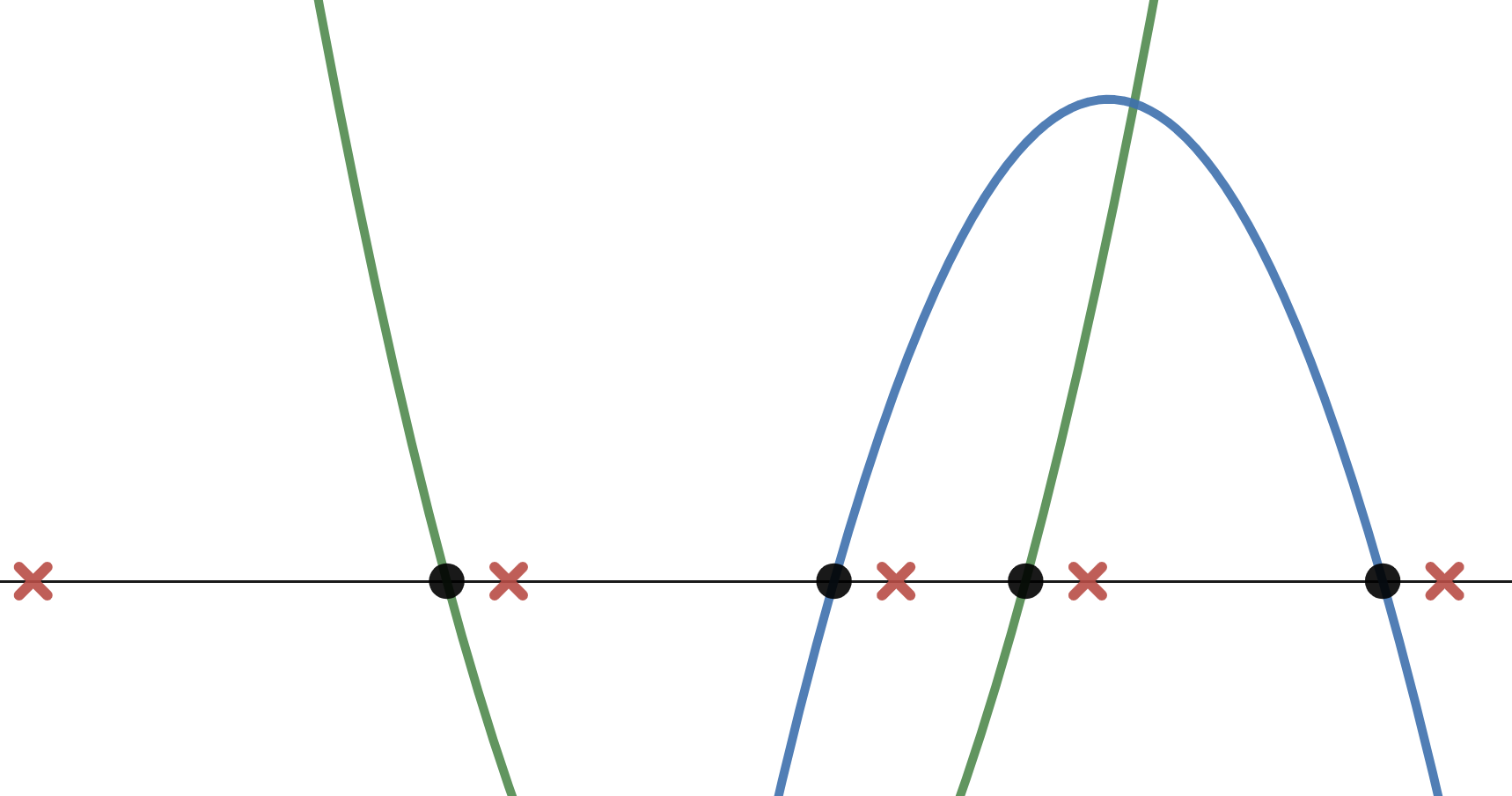}
\vspace{-0.5\baselineskip}
\caption{\label{graph}Two quadratics, their roots (black dots) and off-roots (red \textsf{x}'s)}
\end{wrapfigure}

As a simple example, consider the formula $\exists x. (p<0 \land q<0)$, where $p$ and $q$ are the univariate quadratic polynomials (in variable $x$) depicted in Fig. \ref{graph}.
Noting that the roots of $p$ and $q$ cannot possibly satisfy the strict inequalities, we partition the number line into ranges in between these zeros.

We recognize a key property: In each of the ranges between the roots of $p,q$, the signs of both $p$ and $q$ do not change.
Since the ranges cover all roots of $p,q$, the truth value of the formula at a single point in a range is representative of the truth value of the formula on the entire range.
To discretize the QE problem, we need only pick one sample point for each range.

However, we want to pick appropriate sample points \textit{for any} possible $p$ and $q$.
The points we pick as representatives are called the off-roots, which occur $\epsilon$ units away from the roots, where $\epsilon > 0$ is arbitrarily small.
We additionally need a representative for the leftmost range, which we represent with the point $-\infty$, where $-\infty$ is arbitrarily negative.
Of course, we cannot directly substitute $\epsilon$ and $-\infty$: they are not real numbers!
However, we can \textit{virtually} substitute them.

\subsubsection{Negative Infinity.}
Given any formula $F$, the VS of $-\infty$ should satisfy the equivalence $F_x^{-\infty} \longleftrightarrow \exists y.~\forall x{<}y.~F(x)$ (where $y$ does not occur in $F$).
Intuitively, this says that $-\infty$ acts as if it is arbitrarily negative (so less than the $x$ component of all roots of the polynomials in $F$) and captures information for the leftmost range on the real number line in any valuation of the non-$x$ variables.
If formula $\exists y.~\forall x{<}y.~ax^2 + bx + c = 0$ is true, where $a, b, c$ are polynomials that do not involve $x$, then $ax^2 + bx + c = 0$ holds at infinitely many $x$; since nonzero polynomials have finitely many roots, this can only happen if $ax^2+bx+c$ is the zero polynomial in $x$, i.e., it holds that:
\begin{equation}\label{eq:eqatoms}
(ax^2+bx+c=0)_x^{-\infty}  \longleftrightarrow a=0\land b=0 \land c=0
\end{equation}

The negation of \rref{eq:eqatoms} captures the behavior of $\neq$ atoms. For $<$ atoms, note that the sign value at $-\infty$ is dominated by the leading coefficient, so:
\begin{equation*}
(ax^2+bx+c<0)_x^{-\infty} \longleftrightarrow a<0 \lor (a=0\land (b>0\lor (b=0\land c<0)))
\end{equation*}
Finally, \({(ax^2+bx+c\leq 0)_x^{-\infty}} {\longleftrightarrow} {(ax^2+bx+c= 0)_x^{-\infty}} \lor {(ax^2+bx+c< 0)_x^{-\infty}}\).

In Isabelle/HOL, we formalize that our virtual substitution of $-\infty$ satisfies the desired equivalence (on $\reals$ using \rref{rem:trick}) in the following lemma:
\begin{isabelle}
\infinityevalLem
\end{isabelle}

To explain this lemma, we need to take a slight detour and discuss a few structural details of our framework (which is discussed in greater detail in \rref{sec:framework}).
The datatype \isa{atomUni} contains a triple of real numbers (which represent the coefficients of a univariate quadratic polynomial) and a sign condition:
\begin{isabelle}
\datatypeatomprime
\end{isabelle}
The \isa{aEvalUni} function has type \isa{atomUni $\Rightarrow$ real $\Rightarrow$ bool}; that is, it takes a sign condition with a triple of real numbers $(a,b,c)$ and a real number $x$ and evaluates whether $ax^2 + bx + c$ satisfies the sign condition.
The \isa{evalUni} function has type \isa{atomUni fmUni $\Rightarrow$ real $\Rightarrow$ bool}, where an \isa{atomUni fmUni} is a formula that involves conjunctions and disjunctions of elements of type \isa{atomUni} (and ``True'' and ``False'').
That is, the  \isa{evalUni} function takes such a formula and a real number and evaluates whether the formula is true at the real number.
Thus, \isa{infinity\isacharunderscore evalUni} states that, given \isa{At} of type \isa{atomUni}, with tuple $(a,b,c)$ and sign condition $\sim\ \in \{<,=,\leq, \neq\}$, $\isa{At}_x^{-\infty}$ holds iff $\exists y. \forall x{<}y. ax^2 + bx + c \sim 0$.
This captures the desired equivalence.

\iflongversion
Note that these definitions are set up for univariate polynomials (coefficients are assumed to be real numbers).
This is deliberate.
In \rref{sec:eval}, we will discuss how our framework reduces correctness results for multivariate polynomials to univariate lemmas like \isa{infinity\isacharunderscore evalUni}.
\fi

\subsubsection{Infinitesimals.}\label{sec:Infinitesimals}Given arbitrary $r$ (not containing $x$), VS for $r + \epsilon$ for variable $x$ should capture the equivalence
$F_x^{r + \epsilon} \longleftrightarrow \exists y{>}r. \forall x{\in}(r, y].\ F(x)$, where $F$ does not contain $y$.
Intuitively, this says that (in any valuation of the non-$x$ variables) $r + \epsilon$ captures information for the interval between $r$ and the next greatest $x$-root.

For $=$ and $\neq$ atoms, we proceed in the same manner as we did with $-\infty$, as $(r,y]$ contains infinitely many points and only the zero polynomial has infinitely many solutions.
As before, $\leq$ atoms turn into disjunctions of the inequality and equality representations at $r + \epsilon$.
We are left only to consider $<$ atoms.

Consider $ (p{<}0)_x^{r + \epsilon}$ where $p = ax^2 + bx + c$ with polynomials $a,b,c$ not containing $x$, and an arbitrary $r$ not containing $x$.
Notice that if $(p{<}0)_x^r$, then because polynomials are continuous, we can choose a small enough $y$ so that $\forall x{\in}(r, y].\ p{<}0$.
If instead $(p=0)_x^r$, then consider the partial derivative of $p$ evaluated at $r$.
If $\frac{\partial p}{\partial x}(r)$ is negative, then $\exists y{>}r. \forall x{\in}(r, y].\ p{<}0$  holds, because $p$ is decreasing in $x$ locally after $x{=}r$.
If $\frac{\partial p}{\partial x}(r)$ is positive, then $ \exists y{>}r. \forall x{\in}(r, y].\ p{<}0$  \textit{does not} hold, because $p$ is increasing in $x$ after $x{=}r$.
If $\frac{\partial p}{\partial x}(r)$ is zero, then to ascertain whether $ \exists y{>}r. \forall x{\in}(r, y].\ p{<}0$, we will need to check higher  derivatives.

This pattern forms the following recurrence, with the base case $(p<0)_x^{r + \epsilon} = (p<0)_x^r$ for polynomials $p$ of degree zero:
\begin{equation*}
(p<0)_x^{r + \epsilon} ~\stackrel{\text{def}}{=}~ (p<0)_x^r \lor \big((p=0)_x^r\land (({\partial p}/{\partial x})<0)_x^{r + \epsilon}\big)
\end{equation*}
We use the VS algorithm from Section~\ref{sec:eqVsubst} to characterize $(p<0)_x^r$ and $(p=0)_x^r$.

In Isabelle/HOL, we show that given a quadratic root $r$, the virtual substitution of $r + \epsilon$ satisfies the desired equivalence in the following theorem (on $\reals$ using Remark \ref{rem:trick}; we have an analogous lemma for linear roots $r$): \begin{isabelle}
\infinitesimalquad
\end{isabelle}

Note that \isa{{\isacharbraceleft}{\kern0pt}{r}{\kern0pt}{\isacharless}{\kern0pt}{\isachardot}{\kern0pt}{\isachardot}{\kern0pt}y{\kern0pt}{\isacharbraceright}} in Isabelle stands for the range $(r,y]$. This says that, given \isa{At} of type \isa{atomUni}, with tuple $(a,b,c)$ and sign condition $\sim\ \in \{<,=,\leq, \neq\}$, $\isa{At}_x^{r+\epsilon}$ holds iff $\exists y > r.\forall x \in (r, y]. ax^2 + bx + c \sim 0$, which is the desired equivalence.

\subsubsection{The General VS Theorem.}\label{sec:genVS}
Now that we have explained virtually substituting $-\infty$ and infinitesimals, we are ready to state the general VS theorem.

Let $F$ be a formula of the following shape, where each $a_i,b_i,c_i$, and $d_i$ is a polynomial that is at most quadratic in variable $x$:
\begin{equation*}
    F = \left(\bigwedge a_i =0\right) \land \left(\bigwedge b_i < 0\right) \land \left(\bigwedge c_i \leq 0\right) \land \left(\bigwedge d_i \neq 0\right).
\end{equation*}

Let $R(p)$ denote the set of symbolic expressions of the form $(g_1 + g_2\sqrt{g_3})/g_4$ that, as in \rref{sec:eqVsubst}, are roots of the polynomial $p$ in $x$, where the $g_i$'s are polynomials not involving $x$.
For the zero polynomial, let \(R(0)=\emptyset\).
Note that, as in \rref{sec:eqVsubst}, the $g_i$'s come with certain well-definedness checks that we retain implicitly in the construction (for example, $g_4 \neq 0$ and $g_3 \geq 0$).
We now define:
\begin{equation*}
A = \bigcup R(a_i)\quad B = \bigcup R(b_i)\quad C = \bigcup R(c_i)\quad D = \bigcup R(d_i)
\end{equation*}
Then we obtain the following QE equivalence, where for simplicity we elide the relevant crucial well-definedness checks (cross-reference \cite[Theorem 21.1]{Platzer18}):
\begin{equation}\label{eq:QEequiv1}
(\exists x. F) \longleftrightarrow F_x^{-\infty} \lor \bigvee_{\mathclap{r\in A\cup C}}F_x^{r}\lor \bigvee_{\mathrlap{r\in B\cup C\cup D}}F_x^{r+\epsilon}
\end{equation}

Intuitively, this formula states that if there is a particular $x$ that satisfies $F$, then it must be the case that $x$ is one of the equality roots from $A \cup C$, or that $x$ falls in one of the particular ranges (including $-\infty$ as a range) obtained by partitioning the number line by the roots in $B\cup C \cup D$.

Equation \rref{eq:QEequiv1} can be optimized further by eliding $C$ from the off-roots:
\begin{equation}\label{eq:QEequiv2}
(\exists x. F) \longleftrightarrow F^{-\infty} \lor \bigvee_{\mathclap{r\in A\cup C}}F_x^{r}\lor \bigvee_{\mathclap{r\in B\cup D}}F_x^{r+\epsilon}.
\end{equation}
Intuitively, this optimization holds because polynomials are continuous.
More precisely, if $F$ has the shape $F = (p{\leq}0 \land G)$, and if $r$ is an $x$-root of $p$, then $r$ already satisfies $p{\leq}0$ in any valuation of the non-$x$ variables, so including $r+\epsilon$ as a sample point on account of $p{\leq}0$ is redundant.
It is possible that $G$ contains some atom $q < 0$ or $q \neq 0$ where $r$ is an $x$-root of $q$.
In this case, $r + \epsilon$ will already be a sample point on account of $q$, and we do not need to add it in on account of $p$.
Alternatively, if $G$ does not contain such a $q$, then, in any valuation of the non-$x$ variables, it is impossible for $G$ to be satisfied by $r + \epsilon$ and not $r$, meaning that it is redundant to include $r+\epsilon$ as a sample point on account of $G$.

The general QE theorem is proved in Isabelle/HOL as the following, using Remark \ref{rem:trick} to restrict to the univariate case and avoid well-definedness formulas:
\begin{isabelle}
\generalqe
\end{isabelle}

Here, \texttt{`} is the Isabelle/HOL syntax for mapping a function over a set.
This theorem says that if a finite-length formula \isa{F} is of the requisite shape, then there exists an $x$ satisfying \isa{F} iff \isa{F} is satisfied at $-\infty$ (captured by $\isa{F}_{\isa{inf}}$), or there is a root \isa{r} of one of the $=$ or $\leq$ atoms where \isa{F} \isa{r} holds, or if there is a root \isa{r} of one of the $<$ or $\neq$ atoms where \isa{F}$_{\varepsilon}$ \isa{r} holds.
The proof is quite lengthy and involves a significant amount of casework; however, because we are working with univariate polynomials thanks to \rref{rem:trick}, this casework mostly reduces to arithmetic computations and basic real analysis for univariate polynomials, and some of what we need, such as properties of discriminants and continuity properties of polynomials, is already formalized in Isabelle/HOL's standard library.

\subsection{Top Level Algorithms}\label{sec:toplevel}
We develop several top-level algorithms that perform these VS procedures on multivariate polynomials; these are described in more detail in \rref{app:exported}.
Crucially, each features its own proof of correctness.
For example, for the \isa{VSEquality} algorithm, which performs equality VS repeatedly, we have:

\begin{isabelle}
  \correctnessEquality
\end{isabelle}

Here, the \isa{eval} function expresses the truth value of the (multivariate) input formula given a valuation \isa{xs}, represented as a list of real numbers.
Since we quantify over all possible valuations and express that they are the same before and after running the algorithm, we prove the soundness of \isa{VSEquality}.
The correctness of this theorem only relies on Isabelle/HOL's trusted core.

As our algorithms are general enough to handle formulas with high degree polynomials where VS does not apply, we cannot assert that the result is quantifier free (it might not be).
To demonstrate the practical usefulness of these algorithms, we export our code to SML and experimentally show that these algorithms solve many benchmarks.
The code exports rely on the correctness of Isabelle/HOL's code export, which ongoing work is attempting to establish \cite{DBLP:conf/esop/HupelN18}.

\section{Framework}\label{sec:framework}
We turn to a discussion of our framework, which is designed with two key goals in mind: First, perform VS as many times as possible on any given formula.
Second, reduce unwieldy multivariate proofs to more manageable univariate ones.

\subsection{Representation of Formulas}\label{sec:representation}
We define our type for formulas in the canonical datatype \isa{fm}:
\begin{isabelle}
\datatypefm
\end{isabelle}

As in Nipkow's previous work\cite{nipkow2010linear}, we use De Bruijn indices to express the variables: That is, the 0th variable represents the innermost quantifier, and variables greater than the number of quantifiers represent the free variables.

We have two constructors for each type of quantifier: \isa{ExQ F} (resp. \isa{AllQ F}) indicates a single existential (universal) quantifier, and \isa{ExN n F} (resp. \isa{AllN n F}) represents a \emph{block} of \isa{n} existential (universal) quantifiers. These representations are interchangeable and converted back and forth in our algorithm; we include the block representation for variable ordering heuristics (see \rref{app:VariableOrdering}). 

We utilize the multivariate polynomial library\cite{sternagel2010executable} to define our atoms:
\begin{isabelle}
\datatypeatom
\end{isabelle}
Each atom is normalized without loss of generality, so that the atom \isa{Less p} means $p<0$, \isa{Eq p} means $p=0$, and so on.

For example, the \FOLR~formula $\forall x.((\exists y. x a=y^2 b) \land \neg(\forall y. 5x^2\leq y))$ is represented in our framework as follows, where \isa{Const n} represents the constant $n \in \reals$, and \isa{Var i} represents the $i$th variable:
\begin{verbatim}
AllQ (And (ExQ (Atom (Eq (Var 1 * Var 2 - (Var 0)^2 * Var 3))))
         (Neg (AllQ (Atom (Leq (Const 5 * (Var 1)^2 - Var 0)))))).
\end{verbatim}

Note that we could restrict ourselves to the $\top, \neg, \lor, \exists$ connectives and normalize $\leq$ and $\neq$ atoms to combinations of $<$ and $=$ atoms, and we could still express all of \FOLR.
We avoid this for two reasons: because it would linearly increase the size of the formula, and because we want to handle $\leq$ atoms in the optimized way discussed in \rref{sec:genVS} (see \rref{eq:QEequiv2}).
We do, however, allow for the normalization of $p=q$ into $p-q=0$.
This does not affect the size of the formula, and can afford simplifications: For example, $x^3+x^2+x+1=x^3$ becomes $x^2+x+1=0$.

\subsection{Modified Disjunctive Normal Form}\label{sec:dnf}
Nipkow's prior work \cite{nipkow2010linear} avoided incurring cases where linear VS does not apply by constraining atoms to be linear.
In order to develop a general-purpose VS method which can be used, e.g., as a preprocessing method for CAD, we must reason about cases where VS fails to perform QE for a specific quantifier, and still continue the execution of the algorithm to the remaining quantifiers to simplify the formula as much as possible.
To help with this, we implement a modified disjunctive normal form (DNF) that allows expressions to involve quantifiers.

\subsubsection{Contextual Awareness.}
Let us analyze how to increase the informational content in a formula with respect to a quantified variable of interest.

Say we wish to perform VS to eliminate variable $x$ in the formula $\exists x. F$, where $F$ is not necessarily quantifier free.
In linear time, we remove all negations from the formula by converting it into negation normal form.
We can then normalize $\exists x. F$ into the following form, where the $A_{n,i}$'s are (quantifier-free) atoms:
\begin{equation*}
\exists x. \bigvee_n \Big(\bigwedge_i A_{n,i}\land \bigwedge_j \big(\forall y. F_{n,j}\big) \land \bigwedge_k\big(\exists z. F_{n,k}\big) \Big).
\end{equation*}

This normalization procedure is similar to standard DNF, as it handles quantified formulas as if they were atomic formulas.
We can distribute the existential quantifier across the disjuncts, which results in the equivalent formula:
\begin{equation}\label{dnf}
    \bigvee_n \exists x.\Big(\bigwedge_i A_{n,i} \land \bigwedge_j \left(\forall y. F_{n,j}\right) \land \bigwedge_k\left(\exists z. F_{n,k}\right) \Big).
\end{equation}
Now we run the VS algorithm, i.e. the input to VS is a conjunction of atomic formulas and quantified formulas in the shape of (\ref{dnf}).
Notice that if equality VS applies to atom $A_{n,i}$, then the relevant roots can be substituted into the quantified formulas $F_{n,j}$ and $F_{n,k}$, but roots from $F_{n,j}$ or $F_{n,k}$ cannot be substituted into $A_{n,i}$ since they feature quantified variables which are undefined in the broader context.
So, our informational content is greatest when the number of $A_{n,i}$ atoms is maximized and the sizes of the $F_{n,j}$ and $F_{n,k}$ are minimized.

\subsubsection{Innermost Quantifier Elimination.}\label{sec:innermosttooutermost}
The innermost quantifier has an associated formula which is entirely quantifier free (and thus has no $F_{n,j}$ and $F_{n,k}$).
As such, we opt to perform VS recursively, starting with the innermost quantifier and moving outwards, hoping that VS is successful and the quantifier-free property is maintained.
This is not always optimal.
Consider the following formula:
\begin{equation*}
   \exists x. (x=0\land \exists y.\ xy^3+y=0).
\end{equation*}
If we attempt to perform quadratic VS on the innermost $y$ quantifier, it is cubic and will fail.
However, performing VS on the $x$ quantifier first fixes $x=0$, which converts the cubic $xy^3+y=0$ equality into the linear $y=0$. 
So, an (unoptimized) run of inside-out VS would produce $\exists y. y = 0$, and we could completely resolve the QE query by running VS again.

\subsubsection{Reaching Under Quantifiers.}\label{sec:reach}
We would like to recover usable information from the $F_{n,k}$ formulas to increase the informational content going into our QE algorithm.
It would be ideal if we could ``reach underneath" the existential binders and ``pull out'' the atoms from the formulas.
We can achieve this through a series of transformations. Let $k$ range from $0$ to $K_n$.
If we pull out each existential quantifier one by one, we get the following formula, which is equivalent to formula (\ref{dnf}):
\begin{equation*}
    \bigvee_n \exists z_0.\cdots \exists z_{K_n}.\exists x. \Big(\bigwedge_i A_{n,i} \land \bigwedge_j \left(\forall y. F_{n,j}\right) \land \bigwedge_k F_{n,k} \Big)
\end{equation*}
This works because the rest of the conjuncts do not mention the quantified variable $z_k$ and adjacent existential quantifiers can be swapped freely (without changing the logical meaning of the formula).

We can then recursively unravel the formulas $F_{n,k}$, moving as many existential quantifiers as possible to the front.
Our implementation does this via a bottom-up procedure, starting underneath the innermost existential quantifier and building upwards, normalizing the formula into the form:
\begin{equation*}
    \bigvee_n \exists z_0.\cdots \exists z_{K_n}.\exists x. \Big(\bigwedge_i A_{n,i} \land \bigwedge_j \forall y. F_{n,j} \Big)
\end{equation*}

On paper, these transformations are simple as they involve named quantified variables; however, because our implementation uses a locally nameless form for quantifiers with de Bruijn indices, shifting an existential quantifier requires a ``lifting'' procedure $A\negmedspace\uparrow$ which increments all the variable indices in $A$ by one.
This allows for the following conversion: $A\land \exists z. F \longleftrightarrow \exists z. ((A\negmedspace\uparrow) \land F)$.

\subsection{Logical Evaluation}\label{sec:eval}

Our proofs show that the input formula and the output formula (after VS) are \textit{logically equivalent}, i.e., have the same truth value under any valuation.
This needs a method of ``plugging in'' the real-valued valuation into the variables of the polynomials.
Towards this, we define the \isa{eval} function, which accumulates new values into the valuation as we go underneath quantifiers, and the \isa{aEval} function, which homomorphically evaluates a polynomial at a valuation.

When proving correctness, we focus our attention on one quantifier at a time.
By \rref{rem:trick}, correctness of general VS follows when considering a formula $F$ with a single quantifier, where $F$ contains only polynomials of at most degree two (otherwise general VS does not apply).
With these restrictions, we can substitute a valuation into the non-quantified variables, transforming multivariate polynomials into univariate polynomials.
For example, let $a, b,$ and $c$ be arbitrary multivariate polynomials that do not mention variable $x$.
Let $\hat{p} = \gamma(p)$ denote the evaluation of polynomial $p$ at valuation $\gamma$ ($\hat{p}$ is a real number).
We obtain the following conversion between multivariate and univariate polynomials:
\begin{equation*}
\textbf{eval}\ (ax^2+bx+c = 0)\ \gamma \longleftrightarrow \textbf{evalUni}\ (\hat{a}x^2+\hat{b}x+\hat{c} = 0)\ \hat{x}
\end{equation*}

As such, we develop an alternative VS algorithm for univariate polynomials, where atoms are represented as triples of real-valued coefficients (as seen in \rref{sec:genVsubst}), and show that under this specific valuation, the multivariate output is equivalent per valuation with the output of the univariate case. Thus, we finish the proof of the multivariate case by lifting the proof for the univariate case.

\subsection{Polynomial Contributions}\label{sec:polynomials}
We build on the polynomials library \cite{sternagel2010executable}, which was designed to support executable multivariate polynomial operations.
This choice naturally comes with tradeoffs, and a number of functions and lemmas that we needed were missing from the library.
For example, we needed an efficient way to isolate the coefficient of a variable within a polynomial, which we define in the \isa{isolate\isacharunderscore variable\isacharunderscore sparse} function.
The following particularly critical lemma rewrites a multivariate polynomial in $\reals[a_1,\dots, a_n, x]$ as a nested polynomial $\reals[a_1, \dots, a_n][x]$, i.e., a univariate polynomial in x with coefficients that are polynomials in $\reals[a_1, \dots, a_n]$:

\begin{isabelle}
\sumOverDegree
\end{isabelle}
This is needed rather frequently within VS, as we often seek to re-express polynomials with respect to a single quantified variable of interest, and although it is mathematically quite obvious, its verification was somewhat involved.

Additionally, to utilize the variables within polynomials as de Bruijn indices, we implemented various lifting and substitution operations.
These include the \isa{liftPoly} and \isa{lowerPoly} variable reindexing functions.
These and other contributions to the polynomials library are discussed in \rref{app:polynomials}.

\section{Experiments}\label{sec:experiments}

The benchmark suite consists of 378 QE problems in category \caderwv collected from 94 examples \cite{DBLP:conf/cade/PlatzerQR09}, and category \economics with 45 QE problems \cite{DBLP:journals/corr/abs-1804-10037}.
\caderwv and \economics examples were converted into decision problems, powers were flattened to multiplications, and \caderwv were additionally rewritten to avoid polynomial division.
For sanity checking, we also negated the \caderwv examples \cite{DBLP:conf/cade/PlatzerQR09}.
We run on commodity hardware.\footnote{\label{hardware} MacBook Pro 2019 with 2.6GHz Intel Core i7 (model 9750H) and 32GB memory (2667MHz DDR4 SDRAM).}
The benchmark examples, as well as all scripts to rerun the experiments are in \cite{VSBenchmarks_Zenodo}.

\paragraph{Tools.}
We compare the performance of 
\begin{enumerate*}[label=\emph{\alph*})]
\item our \equalityvs (\textbf{\eqvs}), \generalvs (\textbf{\genvs}), \luckyvslong (\textbf{\luckyvs}), and \legvslong (\textbf{\legvs}) algorithms (\rref{app:exported}) to 
\item Redlog~\cite{DBLP:journals/cca/Dolzmann097} snapshots 2021-04-13\footnote{\tiny\url{https://sourceforge.net/projects/reduce-algebra/files/snapshot_2021-04-13/}} (\textbf{\redlogold}) and 2021-07-16\footnote{\tiny\url{https://sourceforge.net/projects/reduce-algebra/files/snapshot_2021-07-16/}} (\textbf{\redlognew}, which includes bug fixes for contradictions we reported to Redlog developers), 
\item SMT-RAT~21.05\footnote{\tiny\url{https://github.com/ths-rwth/smtrat/releases/tag/21.05}}~\cite{DBLP:conf/sat/CorziliusKJSA15} quantifier elimination (\textbf{\smtratcad}) and satisfiability checking (\textbf{\smtratsat}), 
\item the SMT solver Z3~4.8.10\footnote{\tiny\url{https://github.com/Z3Prover/z3/releases/tag/z3-4.8.10}}~\cite{DBLP:conf/tacas/MouraB08} (\textbf{\zthree}), and 
\item Wolfram Engine~12.3.1(\textbf{\mathematicavs}, \textbf{\mathematicacad}).
\end{enumerate*}
All tools were run in Docker containers on Ubuntu 18.04 with 8GB of memory and 6 CPU cores.
Tool syntax translations from SMT-LIB format were done prior to benchmarking: For our VS algorithms, examples were translated to SML data structures and compiled with MLton\footnote{\tiny\url{http://mlton.org/}}; as a result, measurements do not include parsing.
For \mathematicavs and \mathematicacad, examples were translated into Wolfram syntax, including configuration options restricting QE to quadratic virtual substitution in \mathematicavs.
For \smtratcad, \texttt{check-sat} was replaced with \texttt{eliminate-quantifiers}.

\begin{figure}
\includegraphics[height=17em]{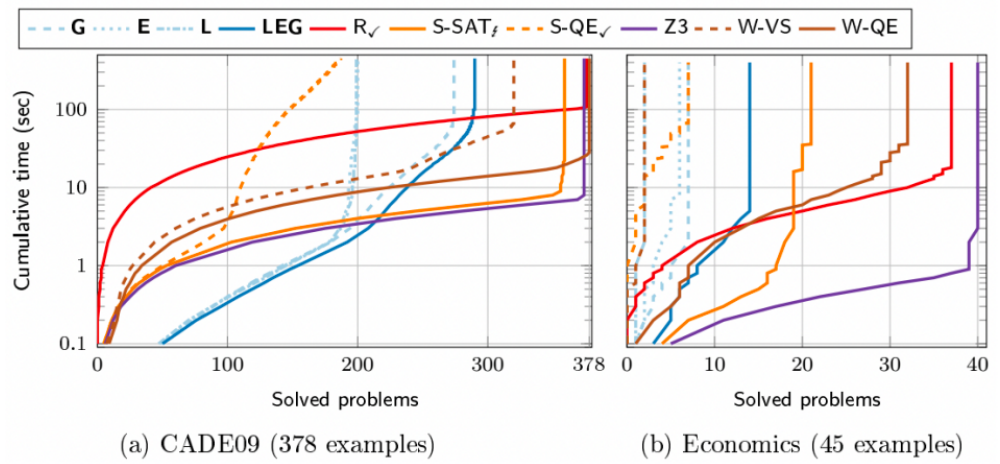}
\caption{Cumulative time to solve fastest $n$ problems
(flatter and more is better)
}
\label{fig:evaluationsummary}

\end{figure}

\paragraph{Results.}
Each example has a timeout of 30s.
Figure~\ref{fig:evaluationsummary} summarizes the performance on the \caderwv and \economics examples in terms of the cumulative time needed to solve (return ``true'', ``false'', ``sat'', or ``unsat'') the fastest $n$ problems with a logarithmic time axis: more problems solved and a flatter curve is better.

Wolfram Engine solves all problems in the \caderwv category, closely trailed by Redlog, Z3.
The near constant computation time offset of Redlog in comparison to Z3, SMT-RAT, and Wolfram Engine may be attributable to the additional step of entering an SMT REPL.
Our verified \equalityvs (\eqvs), \generalvs (\genvs), \luckyvslong (\luckyvs), and \legvslong (\legvs) algorithms rank in performance between the basic quantifier elimination implementation in SMT-RAT (\smtratcad), virtual substitution in Wolfram Engine (\mathematicavs), full SMT approaches (\smtratsat, \zthree), and combined virtual substitution plus CAD implementations (\redlognew, \mathematicacad).
The reduced startup time of our algorithms is attributable to the omitted parsing step.
Overall, \equalityvs and \luckyvslong solve examples fast, but the wider applicability of \generalvs and \legvslong allows them to solve considerably more examples.
Though we have already implemented a number of optimizations for VS (\rref{app:optimiz}) we do not expect to outperform prior tools at this stage, as many of them have been optimized over a period of many years.

\begin{figure}
\includegraphics[height=10.5em]{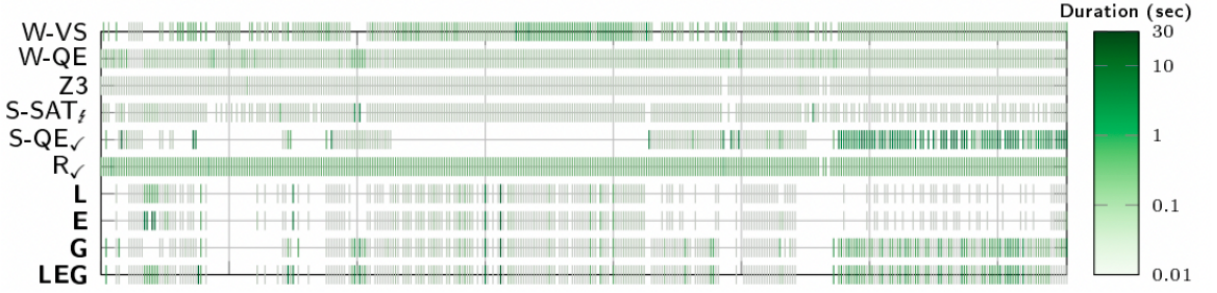}
\caption{\caderwv duration per problem (color indicates duration, lighter is better)}
\label{fig:detailedevaluation}
\end{figure}

A comparison of duration per problem is in \rref{fig:detailedevaluation}.
Though there is considerable overlap between \equalityvs, \generalvs, and \luckyvslong, mutually exclusive sets of solved examples (and considerable performance differences on a number of examples) foreshadow the performance achievable with the combined \legvslong algorithm.

\begin{figure}[b!tp]
\includegraphics[height=8.5em]{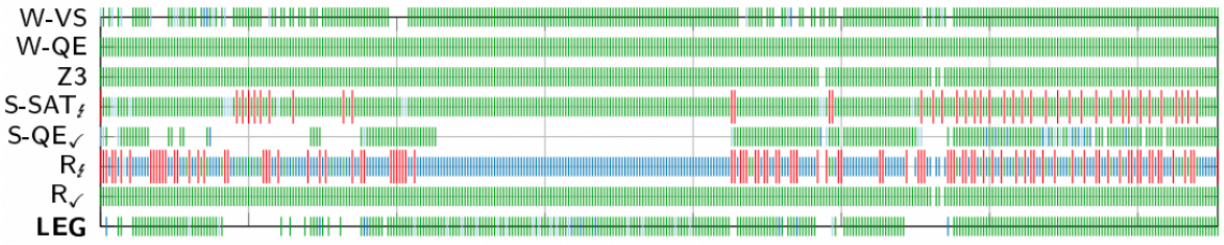}
\caption{\caderwv consistency comparison between original and negated formula: color indicates discrepancies within tools (green \captioncolorsquare{3}{Paired}: answer on original and negated formula agree, dark-blue \captioncolorsquare{1}{Paired}: only original solved, light-blue \captioncolorsquare{0}{Paired}: only negated solved, red+long \captioncolorsquare{5}{Paired}: \textbf{contradictory} answers (both formulas unsat/proved or both sat/disproved), empty: both timeout/unknown)}
\label{fig:falseevaluation}
\end{figure}

\paragraph{Contradictions.}
In \rref{fig:falseevaluation}, we compare the \caderwv results to the results on negated \caderwv examples to highlight \emph{contradictions} between answers (e.g., both $A$ and $\lnot A$ are claimed to be true).
Wolfram Engine and Z3 answer consistently on both formula sets, and solve (almost) all examples.
Redlog, the main VS implementation, in \redlogold and previous versions in general does not perform well on the negated formulas and reports 96 contradictory answers; the contradictory examples were shared with the developers and triggered several bug fixes that are now available in \redlognew (no contradictions found on the benchmark set).
SMT-RAT performs better than \redlogold on the negated formulas, but in satisfiability mode contradicts itself on 41 examples by silently ignoring quantifiers in the input; in quantifier elimination mode, SMT-RAT supports quantifiers and does not report contradictions, but SMT-RAT then incurs a significant performance loss
(\smtratsat reports 359 answers while \smtratcad only solves 187).
No contradictions were found across tools, i.e., whenever a tool's answers were consistent internally, the answers agreed with those of other tools.
Our \legvslong algorithm has similar performance for proving and disproving in terms of absolute number of solved examples, but combining proving and disproving would still solve more examples than just one question individually (as for \smtratcad and \mathematicavs).

In summary, the performance of our verified virtual substitution QE on the benchmark set is encouraging. 
The number of solved examples is close to other VS implementations (304 examples by our \legvslong vs. 322 by \mathematicavs) and the cumulative solving time reveals that the majority of examples are solved fast.

\section{Conclusion and Future Work}\label{sec:future}
We verify linear and quadratic virtual substitution for real arithmetic; our algorithms are \textit{provably correct} up to Isabelle/HOL's trusted core and code export.
Developing practical verified VS in Isabelle/HOL required significant low-level improvements and extensions to Isabelle's multivariate polynomials library.
Our extensive experiments both reveal the benefits of our current optimizations and indicate room for future improvements.
Further optimizations to the polynomial libraries, such as efficient coefficient lookup for polynomials using red black trees, would be welcome.
Expanding our framework to handle formulas that involve polynomial division would also be of practical significance.
Continuing to develop our formalization with such improvements is of especial significance given that our experiments found long-standing errors in existing unverified real arithmetic tools.
This demonstrates that, even if verification were not a virtue in and of itself, real arithmetic is so subtle that formal verification is the best way toward an implementation that is both useful and correct in practice.

\paragraph{Acknowledgments.}
We wish to thank Fabian Immler for his substantial contributions at CMU to the polynomial theories of Isabelle/HOL and regret that his current industry position precludes our ability to include him as a coauthor.
Thank you also to the anonymous FM reviewers for their useful feedback.

This material is based upon work supported by the National Science Foundation under Grant No. CNS-1739629, a National Science Foundation Graduate Research Fellowship under Grants Nos. DGE1252522 and DGE1745016, and by the AFOSR under grant number FA9550-16-1-0288. Any opinions, findings, and conclusions or recommendations expressed in this material are those of the author(s) and do not necessarily reflect the views of the National Science Foundation or of AFOSR.

%
%
\let\oldbibliography\thebibliography
\renewcommand{\thebibliography}[1]{%
  \oldbibliography{#1}%
  \addtolength{\itemsep}{-6pt}%
}

\bibliography{VirtualSubstitution}

\iflongversion
\appendix
\section{Isabelle/HOL Code}
This appendix contains some of the key Isabelle/HOL code used within the Equality VS algorithm (see \rref{sec:eqVsubst}) and the General VS algorithm (see \rref{sec:genVsubst}).

\subsection{Linear Substitution}\label{app:linsubst}
In Isabelle/HOL, we formalize linear substitution in the \isa{linear\isacharunderscore substitution} function.
Here, the call $\isa{linear\isacharunderscore substition\ i\ a\ b\ A}$ returns the \isa{atom\ fm} that is the result of virtually substituting the fraction \isa{a/b} for variable with de Bruijn index \isa{i} into atom \isa{A} (the \isa{atom} and \isa{atom\ fm} datatypes are explained in \rref{sec:representation}).
More precisely, the \isa{linear\isacharunderscore substition} function is taking as input a natural number, which indicates which variable is of interest, two real multivariate polynomials $a$ and $b$, and an \isa{atom} $A$.
It cases on the structure of $A$ and follows the casework described in \rref{sec:eqVsubst}.

\begin{isabelle}
\linearsubstitution
\end{isabelle}

In each case, the first step of the algorithm is to recover the degree of the polynomial we wish to substitute into.
We store this in \isa{d}.
Then, we can express the polynomial as a summation of monomials with respect to the variable we are eliminating on.
To read off the $i$th coefficient of the polynomial \isa{p} with respect to \isa{var}, we use the \isa{isolate\isacharunderscore variable\isacharunderscore sparse\ p\ var\ i}  function.

For the $=$ and $\neq$ cases, we multiply the $i$th coefficient by \isa{a{\isacharcircum}i * b {\isacharcircum}(d-i)}, to reflect that we have normalized by multiplying everything by \isa{b{\isacharcircum}d} (more specifically, this reflects that $(a^i/b^i)\cdot b^d = a^i\cdot b^{d-i}$).
For the $<$ and $\leq$ cases, we additionally need to check the parity of the degree \isa{d}.
When \isa{d} is odd, we must multiply the whole polynomial by an additional factor of \isa{b}, as explained in \rref{sec:eqVsubst}; in the code, we accomplish this by multiplying by the extra factor of \isa{b\isacharcircum(d\ mod\ 2)}  in the $<$ and $\leq$ cases.

\subsection{Quadratic Substitution}\label{app:quadsubst}
In Isabelle/HOL, we formalize quadratic substitution in the \isa{quadratic\isacharunderscore sub} function.
Here, \isa{quadratic\isacharunderscore sub\ i\ a\ b\ c\ d\ A} returns the result of virtually substituting $(a + b \sqrt{c})/d$ for the variable with de Bruijn index \isa{i} in atom \isa{A}.
All cases of the different atoms utilize the same helper functions.
As explained in \rref{sec:eqVsubst}, they differ in the final arrangement of the polynomial $A+B\sqrt{c}$, where $A$ and $B$ are multivariate polynomials that do not mention the variable we are eliminating on.

\begin{isabelle}
\quadraticsub
\end{isabelle}

The helper function \isa{quadratic\isacharunderscore part\isacharunderscore 1} formalizes the normalization of denominators discussed in \rref{sec:eqVsubst}.
It behaves identically to the \isa{linear\isacharunderscore substitution} function discussed in \rref{app:linsubst}, except that the fractional polynomial that \isa{quadratic\isacharunderscore part\isacharunderscore 1} is substituting into variable \isa{var} is \isa{(a+b*(Var\ var))/d}.

\begin{remark}\label{rem:samesubs}
Notice that in the above, we are actually substituting a polynomial mentioning \isa{var} \textit{for the variable} \isa{var}.
This situation arises as a consequence of our multiple-step substitution process for quadratic roots: If we associate the polynomial variable \isa{var} with the meta-variable $x$ and create a second meta-variable $y$, then we want to perform the substitutions $x=(a+by)/d$ and $y=\sqrt{c}$.
After performing the first substitution for meta-variable $x$, the polynomial variable \isa{var} representing the meta-variable $x$ gets eliminated.
As such, it is okay for us to repurpose the same polynomial variable \isa{var} to represent our new meta-variable $y$ when performing the substitution $y=\sqrt{c}$.
Further, reusing this specific variable \isa{var} allows us to avoid potential conflicts with other variables within the polynomial. 
\end{remark} 

The helper function \isa{quadratic\isacharunderscore part\isacharunderscore 2} handles the parity check on the normalization factor, as described in \rref{sec:eqVsubst}.
Here, we simply collect the summation of all the even-degree monomials (in our variable of interest) into one polynomial $A$ and all the odd-degree monomials into another polynomial $B$, so that ultimately our polynomial is of the form $A + B\sqrt{c}$.
\begin{isabelle}
\quadraticparttwo
\end{isabelle}

In this, \isa{of\isacharunderscore nat} casts natural numbers to real numbers.
For each $\isa{0} \leq \isa{i} \leq \isa{deg}$, \isa{Const(of\isacharunderscore nat(i mod 2))} is 0 when \isa{i} is even and 1 when \isa{i} is odd.
Similarly, \isa{Const(1 - of\isacharunderscore nat(i mod 2))} is 1 when \isa{i} is even and 0 when \isa{i} is odd.
So, 
\begin{align*}\isa{(isolate}&\isa{\isacharunderscore variable\isacharunderscore sparse p var i)*(sq\isacharcircum(i div 2)) *}\\
&\isa{(Const(of\isacharunderscore nat(i mod 2)))*(Var var)}
\end{align*}
collects the summation of the odd-degree monomials (in \isa{var}) in the form $a_ic^{\left\lfloor i/2 \right\rfloor}\sqrt{c}$ where \isa{Var var} represents $\sqrt{c}$, while
\begin{align*}
\isa{(isolate}&\isa{\isacharunderscore variable\isacharunderscore sparse p var i)*(sq\isacharcircum(i div 2)) *}\\
&\isa{(Const(1 - of\isacharunderscore nat(i mod 2)))}
\end{align*}
collects the summation of the even-degree monomials in \isa{var} in the form $a_ic^{i/2}$.
Notice that in the even cases, we are able to completely eliminate the square root, whereas the odd cases leave us with a variable of degree one, for which we will need to substitute $\sqrt{c}$.

We use the technique explained in \rref{rem:samesubs} to handle substituting a polynomial mentioning \isa{var} in for the variable \isa{var}; this time, our first polynomial variable \isa{var} represents the meta-variable $y=\sqrt{c}$, and after the substitution, the polynomial variable \isa{var} represents the meta-variable $z=\sqrt{c}$.

\section{Top-Level Algorithms}\label{app:exported}
We develop (and export to SML for experimentation) several top-level algorithms that use various combinations of VS procedures and optimizations; these are briefly detailed here.

The \isa{VSLucky} algorithm recursively searches through the available atom conjunct list, searching for a quadratic equality atom that features a constant coefficient with respect to the variable $x$ that we are eliminating. 
If it finds some $ax^2+bx+c$ where $a$, $b$, or $c$ is a nonzero constant, we are guaranteed that $ax^2+bx+c$ is not the zero polynomial, and thus we have the following full elimination of the quantifier on $x$, without the remaining zero case (cross reference \rref{sec:eqVsubst}):
\begin{align*}
&\left(\exists x. (ax^2+bx+c=0 \land F)\right)\longleftrightarrow \\
&\Big((a=0 \land b\neq0 \land F_x^{-c/b})~\lor \\
&(a\neq 0\land b^2-4ac\geq 0\land (F_x^{(-b+\sqrt{b^2-4ac})/(2a)}\lor F_x^{(-b-\sqrt{b^2-4ac})/(2a)}))  \Big).
\end{align*}
Then we can use VS to expand the RHS of the equivalence.
So, \isa{VSLucky} is indeed lucky, because it fully removes the quantifier on the variable in question.

Next, the \isa{VSEquality} algorithm performs the equality version of VS (\rref{sec:eqVsubst}) iteratively for all equality atoms of at most quadratic degree, and the \isa{VSGeneral} algorithm performs the general version of VS (\rref{sec:genVsubst}).\footnote{Additionally, we have \isa{VSEquality\isacharunderscore 3\isacharunderscore times}, which performs the equality algorithm three times, and \isa{VSGeneral\isacharunderscore 3\isacharunderscore times}, which performs the general algorithm (respectively) three times. Although these account for the kind of potential edge cases discussed in \rref{sec:innermosttooutermost}, they did not realize significant experimental benefits over \isa{VSEquality} and \isa{VSGeneral}.}

Finally, we have \isa{VSLEG} which performs all three of \isa{VSLucky}, \isa{VSEquality}, \isa{VSGeneral} (in that order, as the first is the quickest while the last is the most general); this algorithm was the most competitive one on our benchmarks.

We now discuss the correctness theorem for these algorithms.

\subsection{Correctness of Top-Level Algorithms}
The correctness of our top-level algorithms is established by the following theorem, which expresses that any initial formula \isa{$\varphi$} is logically equivalent to the formula obtained by running the algorithm on \isa{$\varphi$}; in other words, the two formulas (initial and final) have the same truth-value in every state, i.e. for every valuation of free variables.

\begin{isabelle}
\correctness
\end{isabelle}

In this theorem, our \isa{eval} function captures the semantics of substituting a valuation into a formula: For a valuation characterized by a list $\nu$ of real numbers and a formula $\varphi$, \isa{eval $\varphi$ $\nu$} is true whenever $\varphi$ is true at $\nu$, i.e. when the $i$th entry of $\nu$ is plugged in for the $i$th free variable of $\varphi$ (for all $i$).
More precisely, the exact definition of \isa{eval} is as follows:

\begin{isabelle}
\eval
\end{isabelle}

This is the canonical semantics for evaluating atom formulas.
The interesting cases are the \isa{Atom}, \isa{ExQ}, and \isa{AllQ} cases.
We discuss each.  

In the \isa{Atom} case, the \isa{aEval} function is as follows:
\begin{isabelle}
\aEval
\end{isabelle}

Here, we are using the \isa{insertion} function from the multivariate polynomials library \cite{sternagel2010executable} to insert the $n$ values from list \isa{$\nu$} for variables $0, \dots, n-1$.
This captures evaluation by substituting values for the free variables in the formula.

Because the first argument of the \isa{insertion} function has type \isa{(nat $\Rightarrow$  \texttt{'}a)}, where in our case \isa{\texttt{'}a} is \isa{real}, we use Isabelle/HOL's standard library function \isa{nth\isacharunderscore default} to expand \isa{$\nu$} into a mapping from nats to reals.
That is, \isa{(nth\isacharunderscore default\ 0\ $\nu$)\ i}  is the $i$th element of \isa{$\nu$} when $i$ is less than the length of \isa{$\nu$} (lists are 0-indexed in Isabelle/HOL), and 0 otherwise.
Note that this means that if our list \isa{$\nu$} is not long enough to cover the valuations for all the free variables in the polynomial, the \isa{nth\isacharunderscore default} function will assign it a value of $0$.
This is merely a convenience; our correctness lemmas are unaffected by this, as they quantify over all possible valuations \isa{$\nu$} (and thus over all valuations of the correct length).

The \isa{ExQ \isasymphi} case of a new existential quantifier is handled by embedding into Isabelle's built-in notion of existential quantification over the real numbers.
More precisely, in \isa{{\isasymexists}x{\isachardot}{\isacharparenleft}{\kern0pt}eval\ {\isasymphi}\ {\isacharparenleft}{\kern0pt}x{\isacharhash}{\kern0pt}{\isasymnu}{\isacharparenright}{\kern0pt}{\isacharparenright}}, Isabelle treats the quantified variable $x$ as a new free variable, which is added to the front of the valuation $\nu$, written \isa{x\isacharhash $\nu$}.
The \isa{AllQ} case is similar.

As an example, consider evaluating $$\exists x. \forall y.\ (x + y\cdot y + z > 0)$$ in a state $\nu$ where $\nu(z) = 1$.
In our framework in Isabelle/HOL, this translates as:
\begin{verbatim}
eval (ExQ (AllQ (Var 1 + Var 0*Var 0 + Var 2 > 0)) 1
\end{verbatim}
Here we have two quantified variables, Var 0 and Var 1, and one free variable, Var 2 (this is because we are using de Bruijn indices).
Var 0 matches the \isa{AllQ} quantifier and Var 1 matches the \isa{ExQ} quantifier.
We are considering the valuation where Var 2 is set to 1.

In the first step, we expand to:

\begin{verbatim}
Exists x.
    (eval (AllQ (Var 1 + Var 0*Var 0 + Var 2 > 0)) (x#1)),
\end{verbatim}
and in the second step, we achieve:

\begin{verbatim}
Exists x. Forall y.
	(eval (Var 1 + Var 0*Var 0 + Var 2 > 0) (y#(x#1))).
\end{verbatim}
This asks whether there exists an $x$ such that for all $y$, $x + y^2 + 1 > 0$, which matches the desired semantics.

We additionally include \isa{ExN} and \isa{AllN}, which take in two inputs $i$ and $\varphi$ and are equivalent to the single quantifier forms \isa{ExQ} and \isa{AllQ} repeated $i$ times on $\varphi$ (which, thanks to de Bruijn indices, are quantifiers for $i$ different variables).
In the \isa{eval} function, these are represented via a quantified list $\isa{l}$ of real number valuations of length $i$.
Having these \isa{ExN} and \isa{AllN} in the representation allows for specialized algorithms like our block quantifer heuristics (see \ref{app:VariableOrdering}).

Now that we have established that our correctness theorem is correctly stated, we discuss its proof.

\subsection{Proving Correctness: Extensibility}\label{app:qednf}
It would be very tedious to independently prove the correctness theorems for all of our top-level algorithms.
Instead, our framework is designed so that any optimization \isa{opt} can easily be incorporated into our framework, as long as \isa{opt} does not change the truth-value of any formula; i.e. our framework is designed to be \textit{extensible}.
This allows us to cleanly substitute different combinations of optimizations into our top-level QE algorithms without incurring the burden of significant reproving.

This extensibility can be seen in the following correctness lemma for our \isa{QE\isacharunderscore dnf} function, which lifts QE algorithms defined for a single quantifier to apply to all quantifiers using the modified DNF transformation discussed in \rref{sec:dnf}:
\begin{isabelle}
  \overallAlgo
\end{isabelle}

Here, the \isa{step} function is intended to be a function that performs virtual substitution.
Its behavior is governed by the \isa{steph} hypothesis.
The \isa{opt} function is intended to be a function that performs various optimizations, and its behavior is governed by the \isa{opt} hypothesis.
Intuitively, \isa{QE\_dnf\_eval} proves that any function that obeys the \isa{steph} hypothesis (as we prove that our VS procedures do) and any functions that obey the \isa{opt} hypothesis (as we prove that our optimizations do) can be combined to create an overall top-level QE procedure.
We now discuss the characteristics of \isa{opt} and \isa{step} further. 

The \isa{opth} hypothesis assumes that we have a procedure \isa{opt} which preserves the truth value of the \isa{eval} function for every valuation on every formula.
In our QE framework, this function is called before performing the DNF transformation of \rref{sec:reach}, and then \isa{step} is called after performing the DNF transformation.

As specified by the \isa{steph} hypothesis, the \isa{step} function receives as input two natural numbers \isa{amount} and \isa{var}, a conjunct list of atoms, \isa{L} and a conjunct list of atom formulas, \isa{F}.
Here, \isa{amount} is optional information that indicates for how many of the \isa{var+1} existential quantifiers in the prefix QE has not yet been attempted.
Since the modified DNF transformation is performed before \isa{step}, some quantifiers will result from this algorithm, and for these we have already attempted VS.
Tracking \isa{amount} allows us to stop computation early, rather than re-attempting futile VS on quantifiers where it previously did not apply.

The left-hand side of the equality in \isa{steph} represents the form that our formula takes in one of the disjuncts of the DNF transformation (cross reference \rref{sec:reach}): there are \isa{var+1} existential quantifiers in this disjunct (we use zero-indexing for the variable), and everything in \isa{L} and \isa{F} is conjuncted.
Intuitively, this is what we pass into virtual substitution.
The right-hand side of the equality in \isa{steph} represents the result of calling the \isa{step} function.
The equality captures that the original formula and the formula we obtain by applying \isa{step} have the same truth value in any valuation \isa{$\nu$}.
So, overall, the \isa{steph} equality captures that \isa{step} can perform any arbitrary manipulation of the inputs as long as it preserves logical equivalence.

The \isa{QE\_dnf\_eval} lemma showcases the extensible nature of our framework and allows us to determine the soundness of the top level algorithms described in \rref{app:exported}, but it makes no claim about whether VS is actually simplifying our formula.
We are unable to make such claims for this top-level procedure, as we must allow for our VS algorithms to fail to make progress in cases where VS does not apply (e.g. in the presence of high degree polynomials), but we now discuss the specific cases where our VS \isa{step} procedures successfully eliminate quantifiers, as well as the important correctness lemmas for these procedures.

\subsection{Proving Correctness: VS}
Our \isa{elimVar} function is the multivariate analog of the univariate general VS procedure discussed in \rref{sec:genVS}) (and it closely resembles this procedure).
More specifically, \isa{elimVar} is an overarching proof procedure that analyzes the roots of a polynomial and applies the appropriate VS algorithms for both the equality and the off-root cases.
It checks whether we wish to substitute the exact root (for $=$ and $\leq$ atoms) or the off-roots (for $<$ and $\neq$ atoms).

As input, \isa{elimVar} receives a variable \isa{var} we are eliminating on, a list of atoms \isa{L} and formulas \isa{F} which are joined by conjunction, and the atom we wish to substitute, which is guaranteed to be at most quadratic (cross-reference \rref{sec:reach}).
Each of the top-level VS algorithms utilizes \isa{elimVar} as a helper function to substitute particular atoms.
Significantly, \isa{elimVar} has the property that it removes the variable it is substituting.
This is expressed in the following lemma, where our \isa{variableIsRemoved} function is true when the input variable \isa{var} is not present in a formula:
\begin{isabelle}
  \freeInElimVar
\end{isabelle}

An important lemma for the equality case of VS can be proven when we specify that the atom \isa{At} input to the \isa{elimVar} function is an equality atom with a quadratic or linear polynomial with respect to the variable being eliminated:
\begin{isabelle}
  \elimVarEq
\end{isabelle}

Here, we are assuming that we have the equality atom \isa{a*(Var var){\isacharcircum}{\kern0pt}{\isadigit{2}}+b*(Var var)+c} within our list of atoms \isa{L}, where the variable \isa{var} does not occur in the coefficients \isa{a}, \isa{b}, and \isa{c}; this is captured by the \isa{inList} and \isa{noVariable} hypotheses.
We also assume that this polynomial is quadratic or linear in the particular variable of interest (in the \isa{nonzero} hypothesis).
Performing \isa{elimVar} on this atom allows us to remove the quantifier completely: in the linear case, when \isa{Eq(a*(Var var)\isacharcircum 2 + b*(Var var) + c)} is \isa{Eq(b*(Var var) + c)}, \isa{elimVar} virtually substitutes \isa{-c/b} for \isa{var} into every formula; in the quadratic case, \isa{elimVar} produces the disjunct of virtually substituting the two possible quadratic roots into every formula.
This lemma establishes the correctness of the Equality VS algorithm as described in Section \ref{sec:eqVsubst}.
It notably only establishes equivalence of existence of a value for the variable $x$, rather than equivalence of truth-value.

Additionally, we use \isa{elimVar} to formulate the lemma for the general VS case:

\begin{isabelle}
  \genQeEval
\end{isabelle}

This lemma regards the executable multivariate procedure that corresponds to the univariate lemma explained in Section \ref{sec:genVS}.

Here, we assume that our formula \isa{F} is a conjunction of atoms \isa{L} as expressed in the first hypothesis.
Additionally, the second hypothesis expresses that \isa{F} has the requisite shape for general VS to apply; this means that every polynomial in \isa{L} has at most degree two with respect to the variable we are eliminating, \isa{var}.
The third assumption states that our valuation \isa{\isasymnu} is long enough to cover the variable in question (which makes it a valid valuation).

Under these assumptions, we show that there exists an $x$ that can be substituted for \isa{var} to make formula \isa{F} true iff substituting one of the sample points prescribed by virtual substitution (which are negative infinity, the roots of the atoms with respect to \isa{var}, and the corresponding off-roots) for \isa{var} makes \isa{F} hold.
Notice that the $\forall$ quantification on the $x$ on the RHS of the equality is permissible because we have actually eliminated the variable $x$ from the formula after applying virtual substitution (this is established by the \isa{elimVar\isacharunderscore removes\isacharunderscore variable} lemma).

\subsection{More Polynomial Library Contributions}\label{app:polynomials}
In order to prove correctness of VS, we needed to formalize a large number of additions to the Isabelle/HOL multivariate polynomials library.
For example, we added a partial derivative function (needed to formalize VS for infinitesimals, as discussed in \rref{sec:Infinitesimals}); and prove its correctness for polynomials of degree at most two (because more is not needed for quadratic virtual substitution).

For execution, we also needed to extend code theorems for the generically defined \isa{insertion} function, which inserts a valuation into a polynomial, in order to be able to compute the result of valuations and export our program.
This required a number of lemmas that rely on the \isa{monomials} function, which separates a polynomial into a sum of monomials, and the \isa{degree} function, which computes the degree of a multivariate polynomial with respect to a single variable.

To make these new functions usable in proofs, we formulated a large collection of lemmas for polynomials with real-valued coefficients.
These include a variety of simplification lemmas for the interaction of the \isa{isolate\isacharunderscore variable\isacharunderscore sparse} function discussed in \rref{sec:polynomials} and the \isa{insertion}, and \isa{degree} functions across summations and products of polynomials.

To utilize the variables within polynomials as de Bruijn indices, we implemented various lifting and substitution operations.
Our \isa{liftPoly} function takes in a lower limit $d$, a lifting amount $a$, and a polynomial $p$ and returns a polynomial which reindexes variables within $p$ such that every variable greater or equal to $d$ is increased by $a$.
This is commonly denoted by $p\negmedspace\uparrow^a_d$.
This \isa{liftPoly} is needed in cases where we want to reshuffle formulas to increase the number of quantifiers surrounding a formula.
For example, $\forall. ((\exists A) \land B)$ is equivalent to $\forall. \exists. (A \land (B\negmedspace\uparrow^1_1))$.

We also need an inverse function to \isa{liftPoly}, which we call \isa{lowerPoly}.
Whenever we have eliminated a quantified variable with our QE procedure, we can drop that quantifier and use \isa{lowerPoly} to reindex all the other variables accordingly.
Our \isa{clearQuantifiers} procedure implements this.

\section{Optimizations}\label{app:optimiz}
We formalize a number of optimizations for VS, as optimizations are critical for achieving reasonable performance.
It will take some time to catch up to the highly optimized performance of tools like Wolfram Engine and Redlog.

This is the benefit of an extensible framework: future optimizations can be easily integrated.
From \rref{app:qednf}, we see that any optimization function \isa{opt} that satisfies the truth-preservation lemma \isa{lemma "eval (opt F) xs = eval F xs"} can be cleanly integrated into our algorithm (using the \isa{QE\isacharunderscore dnf\isacharunderscore eval} theorem), and the composition of several of these optimization functions directly preserves this property.
We discuss our current optimizations in this appendix.

\subsection{Unpower}

As our algorithm performs VS on quadratic and linear polynomials, it is critical to reduce the degree of polynomials whenever possible.
The most natural simplification we can perform is to factor out a common $x^n$ from every monomial:
\begin{equation*}
\sum a_i x^{n+i} = x^n \sum a_i x^i = px^n
\end{equation*}
From here, we can split the atom $px^n \sim 0$, where $\sim\ \in \{=,<. \leq, \neq\}$, into lower-degree atoms that involve $p$ and $x^n$ separately.

In the equality case, $px^n=0$ reduces to $x=0 \lor p=0$, as the product of the components is zero if and only if at least one of the components is zero.
For inequalities, we case on the parity of $n$.
If we have an even exponent, we can reduce $px^n<0$ into $p<0 \land x\neq 0$, since $x^n$ is nonnegative.
Otherwise, we must assert that the sign values differ: $(p<0\land x>0) \lor (p>0\land x<0)$.

All $\neq$ are treated as negated equality atoms.
For $\leq$ atoms, we follow a similar structure to $<$: when $n$ is even, we result in $p\leq 0 \lor x=0$, and when $n$ is odd we result in $p=0\lor (p<0\land x\geq 0) \lor (p>0\land x\leq 0)$.

\subsection{Simplifying Constants}

It is clear that the atom $5=0$ is always false and can be replaced by the \isa{FalseT} formula.
Our \isa{simpfm} function replaces constant polynomial atoms with their respective \isa{TrueT} or \isa{FalseT} evaluations and performs shortcut optimizations for the $\lor, \land, \neg$ connectives.
This is especially useful when our QE algorithm is successful on closed formulas: since no more variables are present, the formula becomes a collection of constant polynomials atoms and connectives which means the whole formula can be reduced to either \isa{TrueF} or \isa{FalseF} with this simplification.

This constant identification and constant folding is also crucial for effiency within the QE procedure to cut down the expansion of the formula.
Recall that we transform QE problems to have the form 
\begin{equation}\label{eq:transformedformat}\bigvee \exists z_0.\cdots \exists z_n.\exists x. \left(\left(\bigwedge A_i\right) \land \left(\bigwedge \forall y. F_j\right) \right),
\end{equation} 
where the $A_i$ are atoms and the $F_i$ are formulas (see \rref{sec:reach}).
Now consider a QE problem of the form $\exists x. (ax^2+bx+c = 0 \land F)$.
In the event that at least one of $a$, $b$, or $c$ is a nonzero constant polynomial, we can immediately determine the \emph{only} possible values of $x$ in the whole formula, which are just the roots of this specific polynomial.
As such, in both our general QE and equality QE algorithms, for each disjunct in a formula of the form of \rref{eq:transformedformat}, our algorithm performs a linear scan for these ``lucky'' atoms within the conjunct list of atoms before proceeding with the VS algorithm.
If a lucky atom is found, we immediately eliminate its associated quantifier and then proceed with other steps of the algorithm.
This optimization demonstrates significant experimental benefits, as it eliminates quantifiers without greatly increasing the size of the formula.
It also utilizes the DNF form optimizations discussed in Section \ref{sec:reach} to reach underneath existential quantifiers to find more ``lucky'' atoms.
It could be further adapted in the future to, for example, allow for simplifications underneath universal quantifiers.

Even better than the ``lucky'' atoms, where a single coefficient is constant, are ``luckiest'' atoms, in which all coefficients are constants: with luckiest atoms, we are substituting real numbers rather than polynomials.
We found that ``luckiest" atoms are empirically very significant (identifying and cleverly utilizing them yields significant speedup), so our algorithms preprocess input formulas for these kinds of atoms and perform virtual substitution on them first.

\subsection{Variable Ordering Heuristics}\label{app:VariableOrdering}
Variable ordering heuristics are of great practical significance in QE \cite{DBLP:journals/corr/abs-1804-10037}.
We proved that one can freely swap the ordering of quantifiers in a homogeneous \emph{block} (a number of existential or of universal quantifiers that occur in a row) without affecting a formula's truth value.
To capitalize on this, our \isa{groupQuantifiers} function locates instances of homogeneous blocks in a formula and converts them into an equivalent block quantifier representation; for example, \isa{ExQ (ExQ F)} is converted into \isa{ExN 2 F} (see \rref{sec:representation}).
This allows later algorithms to identify blocks by pattern matching.
When the VS algorithm reaches a block \isa{ExN n F}, it uses the modified DNF algorithm on \isa{F} and focuses on a single disjunct, at which point we invoke a heuristic function to choose which of the quantifiers in the innermost block to eliminate first.

Since DNF yields a disjunct list of conjuncts, the input information to the heuristic function (at each disjunct) is a conjunct list of atoms \isa{L} and formulas \isa{F}.
As such, a heuristic function \isa{H} is of the type \isa{H\ {\isacharcolon}{\kern0pt}{\isacharcolon}{\kern0pt}\ nat\ {\isasymRightarrow}\ atom\ list\ {\isasymRightarrow}\ atom\ fm\ list\ {\isasymRightarrow}\ nat}, where \isa{H\ (N-1)\ L\ F} analyses the input conjuncts \isa{L} and \isa{F} and determines the optimal variable ranging from \isa{0} to \isa{N-1} to eliminate on (remember that we use 0-indexing for variables).
After selecting a variable \isa{var} to eliminate, we swap \isa{var} with variable \isa{N-1}, perform VS on the newly reindexed \isa{var}, optimize the result, and then run DNF again.
We then recursively proceed on each new disjunct until we have attempted to eliminate every quantifier.

The only property that the heuristic function must satisfy for correctness purposes is that it must suggest a variable within the block of interest; this allows users to create their own heuristics without incurring significant proof burden.
Verifying this property is trivial for heuristics that check their result to disallow results outside of the desired range.

We implement three heuristic functions: one of our own design, one based on the literature, and the identity heuristic.
The identity heuristic always returns the innermost variable, which yields experimental results comparable to an earlier version of the framework which did not implement block quantifiers; this demonstrates that we do not incur significant overhead by supporting block quantifiers.
We also implement a heuristic based on Brown's heuristic for quantifier ordering for CAD, as presented in \cite{DBLP:journals/corr/abs-1906-01455}; this yielded promising experimental results.
Lastly, the heuristic that we designed both chooses which variable to eliminate first and also chooses which VS algorithm to use at each step; this heuristic was the most experimentally successful of the three.
Overall, our results indicate that block quantifiers do not introduce significant overhead and confirm that variable ordering heuristics are of practical significance in virtual substitution.
\fi
\end{document}